الجمهورية الجزائرية الديمقراطية الشعبية
وزارة التعليم العالي والبحث العلمي

**University Ferhat Abbas Setif 1**
**Faculty of Natural and Life Sciences**

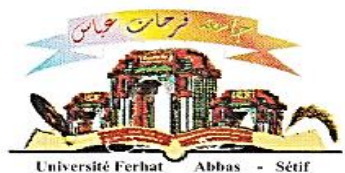

جامعة فرحات عباس سطيف 1
كلية علوم الطبيعة والحياة

DEPARTEMENT OF BIOCHIMESTRY                N°…………/SNV/**2020**

# THESIS

Presented by

# KARBAB Ahlem

For the fulfillment of the requirements for the degree of

## 3rd CYCLE DOCTORATE

**Biology**

**Special filed: Biochemistry**

## TOPIC

### EXTRACTION, ISOLATION, STRUCTURE ELUCIDATION AND EVALUATION OF TOXICITY, ANTI-INFLAMMATORY AND ANALGESIC ACTIVITY OF *PITURANTHOS SCOPARIUS* CONSTITUENTS

**Presented publically in : *23 /12 / 2020***

**JURY:**

| | | |
|---|---|---|
| **Chair:** | **KHENNOUF Seddik** | Pr. UFA Sétif 1 |
| **Supervisor:** | **CHAREF Noureddine** | Pr. UFA Sétif 1 |
| **Co-Supervisor:** | **ABU-ZARGA Musa H.** | Pr. The University of Jordan, Amman |
| **Examiners:** | **ARRAR Lekhmici** | Pr. UFA Sétif 1 |
| | **GHERRAF Noureddine** | Pr. University of Oum El Bouaghi |
| | **ZELLAGUI Amar** | Pr. University of Oum El Bouaghi |

***Laboratory of applied biochemistry***

بسم الله الرحمن الرحيم

قالوا

سبحانك لا علم لنا إلا ما علمتنا

إنك أنت العليم الحكيم

سورة البقرة : الآية 31

صدق الله العظيم

# Dedication
# And
# Acknowledgements

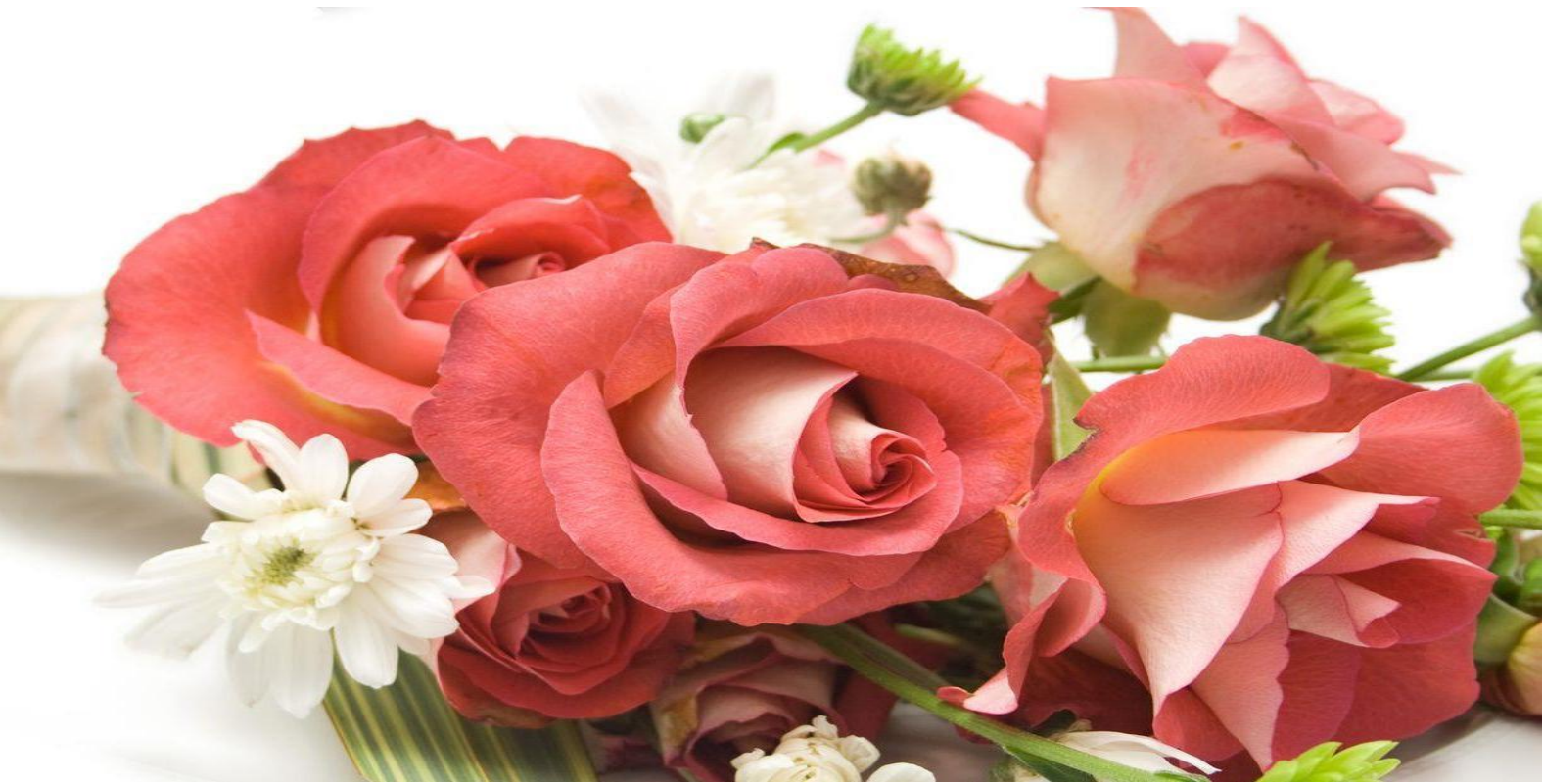

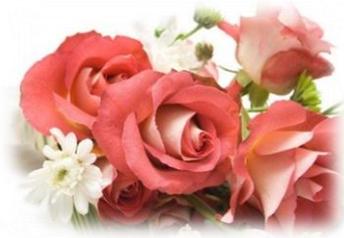

# **DEDICATION**

My very special thanks and dedicate this work should go to the one person to whom I owe everything I am today, my father **Saïd**, and to my dear mother **Aziza** for her helpful advice, care, and affection.

**I dedicate this work to:**

**My** soul mates **Grandfathers: TOUIL Mokhtar** and **KARBAB Messoud**

**My** dear**s Grandmothers: BALLOUT Djannet** and **DOUIBI Halima**

**M**y best brother **El Mouatez Billah**, his wife **Fatima** and their daughters: **Yousr**, **Youmn**, **Mozen** and **Guaim**

**M**y best sister **Rima**, her husband **Hamza** and children **Noussaiba and Abdallah**

**M**y soul mate brother **Hamza** and my best sisters **Manar** and **Mountaha**

**To** all my family: **KARBAB** and **TOUIL**, near or far

**M**y bests and soul mate teachers **Djamila** and **Ezzia**

**M**y best friends **LAWI Kanza**, **AMARI Salima**, **Samira**, **Intisar and Nadjet** who has listened to my complaining and supported me every step of the way in **Algeria**.

**I** am extremely grateful to my dearest friends **Razan**, **Rola**, **Areedj**, **Chorouk**, **Wafaa**, **Souad** and all my best friends they were always on my side during difficult situations and together we shared wonderful moments in **Jordan**.

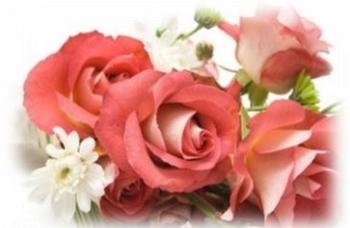

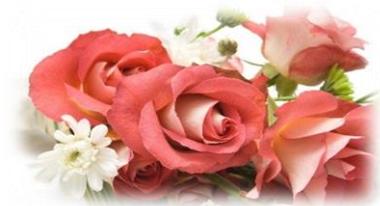

# ACKNOWLEDGEMENTS

Alhamdulillah, all praises to Allah for the strengths and His blessing in completing this thesis.

I would like to express my special appreciation goes to my supervisor **Prof. CHAREF Noureddine**, who has supported me throughout my thesis with his knowledge and his constructive comments and suggestions. His guidance, valuable advice, continuous encouragement, availability, kindness, and fruitful supervision helped me during the whole time of research and writing up the results of this work not only in the scientific matter but also in everyday life.

I would like to extend huge, warm thanks to my co-supervisor **Prof. ABU-ZARGA Musa. H.** faculty of science, The University of Jordan, Amman, for his great support, encouragement, kindness, help and his guidance for isolation, identification and structural elucidation of the pure compounds and for help and provision of research facilities in his laboratory.

I want to offer my sincere gratitude for my committee and their readiness and willingness to read thesis. Special thanks to **Prof. KHENNOUF Seddik**, **Prof. ARRAR Lekhmici, Prof. GHERRAF Noureddine** and **Prof. ZELLAGUI Amar**, for having agreed to participate in the thesis committee, critical reading of the thesis, and for their valuable comments, discussions and helpful suggestions.

Special thanks are due to **Prof. MUBARAK Mohammad S.** for his continuous support, kindness, assistance, advices and guidance during my research in Jordan.

Sincere appreciation to **Prof. ABU-ZARGA Musa. H., Prof. MUBARAK Mohammad, Prof. ARRAR Lekhmici, Prof. BAGUIANI Abderrahmane, Prof. KHENNOUF Seddik, Prof. DAHAMNA Saliha,** for help and provision of research facilities in their laboratories.

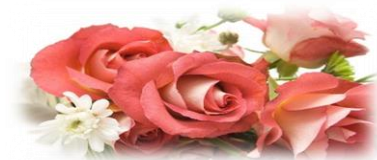

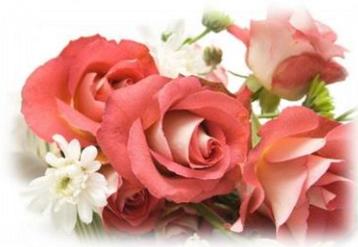

I want to offer my special thanks to **Dr. MOUKHNACHE Kamel** and **Dr. GUEMMAZ Thoria** for their valuable advices, continuous support, assistance, and to all other technical supports.

My special thanks to **Prof. ABDELOUCHE** and the entire team of the laboratory of Anapath, CHU of Setif, for their help in conducting clinical tests. I am also extremely grateful to **Dr. OUHIDA Soraya** for providing all the facilities to realise histological slides and interpreting them. I am also thankful to **Dr. DJABI Farida**, Head of Central laboratory of University Hospital, CHU of Setif and **Dr. Siham** and **Lina** for providing the facilities to evaluate biochemical parameters.

I'm very grateful to the Algerian Ministry of Higher Education and Scientific Research (MESRS), and Directorate General of Scientific Research and Technological Development (DGRSDT) who give me the opportunity and support for the exchange and development of my experiences during my research, one year outside my country (Jordan).

This work was carried out at the department of biochemistry, laboratory of applied biochemistry, University Ferhat Abbas, Setif-1 and the department of chemistry, laboratory of natural product, The University of Jordan, Amman.

I feel very fortunate for having the opportunity to work in this laboratory with excellent research facilities and surrounded by supporting people. I wish to thank all of them.

In the end, my warm thanks for the love and support provided by my friends and professors both in laboratory of applied biochemistry Setif-1, Algeria and laboratory of natural products of the University of Jordan-Amman and anyone who has meant more to me than I can ever put into word.

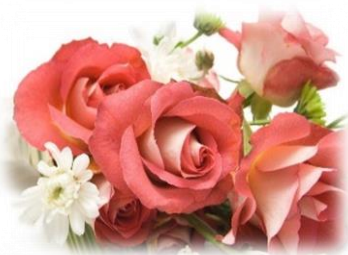

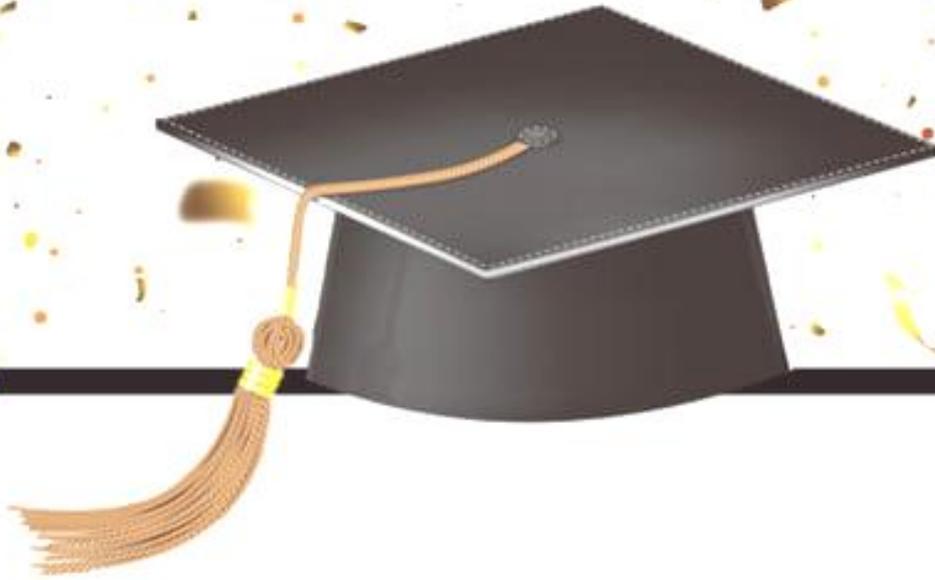

# LIST OF PUBLICATIONS

# List of publications

## I. *Presented research publications*

**KARBAB A**, CHAREF N, ABU-ZARGA M, QADRI M and MUBARAK MS. Ethnomedicinal documentation and anti-inflammatory effects of *n*-butanol extract and of four compounds isolated from the stems of *Pituranthos scoparius*: An in vitro and in vivo investigation. **Journal of ethnopharmacology**. 267; 113488, **2021**.

**KARBAB A**, MOKHNACHE K, OUHIDA S, CHAREF N, DJABI F, ARRAR L and MUBARAK MS. Anti-inflammatory, analgesic activity, and toxicity of *Pituranthos scoparius* stem extract: An ethnopharmacological study in rat and mouse models. **Journal of ethnopharmacology**. 258; 112936, **2020**.

**KARBAB A**, MOKHNACHE K, ARRAR L, BAGHIANI A, KHENNOUF S and CHAREF N. Fractionation, phytochemical screening and free radical scavenging capacity of different sub-fractions from *Pituranthos scoparius* roots. **Journal of drug delivery & therapeutics**. 10(3); 133-136, **2020**.

**KARBAB A**, MOKHNACHE K, ARRAR L and CHAREF N. Total phenolic contents and antioxidant capacity of aqueous extract from *Pituranthos scoparius* (coss. & dur.) Growing in Algeria. **Journal of drug delivery & therapeutics**. 10(3); 125-127, **2020**.

**KARBAB A**, CHAREF N and ARRAR L. Phenolic contents, *in vitro* antioxidant, and *in vivo* anti-inflammatory studies of aqueous extract from *Pituranthos scoparius* (Coss. & Dur.) growing in Algeria. **Iranian journal of pharmacology & therapeutics**. 17; 1-7, **2019**.

## II. *Collaborative research publications*

MOKHNACHE K, **KARBAB A**, SOLTANI E-K, BOUOUDEN W, OUHIDA S, ARRAR L, ESTEBAN MA, CHAREF N and MUBARAK MS. Synthesis, characterization, toxic substructure prediction, hepatotoxicity evaluation, marine pathogenic bacteria inhibition, and DFT calculations of a new hydrazone derived from isoniazid. **Journal of molecular structure**. 1221; 128817, **2020**.

MOKHNACHE K, **KARBAB A,** CHAREF N, ARRAR and MUBARAK MS. Synthesis, characterization, superoxide anion scavenging evaluation, skin sensitization predictions, and DFT calculations for a new isonicotinylhydrazide analog. **Journal of molecular structure**. 1180; 139-150, **2019**.

MOKHNACHE K, **KARBAB A**, MADOUI S, KHITHER H, SOLTANI E-K, CHAREF N and ARRAR L. Topical Anti-inflammatory evaluation of new hydrazone: 4-chloro-3-nitrobenzylidene) isonicotinohydrazide. **Journal of drug delivery & therapeutics**. 9(5-s); 1-3, **2019.**

MOKHNACHE K, **KARBAB A**, MADOUI S, KHITHER H, SOLTANI E-K and CHAREF N. Synthesis, characterization, in vivo acute toxicity and

# ABSTRACTS

**ملخص:**

القزاح هو نبات طبي يستخدم في الطب التقليدي في الجزائر ودول شمال إفريقيا الأخرى لعلاج العديد من الأمراض مثل الربو والروماتيزم والحصبة والأمراض الجلدية واليرقان واضطرابات الجهاز الهضمي. يهدف هذا العمل إلى التحقيق في مسح عرقي نباتي حول *Pituranthos scoparius* وتقييم السمية، ومضادات الالتهاب (في المختبر وفي الكائن الحي) المحتملة في مضادات الأكسدة والتأثيرات المسكنة لسيقان وجذور *Pituranthos scoparius*. علاوة على ذلك، لعزل وتوضيح المكونات الكيميائية لمستخلص البيوتانول من سيقان *P. scoparius* (ButE) وتحديد السمية والتأثيرات المضادة للالتهابات لهذه المركبات المضافة إلى ButE. أظهرت بيانات من دراسة عرقية دوائية أن 24.47 % من الأشخاص استخدموا هذا النبات في الطب الشعبي التقليدي. أظهرت النتائج التي تم الحصول عليها من التحليل الكيميائي النباتي النوعي لمستخلصات الساق (SE)، ومستخلصات الجذور (RE)، و ButE وجود البوليفينول، والفلافونيد، والعفص، والكينون الحر في كل من SE ، و RE ، و ButE ، بينما تم العثور على القلويدات والكومارين في ButE. من الناحية الكمية، أظهرت نتائجنا أنه تم العثور على أكبر كمية من إجمالي محتويات البوليفينول والفلافونيد والعفص في مستخلص الإيثيل أسيتات (EaE) لجزء السيقان بـ 434.34 ± 2.75 ميكروغرام مكافئ حمض الغاليك، 207.49 ± 1.03 ميكروغرام من مكافئ كيرسيتين و126.32 ± 1.32 ميكروغرام مكافئ حمض التانيك / ملغ مستخلص جاف، على التوالي. أربعة مركبات، يعرف بأنه نادر باسم-β-(2→1) isorhamnetin-3-O-β-apiofuranosyl glucopyranoside، بالإضافة إلى ثلاثة مركبات معروفة، وهي : isorhamnetin-3-O-β-glucoside و D-mannitol و isorhamnetin-3-O-β-glucopyranosyl- (1→6)-β-glucopyranoside تم عزلها من ButE. تم تمييز هذه المركبات عن طريق الرنين المغناطيسي النووي وبيانات الطيف الكتلي عالية الدقة (HRMS). لم تسبب المستخلصات الخامة (CrEs) للسيقان والجذور أي وفيات أو تغيرات في سلوك الحيوانات المعالجة؛ تم العثور على قيم $LD_{50}$ أعلى من 5 غ / كغ من وزن الجسم. النتائج من مضادات الأكسدة في المختبر تسلط الضوء على خصائص مضادات الأكسدة الجيدة لمستخلصات مختلفة من السيقان والجذور في اختبارات النماذج المختلفة. أظهرت النتائج من النشاط المضاد للالتهابات في المختبر أن المستخلصات الخام، ButE، والمركبات الأربعة أظهرت مستوى معنويًا من التثبيط ($p < 0.05$) مع التركيز المعتمد (0.5، 1، و 2 مغ / مل). أنتج الإعطاء الفموي لـ CrEs بجرعات 100 و300 و600 مغ / كغ تأثير تثبيط كبير يعتمد على الجرعة في كل من ذمة الأذن الناتجة عن زيلين وزيت كروتون في الفئران. أظهرت إدارة CrEs بجرعة 100 و250 و500 مغ/ كغ بشكل كبير ($p < 0.05$) تأثيرًا مضادًا للتورم في وذمة مخلب الجرذ التي يسببها الكاراجينان بعد 3 ساعات. في نموذج التلوي الناجم عن حمض الأسيتيك، قللت CrEs بشكل كبير ($p < 0.05$) من التلوي بجرعة 500 مغ / كغ. أظهر التأثير الموضعي المضاد للالتهابات أن المركبات الأربعة المعزولة، بالإضافة إلى ButE ، تظهر تأثيرًا كبيرًا ($p < 0.05$) يعتمد على الجرعة (0.5 و1 مغ / أذن) باستخدام وذمة الأذن التي يسببها زيت كروتون في الفئران. أظهرت النتائج من السمية الخلوية في المختبر أن تحلل % كلا من ButE، جنبًا إلى جنب مع المركبات المعزولة، وجد أنها غير سامة. من خلال الدراسة العرقية الطبية، أثبتت النتائج التي توصلنا إليها تؤكد الاستخدام الطبي لـ *P. scoparius* في الطب التقليدي وكمصدر إضافي للعوامل الطبيعية المضادة للالتهابات.

**الكلمات المفتاحية**: *Pituranthos scoparius* ، دراسة عرقية ، سمية ، مركبات طبيعية ، نشاط مضاد للأكسدة ، نشاط مضاد للالتهابات ، تأثير مسكن ، بوليفينول.

## ABSTRACT

*Pituranthos scoparius* is a medicinal plant that is used in traditional medicine in Algeria and other North African nations to treat several diseases such as asthma, rheumatism, measles, dermatoses, jaundice, and digestive disorders. The present work aimed to investigate an ethnobotanical survey about *Pituranthos scoparius* and assess the toxicity, anti-inflammatory (*in vitro*, and *in vivo*) potential, *in vitro* antioxidant, and analgesic effects of stems and roots of *Pituranthos scoparius*. Furthermore; to isolate and elucidate the chemical constituents of the *n*-butanol stem extract of *P. scoparius* (ButE) and determine the toxicity and anti-inflammatory effects of these compounds added to the ButE. Data from an ethnopharmacological study showed that 24.47 % of people used this plant in folk medicine. Results obtained from the qualitative phytochemical analysis of stem extracts (SE), roots extracts (RE), and ButE revealed polyphenols, flavonoids, tannins, and free quinones in both SE, RE, and ButE, whereas alkaloids and coumarins were present in the ButE. Quantitatively, our results showed that the most important highest amount of total polyphenols, flavonoids, and tannins contents were found in ethyl acetate extract (EaE) of stems with 434.34 ± 2.75 μg gallic acid equivalent; 207.49 ± 1.03 μg quercetin equivalent, and 126.32 ± 1.32 μg tannic acid equivalent/ mg dried extract, respectively. Four compounds, identified as the rare isorhamnetin-3-O-β-apiofuranosyl (1→2)-β glucopyranoside, in addition to three well known compounds, namely isorhamnetin-3-O-β-glucoside, D-mannitol, and isorhamnetin-3-O-β-glucopyranosyl-(1→6)-β-glucopyranoside were isolated from ButE. These compounds were characterized by means of NMR and high-resolution mass spectral (HRMS) data. The crude extracts (CrEs) of stems and roots did not cause any deaths or changes in the behavior of treated animals; $LD_{50}$ values were found to be higher than 5 g/kg body weight (BW). Results from the *in vitro* studies highlight good antioxidant proprieties of different extracts from stems and roots in various models. The results from the *in vitro* anti-inflammatory activity showed that crude extracts, ButE, and the four compounds exhibited a significant level inhibition of protein ($p < 0.05$) with dependent concentration (0.5, 1, and 2 mg/mL). Oral administration of CrEs at the doses of 100, 300, and 600 mg/kg produced a significant dose-dependent inhibition effect in both xylene and croton oil-induced ear edema in mice. Administration of CrEs at a dose of 100, 250, and 500 mg/kg significantly ($p < 0.05$) exhibited an anti-edematogenic effect in the carrageenan-induced rat paw edema after 3h. In the acetic acid-induced writhing model, CrEs significantly ($p < 0.05$) reduced writhing at a dose of 500 mg/kg. The topical anti-inflammatory effect showed that the four isolated compounds, as well as ButE, exhibit a significant ($p < 0.05$) dose-dependent (0.5 and 1 mg/ear) anti-inflammatory effect using croton oil-induced ear edema in mice. Results from in vitro cytotoxicity studies showed that the % lysis of the ButE, along with isolated compounds, was found to be virtually non-toxic. Through ethnomedicinal study, our findings prove the medicinal use of *P. scoparius* in traditional medicine and as an additional source for natural anti-inflammatory agents.

**Keywords:** *Pituranthos scoparius*, ethnomedicinal study, toxicity, natural compounds, antioxidant activity, anti-inflammatory activity, analgesic effect, polyphenols.

**RESUME**

*Pituranthos scoparius* est une plante médicinale utilisée en médecine traditionnelle en Algérie et dans d'autres pays d'Afrique du Nord pour traiter plusieurs maladies telles que l'asthme, les rhumatismes, la rougeole, les dermatoses, la jaunisse et les troubles digestifs. Le présent travail visait à enquêter sur une enquête ethnobotanique sur *Pituranthos scoparius* et à évaluer la toxicité, le potentiel anti-inflammatoire (*in vitro* et *in vivo*) des effets antioxydants in vitro et analgésiques des tiges et des racines de *Pituranthos scoparius*. De plus, pour isoler et élucider les constituants chimiques de l'extrait de tige *n*-butanol de *P. scoparius* (ButE) et déterminer la toxicité et les effets anti-inflammatoires de ces composés ajoutés au ButE. Les données d'une étude ethnopharmacologique ont montré que 24.47% des personnes utilisaient cette plante en médecine traditionnelle populaire. Les résultats obtenus à partir de l'analyse phytochimique qualitative d'extraits de tige (SE), d'extraits de racines (RE) et de ButE ont révélé des polyphénols, des flavonoïdes, des tanins et des quinones libres à la fois dans SE, RE et ButE, alors que les alcaloïdes et les coumarines ont été trouvés dans le ButE. Quantitativement, nos résultats ont montré que la quantité la plus importante de polyphénols totaux, de flavonoïdes et de tanins était trouvée dans l'extrait d'acétate d'éthyle (EaE) de la partie des tiges avec 434.34 ± 2.75 µg d'équivalent d'acide gallique; 207.49 ± 1.03 µg d'équivalent quercétine et 126.32 ± 1.32 µg d'équivalent acide tannique / mg d'extrait séché, respectivement. Quatre composés, identifiés comme le rare isorhamnetin-3-O-β-apiofuranosyl (1 → 2) -β glucopyranoside, en plus de trois composés connus, à savoir l'isorhamnetin-3-O-β-glucoside, le D-mannitol et l'isorhamnetine-3-O-β-glucopyranosyl-(1 → 6)-β-glucopyranoside a été isolé à partir de ButE. Ces composés ont été caractérisés au moyen de données RMN et spectrales de masse à haute résolution (HRMS). Les extraits bruts (CrE) de tiges et de racines n'ont causé aucun décès ni changement de comportement des animaux traités; Les valeurs de $DL_{50}$ étaient supérieures à 5 g / kg de poids corporel. Les résultats de l'antioxydante in vitro mettent en évidence les bonnes propriétés antioxydantes de différents extraits de tiges et de racines dans divers modèles de tests. Les résultats de l'activité anti-inflammatoire in vitro ont montré que les extraits bruts, ButE et les quatre composés présentaient un niveau significatif d'inhibition ($p < 0.05$) avec une concentration dépendante (0.5, 1 et 2 mg / mL). L'administration orale de CrE à des doses de 100, 300 et 600 mg / kg a produit un effet d'inhibition dose-dépendant significatif dans l'œdème de l'oreille induit à la fois par le xylène et l'huile de croton chez la souris. L'administration de CrE à une dose de 100, 250 et 500 mg / kg de manière significative ($p < 0.05$) a montré un effet anti-œdématogène dans l'œdème de patte de rat induit par la carraghénane après 3 h. Dans le modèle de contorsions induites par l'acide acétique, les CrEs réduisaient significativement ($p < 0.05$) les contorsions à une dose de 500 mg / kg. L'effet anti-inflammatoire topique a montré que les quatre composés isolés, ainsi que le ButE, présentent un effet anti-inflammatoire dose-dépendant significatif ($p < 0.05$) (0.5 et 1 mg / oreille) en utilisant un œdème auriculaire induit par l'huile de croton chez la souris. Les résultats de la cytotoxicité in vitro ont montré que le % de lyse du ButE, ainsi que des composés isolés, s'est avéré pratiquement non toxique. Grâce à une étude ethnomédecine, nos résultats prouvent l'utilisation médicinale de *P. scoparius* en médecine traditionnelle et comme source supplémentaire d'agents anti-inflammatoires naturels.

**Mots clés:** *Pituranthos scoparius*, étude ethnomédecine, toxicité, composés naturels, activité antioxydante, activité anti-inflammatoire, effet analgésique, polyphénols.

# LIST OF FIGURES AND TABLES

# LIST OF FIGURES

# LIST OF TABLES

# ABBREVIATIONS

# ABBREVIATIONS

| | |
|---|---|
| **5HETE** | 5-Hydroxyeicosatetraenoic acid. |
| **5-LOX** | Inhibit 5-lipoxygenase |
| **AA** | Arachidonic acid |
| **ABTS** | 2,2'-azino-bis(3-ethylbenzothiazoline-6-sulfonic acid |
| **ALP** | Alkaline phosphatise |
| **ALT** | Alanine aminotransferase |
| **AqE** | Aqueous extract |
| **AST** | Aspartate aminotransferase |
| **BHT** | Butylated hydroxytoluene |
| **ButE** | *n*-butanol stems extract |
| **CAT** | Catalase |
| **CC** | Column chromatography |
| **ChE** | Chloroform extracts |
| **CHU** | Hospitalo-universitary centre |
| **COX** | Cyclooxygenase |
| **CrE** | Crude extract |
| **DE** | Dried extract |
| **DecE** | Decoction extract |
| **Dich** | Dichloromethane |
| **Dicl** | Diclofenac |
| **DPPH** | 1,1-diphenyl-2 picrylhydrazyl |
| **EaE** | Ethyle acetate extracts |
| **ESI** | Electrospray ionization |
| **GE** | Gallic acid equivalent |
| **GPx** | Glutathione peroxidase |
| **GRx** | Glutathione reductase |
| **GSH** | Glutathione |
| **HRMS** | High-resolution mass spectra |
| **IFBDO** | International federation of blood donor organizations |
| **IL-10** | Interleukin-10 |
| **Indo** | Indomethacin. |
| **$LD_{50}$** | Lethal dose 50 |
| **LDH** | Lactate dehydrogenase |
| **LOX** | Lipoxygenase |
| **LPS** | Lipopolysaccharides |
| **NF-κB** | Kappa-B nuclear factor |
| **NO** | Nitric oxide |
| **NOS** | Nitric oxide synthase |
| **NQO1** | Quinone oxidoreductase 1 |
| **NSAIDS** | Non-steroidal anti-inflammatory drugs |
| **OECD** | Organisation for economic co-operation and development |
| **P** | Peliosis |
| **PG** | Prostaglandins |
| **$PGE_2$** | Prostaglandin $E_2$ |
| **$PGE_2α$** | Prostaglandin $E_2$ alpha |
| **PKC** | Protein kinases C |
| **Prxs** | Peroxiredoxins |
| **PSB** | *n*-Butanol extract fraction |
| **PSB1** | Isorhamnetin-3-*O*-β-glucoside |
| **PSB2** | Isorhamnetin-3-O-β-apiofuranosyl (1→2)-β glucopyranoside |
| **PSB3** | D-mannitol |
| **PSB4** | Isorhamnetin-3-O-β-glucopyranosyl-(1→6)-β-glucopyranoside |
| **QE** | Quercetin equivalent |
| **R** | Roots |
| **RBCs** | Red blood cells |
| **RE** | Roots extracts |
| **RNS** | Reactive nitrogen species |
| **ROS** | Reactive oxygen species |
| **ROW** | Relative organ Weight |
| **S** | Stems |
| **SAR** | Structure-activity relationship |
| **SD** | Standard deviation |
| **SE** | Stem extracts |
| **SOD** | Superoxide dismutase |
| **TC** | Tannins contents |
| **TCA** | Trichloroacetic acid |
| **TE** | Tannic acid equivalent |
| **TLC** | Thin layer chromatography |
| **TFC** | Total flavonoids contents |
| **TPA** | 12-*O*-tetradecanoyl-13-phorbol acetate |
| **TPC** | Total phenolic contents |
| **VC** | Vascular congestion |
| **VEGF** | Vascular endothelial growth factor |

# LIST OF CONTENTS

## MATERIALS AND METHODS

# INTRODUCTION

# INTRODUCTION

Plants have served humankind for centuries as an essential source of food and medicine. Studies related to plants or plant products for medicinal purposes are gaining attention because many of these products have a strictly local use and may propose an alternative treatment for several diseases (Malheiros et al., 2017). With the evolution of advanced technology, several research papers dealing with medicinal plants' inhibitory effect and isolated natural compounds on inflammation processes have been reported (Malheiros et al., 2017; Zhu et al., 2018). Many plants, particularly medicinal plants with pharmacological properties, have been extensively studied for their antioxidant activity and shown to be rich sources of bioactive compounds with the significant potential to prevent incurable diseases (Do et al., 2014). Investigating medicinal plants for their biological properties as sources of bioactive components is crucial, and the first step in developing effective alternative medications (Singh et al., 2017). Besides, phytochemicals from plants used as pure compounds, crude, or fractionated extracts have been heavily studied for their anti-inflammatory effects (Zhang et al., 2019). In this context, natural products with anti-inflammatory activity have long been used as a folk remedy for inflammatory conditions such as fevers, pain, migraine, and arthritis. They can act by different mechanisms of action, including inhibition of inflammatory mediators' synthesis or expression (Yuan et al., 2006). Additionally, natural products can decrease skin, joint, cardiovascular, lung, neuro, and gastrointestinal inflammation (Maione et al., 2016). For example, quercetin, allicin, terpenes, polyphenols, and flavonoids (Howes, 2018) have been recognized as anti-inflammatory agents. Unfortunately, many plants' toxicological effect, particularly medicinal plants, is not well established (Mordenis, 2019).

As a result, to acquire effective plant-based therapeutic agents for specific diseases or disorders, these agents' toxicological and pharmacological profiles need a thorough evaluation (Olarenwaju et al., 2018).

In the rural populations of tropical Africa, high cost, less access to modern synthetic drugs, and side effects forced many people to depend on traditional herbal medicines. Along this line, the quest for new antioxidant, anti-inflammatory, and analgesic drugs with lesser side effects has drawn people's attention in the health profession, dieticians, scientists, and medicinal chemists all over the world.

In this context, the present research investigates one of the medicinal plants used for various therapeutic purposes in Algerian folk medicine, which is *Pituranthos scoparius*, commonly known as "Guezzah". It belongs to Apiaceae's family, and it is an endemic plant of North Africa and is widespread in Algeria. This plant attracted our attention because it has been used in traditional medicine to treat numerous diseases such as asthma and rheumatism, digestive disorders, measles, postpartum care, jaundice, and firefighting indigestion, hepatitis, spasms, pains, diabetes, and urinary infections. Furthermore, it is used in food as a flavoring agent.

Based on the preceding discussion, the present work aimed to evaluate the ethnopharmacology study, phytochemicals constituents, toxicity profile, and pharmacological activities of *Pituranthos scoparius* (stems and roots). The current study was undertaken on the following objectives:

- ➢ Evaluate the ethnopharmacological survey about *Pituranthos scoparius* from the area of Setif, especially Djebel Zdimm and adjoining areas
- ➢ Extraction procedures, fractionation, purification along to isolation of the chemical constituents from the extract of the n-butanol stem
- ➢ Qualitative and quantitative of phytochemical components of different extracts

- Structure elucidation and identification of isolated compounds from *n*-butanol extract
- An *in vitro* study of antioxidant activity of CrE (stems and roots) and its fractions: 1,1-diphenyl-2 picrylhydrazyl (DPPH) and 2,2'-azino-bis(3-ethylbenzothiazoline-6-sulfonic acid (ABTS) radical scavenging activities, reducing power, hydroxyl radical scavenging assay, β-carotene bleaching assay, and metals chelating activity
- An *in vitro* and *in vivo* of anti-inflammatory activity:
- *In vitro* anti-inflammatory of (CrE of stems and roots along with to *n*-butanol extract and isolated compounds) using the protein denaturation inhibition test
- *In vivo* anti-inflammatory of CrE of stems and roots using xylene-induced ear edema assay, croton oil-induced ear-edema, and carrageenan-induced rat paw inflammation models
- Topical anti-inflammatory activity of *n*-butanol extract along to the isolated compounds using croton oil-induced ear-edema test
- Evaluation of the analgesic activity of CrE from stems and roots by acetic acid-induced writhing in mice model
- An *in vitro* evaluation of toxicity for *n*-butanol extract and isolated compounds used hemolysis assay.
- Evaluation of acute oral toxicity of plant CrE of both stems and roots.

# REVIEW

# OF LITERATURE

# 1. Natural products

## 1.1. Definition

Natural products are organic compounds that are formed by living systems. They have been used for centuries as medicines to treat many diseases. Despite the development and growth in the conventional medicine industry, most of the world population still uses plants to answer their health problems (Judith et al., 2014). Also, plants played an important role in pharmacy and agriculture, and from these plants, new herbal and useful drugs were isolated and synthesized (Sharma and Cannoo, 2013). Natural products are sources of about 90% of newly discovered pharmaceuticals (Judith et al., 2014).

## 1.2. Types of natural products

Naturally occurring compounds may be divided into three broad categories:

> ➤ *Primary metabolites:* these are the compounds that occur in all cells and play an essential role in the metabolism and reproduction of those cells. These compounds include the nucleic acids, the standard amino acids, sugars and the high-molecular-weight polymeric materials that form cellular structures such as cellulose, the lignins, and the proteins.

> ➤ *Secondary metabolites:* which are those compounds that are characteristic of a limited range of species. In general, these compounds do not play a role in growth, development, and reproduction, but they may act as defence chemicals. They include alkaloids, terpenoids and phenolic compouds.

## 1.3. Phenolic compounds and classification

Phenolic compounds are secondary metabolites that plants produce to protect themselves from other organisms. Dietary polyphenols have been shown to play essential roles in human health. High intake of fruits, vegetables, and whole grains rich in polyphenols, has been linked to lowered risks of many chronic diseases including cancer, cardiovascular disease, chronic inflammation, and many degenerative diseases (Tsao, 2010). Phenols are a large group of secondary plant metabolites and can be classified into three major sub-groups:

**1.3.1. Phenolic acids**

Phenolic acids are non-flavonoid polyphenol compounds that can be further divided into two main types, benzoic acid and cinnamic acid derivatives based on C1-C6 and C3-C6 backbones (Figure 1). While fruits and vegetables contain many free phenolic acids, grains, and seeds-particularly in the bran or hull-phenolic acids, are often in the bound form (Chandrasekara and Shahidi, 2010). These phenolic acids can only be freed or hydrolyzed upon acid or alkaline hydrolysis, or by enzymes.

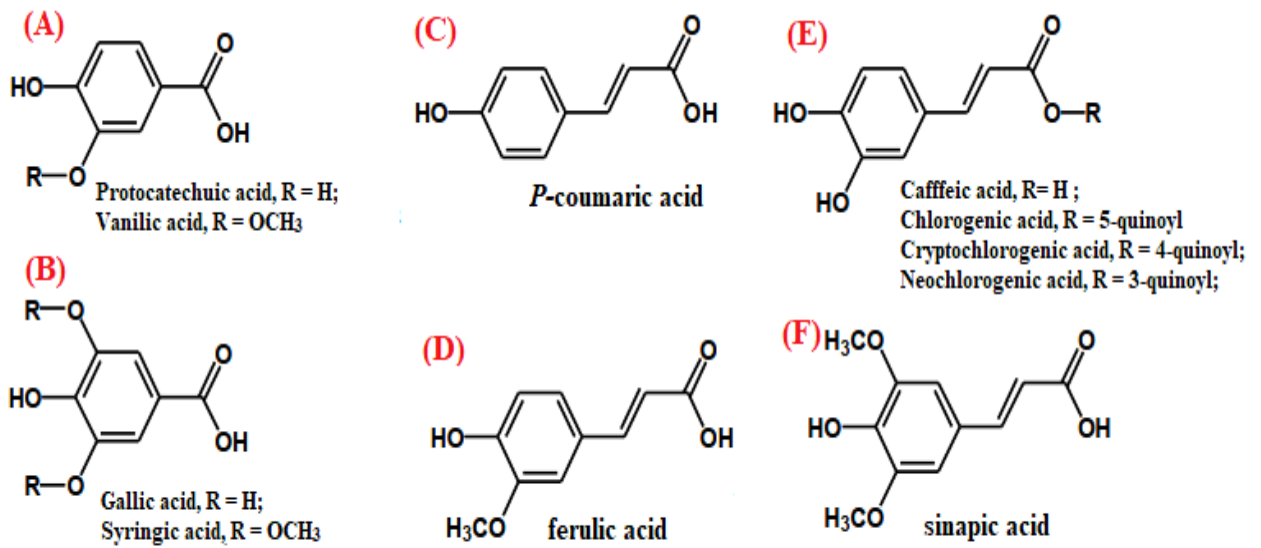

**Figure 1:** Typical phenolic acids in food: benzoic acids and cinnamic acids (Tsao, 2010).

**1.3.2. Flavonoids**

Flavonoids are ubiquitous in plants. One of the exciting questions remains why plants produce such various flavonoid metabolites-up to 10.000 across the plant kingdom (Mathesius, 2018). This special issue highlights some of the many functions that flavonoids have evolved to control plant development, plant-microbe interactions, and plant-animal interactions. Flavonoids have the C6-C3-C6 general structural backbone in which the two C6 units (Ring A and Ring B) are phenolic (Figure 2).

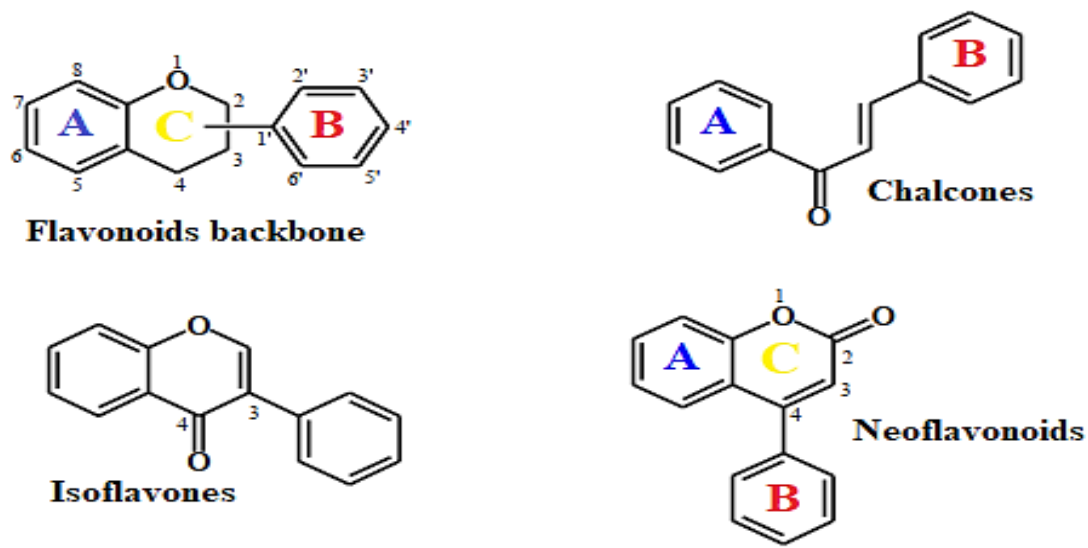

**Figure 2:** Basic structures of flavonoids (Tsao, 2010).

Flavonoids can be additionally divided into various sub-groups such as anthocyanins, flavan-3-ols, flavones, flavanones, and flavonols (Figure 3); due to the hydroxylation pattern and variations in the chromane ring (Ring C). While the vast majority of the flavonoids have their Ring B linked to ring C's C2 position, some flavonoids such as isoflavones and neoflavonoids, whose Ring B is associated to C3 and C4 positions of Ring C, sequentially, are similarly found in plants. Chalcones, though lacking the heterocyclic Ring C, are still classified as branches of the flavonoid family (Tsao, 2010).

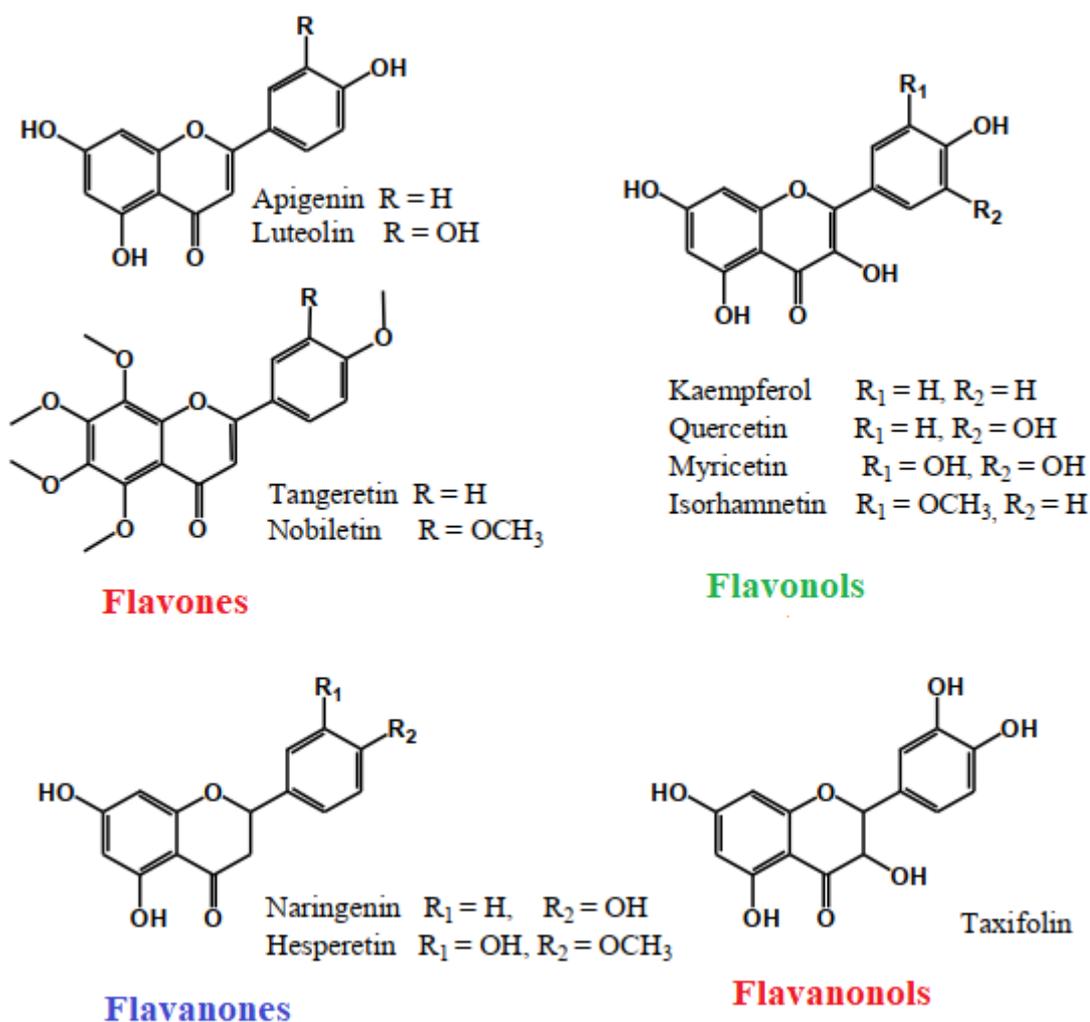

**Figure 3:** Structures of flavones, flavonols, flavanones and flavanonols

**1.3.3. Other polyphenols**

There are numerous non-flavonoid polyphenols found in foods that are considered essential to human health in addition to the phenolic acids, phenolic amides and flavonoids. For exemple, resveratrol is unique to the grapes and red wine; ellagic acid and its derivatives are obtained in berry fruits, strawberries and raspberries, and in the skins of diverse tree nuts. Additionally, lignans exist in the bound forms in flax, sesame and many grains; are hydrolysis products. Curcumin is a potent antioxidant from turmeric. Ellagic acid is a dimer of gallic acid and rosmarinic acid is a dimer of caffeic acid. Moreover, both ellagic acid and gallic acid are found in the free forms, their glucose esters, a group recognized as hydrolysable tannins, too exist in several plants (Tsao, 2010).

## 2. Biological roles and implications

Polyphenols exhibited different pharmacological properties which include: antibacterial activity (Mostafa et al., 2020), anticancer properties (Kikuchi et al., 2019), cardioprotective properties (Speer et al., 2019), antidiabetic properties (Sun et al., 2020), antioxidant activities (Stagos, 2020), anti-inflammatory properties (Abbasi-Parizad et al., 2020) and analgesic activities (Hossain et al., 2019).

### 2.1. Effect of polyphenols on oxidative stress

Oxidative stress leads to the extreme production of reactive oxygen species (ROS) and reactive nitrogen species (RNS) in the cells and tissues and endogenous defense systems cannot be able to neutralize those (Castelli et al., 2018). Imbalance in this protective mechanism can lead to the damage of cellular macromolecules such as DNA, proteins, and lipids (Burgos-Morón et al., 2019). Free radicals act a major part in the development of chronic and degenerative ailments such as cancer, autoimmune disorders, rheumatoid arthritis, cataract, aging, cardiovascular, neurodegenerative diseases and diabetes mellitus (Khettaf et al., 2016; Gowd et al., 2019). Fortunately, antioxidants may prevent and/or relieve oxidative stress-related diseases through delaying or reducing such oxidative damage. The endogenous defense systems are divided into two types:

- **Enzymatic antioxidants:** which maintain the oxidative equilibrium in the first line (Zhang and Tsao, 2016), such as quinone oxidoreductase 1 (NQO1), superoxide dismutase (SOD), catalase (CAT), glutathione peroxidase (GPx), peroxiredoxins (Prxs), and glutathione reductase (GRx).
- **Non-enzymatic antioxidants:** which maintain the oxidative equilibrium in the second line of defense against ROS, which comprise endogenous (metabolic) and exogenous (nutrient) antioxidants (Li et al., 2017).

Polyphenols are one of the most essential groups of exogenous natural antioxidants. The antioxidant roles of polyphenols are as follows:

- ✓ Neutralizing the free radicals via transfer of electron and/or hydrogen atom,
- ✓ Reducing the formation of metal-dependent hydroxyl radicals through chelation mechanism,
- ✓ Decreasing cell apoptosis through modulation of mitochondrial dysfunction, and evoking endogenous antioxidant enzymes (Wu et al., 2020).

**2.1.1. Structure-activity relationships of polyphenols**

Antioxidant activity may depends on the structure of the functional groups of polyphenols. The multiple of hydroxyl groups greatly influences several mechanisms of antioxidant activity such as scavenging radicals and metal ion chelation ability. Additionally, the position and the arrangement of the hydroxyl groups around the phenolic molecule is also important for anti-oxidative activity (Kassim et al., 2018). Polyphenolic compounds may react in plasma membrane with nonpolar compounds existing in the hydrophobic inner membrane layer; so changes in the membrane may influence oxidation degree of proteins or lipids. Some flavonoids in the hydrophobic nucleus of membrane may protect the structure, function of membrane and prevent the access of oxidants. These processes may help to know the basic mechanisms of action of polyphenols including cellular interaction and signal transduction (Hussain et al., 2016).

**2.2. Effect of polyphenols on inflammation**

Inflammation is a vital and complex bodily response to injury, infection, or destruction characterized by heat, pain, swelling, redness, and disturbed physiological functions (Chandra, 2012). It additionally involves an increase of protein denaturation, an increase of vascular permeability, and membrane alteration (Gunathilake et al., 2018). As a result, production of auto-antigens caused by denaturation of protein in some conditions of inflammation included rheumatic arthritis, diabetes, and cancer (Dharmadeva et al., 2018). However, prolongation of inflammation can lead to various diseases, including rheumatoid arthritis, psoriasis, and inflammatory bowel disease. The most commonly used drugs for management of inflammatory conditions are non-steroidal anti-inflammatory drugs (NSAIDs), which have several adverse effects, such as gastric irritation which could lead to the formation of gastric ulcers (Chandra et al., 2012). One of the important anti-inflammatory mechanisms is the inhibition of eicosanoids generating enzymes including phospholipase A2, lipoxygenase and cyclooxygenase thereby reducing the concentration of leukotrienes and prostanoids. Polyphenols can modulate the activity of arachidonic acid metabolizing enzymes such as cyclooxygenase (COX), lipoxygenase (LOX), and NOS. Inhibition of these enzymes reduces the production of arachidonic acid (AA), prostaglandins, leukotrienes, and NO, thus exerting an important anti-inflammatory action. Polyphenols inhibit arachidonic acid pathway is shown in Figure 4 (Hussain et al., 2016).

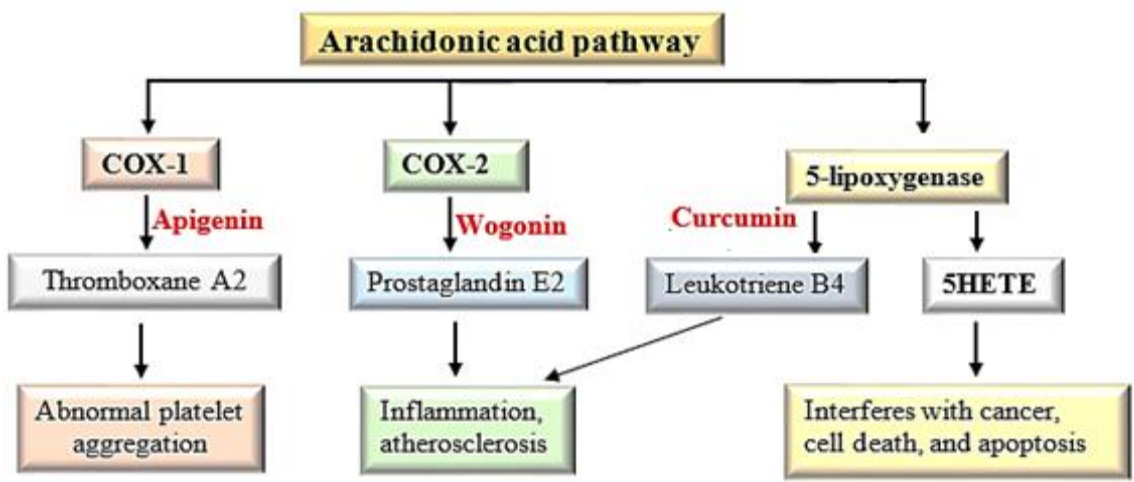

**Figure 4:** Metabolic pathways involved in arachidonic acid metabolism leading to inflammatory diseases. **5HETE:** 5-Hydroxyeicosatetraenoic acid.

However, the effects on the balance between pro- and anti-inflammatory cytokine expression have been shown to be specific for specific cytokines and influenced by polyphenol structures highlighting the complex action exerted by these compounds (Hussain et al., 2016)

Additionally, natural products can decrease skin, joint, cardiovascular, lung, neuro, and gastrointestinal inflammation (Maione et al., 2016). For example, quercetin, allicin, terpenes, polyphenols, and flavonoids (Howes, 2018) have been recognized as anti-inflammatory agents. Summarized in Table 1 are some selected reviews and reports related to the effects of natural products on inflammation pathways.

**Table 1:** Summary of some effects of natural products on inflammation pathways.

| Natural Products | Effects on Inflammation Pathways | Reference |
|---|---|---|
| Caffeic acid | Suppresses ultraviolet B radiation-induced expression of interleukin-10 (IL-10) and the activation of mitogen-activated protein kinases in mouse skin | (Aravindaram and Yang, 2010) |
| Catechin, epigallocatechin-gallate | Inhibit cyclooxygenase 2 ($COX_2$) expression induced by tumor promoters in rat skin, 12-*O*-tetradecanoylphorbol-13-acetate (TPA), and cyclooxygenase (COX) activity in macrophages induced by lipopolysaccharides (LPS) | (Kosala et al., 2018) |
| Quercetin, morin, kaempferol, and myricetin | Inhibit 5-lipoxygenase (5-LOX) activity | |
| Theophylline | Reduces colonic MPO activity and TNF-α, interleukin-1β (IL-1β), and interleukin-6 (IL-6) level in an inflamed colon | (Peng et al., 2019) |
| Caulerpin | Decreases the production of TNF-α, IFN-γ, IL-6, IL-17, increases the production of IL-10, and suppresses NF-κB p65 subunit activation in colon tissues | |
| Celastrol | Regulates protein kinase B, TNF-α, nuclear factor-*κ*B (NF-κB), COX-2, and vascular endothelial growth factor (VEGF) | (Wang and Zeng, 2019) |
| Asiaticoside | Inhibits pro-inflammatory mediators: TNF-α, IL-1β, IL-6, and $PGE_2$ | |

## 2.3. Relationships between oxidative stress and inflammation

Oxidative stress and inflammation are linked with several chronic diseases, including hypertension, diabetic complications, neurodegenerative diseases, cardiovascular diseases, chronic kidney disease, alcoholic liver disease, aging, and cancer (Liguori et al., 2018). Various studies support an interdependent relationship between inflammation and oxidative stress (Biswas et al., 2015; Burgos-Morón et al., 2019). During the inflammatory process, the activated phagocytic cells like macrophages and neutrophils produce large amounts of reactive nitrogen, ROS, and chlorine species, including superoxide, hydrogen peroxide, nitric oxide, hydroxyl free radical, hypochlorous acid, and peroxynitrite to kill the invading agents (Fialkow et al., 2007). There may be the excessive generation of reactive species under pathological inflammatory conditions, and some of those reactive species spread out of the phagocytic cells, and thus they can induce localized oxidative stress and tissue injury (Fialkow et al., 2007). Apart from the direct production of reactive species by the professional phagocytic cells, the nonphagocytic cells can produce reactive species in response to proinflammatory cytokines (Wu et al., 2013; Li et al., 2015).

Number of reactive species liberated by inflammatory cells at the site of inflammation leading to exaggerated oxidative stress. When oxidative stress appears as a primary disorder inflammation develops as a secondary disorder and further enhances oxidative stress. In a similar fashion, inflammation as a primary disorder can induce oxidative stress as a secondary disorder, which can further enhance inflammation (Figure 5). Furthermore, a number of reactive oxygen/nitrogen species can initiate intracellular signalling cascade that enhances proinflammatory gene expression (Biswas, 2015). For example, tripeptide glutathione (GSH) is an intracellular thiol antioxidant; lower level of this GSH causes higher ROS production, which results in imbalanced immune response, inflammation, and susceptibility to infection (Hussain et al., 2016).

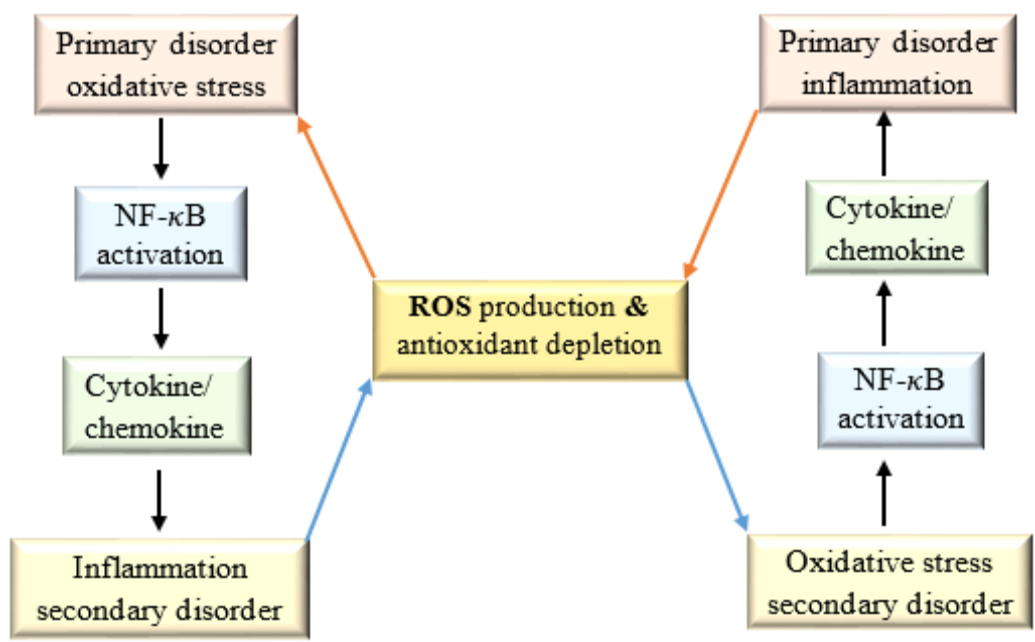

**Figure 5:** Overview of interdependence between oxidative stress and inflammation.

## 3. *Pituranthos scoparius*

*Pituranthos scoparius (*synonym**:** Deverra scoparia (Coss. & Dur.)), regionally named "guezzah" belongs to the Apiaceae family which thrives in North Africa (Figure 6). It is well known in Algeria, particularly in the hills of the Sahara (Chikhoune et al., 2017). This plant is largely used in food as a flavoring agent (Benmekhbi et al., 2008). The genus of *Pituranthos* includes more than 20 spices (Chahrazed et al., 2016). Flowering time is during spring, summer and autumn seasons from February to October (Adida et al., 2014).

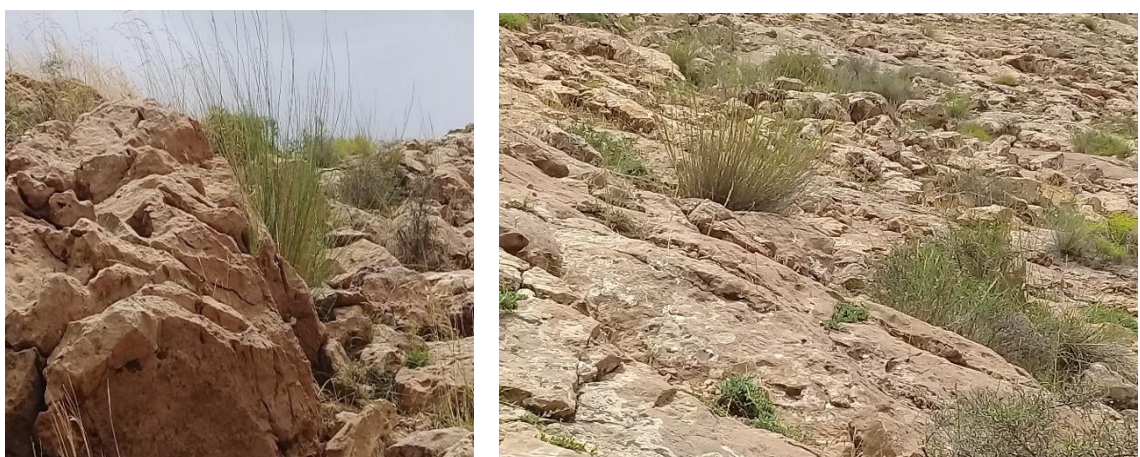

**Figure 6:** *Pituranthos scoparius* from region of Setif, Algeria

## 3.1. Systematics and classification

According to Quezel and Santa (1963), *Pituranthos scoparius* is classified as follows:

- **Kingdom :** Plantae
- **Division :** Spermaphytes
- **Sous-division :** Angiosperm
- **Class :** Eudicotyledons
- **Order :** Apiales
- **Family :** Apiaceae
- **Genus :** *Pituranthos*
- **Specie :** *Pituranthos scoparius*

## 3.2. Geographical spread

*Pituranthos scoparius* is a North African endemic species, and is widespread in Algeria, particularly in the high plateau of most parts of the Sahara. It is however observed very frequently on the plateau of Tassili, Ajjers and in the Hoggar, especially in the stony wadis lily (Benchelah et al., 2011).

## 3.3. Morphological representation

*P. scoparius* is an aphyllous perennial plant; the upper leaves are reduced to their sheath. The stems are erect, 40–80 cm high, and form dense clumps that send out laterally short rigid branches. This plant is characterized by stems in the form of rushes which are often much ramified, without leaves or nearly so, with small fruits of less than 3 mm (Lograda et al., 2013). Flowering with white flowers with lateral umbels and short peduncle. Its fruits are longer than wide, bristling with erect hairs (Figure 7). *Pituranthos scoparius* is a perennial plant, with yellowish stem, in tufts, branched at the top, simple and parallel to each other in their lower half, bearing lateral umbels; peduncles often short; white petals with narrow veins (Hammoudi et al., 2015).

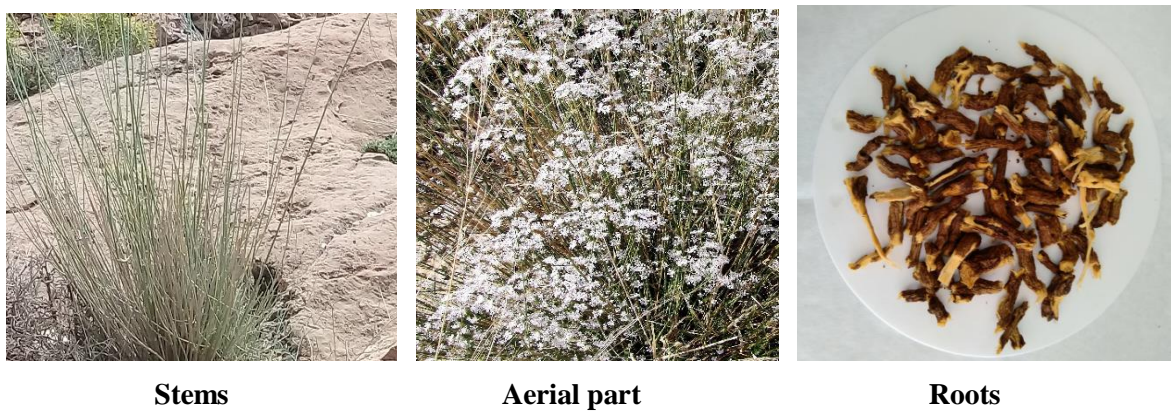

**Stems**          **Aerial part**          **Roots**

**Figure 7:** The different parts of *Pituranthos scoparius*: stems, aerial part and roots

## 3.4. Traditional uses and medicinal properties

*Pituranthos scoparius* has been traditionally used by local healers to cure ailments such as asthma and rheumatism (Benmekhbi et al., 2008; Adida et al., 2015), measles, jaundice, postpartum care, firefighting indigestion, sore stomach, and abdomen (Adida et al., 2015). Furthermore, it was used for headaches (Chikhoune et al., 2017), spasms, pains, diabetes, hepatitis, and urinary infections (Boudjelal et al., 2013). The dry stems of this plant are used in the preparation of powders used against reptile bites (Benchelah et al., 2011). A decoction of the aerial parts is used in treating digestive disorders (Boudjelal et al., 2013).

## 3.5. Reported biological activities

This genus is economically important because it has exhibited various biological and pharmacological activities including: antimicrobial (Benmekhbi et al., 2008; Adida et al., 2014), antioxidant (Adida et al., 2015; Chikhoune et al., 2017), anti-urolithiatic (Benalia et al., 2016), analgesic and anti-inflammatory properties (Harchaoui et al., 2018).

## 3.6. Chemical composition

Previous studies showed that the phytochemical analysis of ethyl acetate extract from *P. scoparius* roots revealed the isolation of two isocoumarins: 3-n-propyl-5-methoxy-6-hydroxy- isocoumarin and 3-n-propyl-5,7-dimethoxy-6-hydroxy isocoumarin (Haba et al., 2004). In addition, the butanolic extract of *Pituranthos scoparius* lead the isolation of five flavonoids identified as: apigenin-6,8-di-C-glucoside, isorhamnetin-3-*O*-glucoside, apigenin-7-*O*-glucoside, apigenin-7-*O*-rhamnoside, isorhamnetin-3-*O*-rutinoside (Benmekhbi et al., 2008). The extracts of the Deverra species studied, revealed that they contain biologically active secondary metabolites specifically coumarins, alkaloids, flavonoids and essential oils (Attia et al., 2011; Gourine et al., 2011). In this respect, Dahia et al. (2009) showed that the *P. scoparius* methanolic extract is rich in flavonoid glycosides, and cinnamic acids, which may explain the traditional use of this plant as an alternative medication to treat fever and rheumatism (Yen et al., 2009). However, since some plants might be toxic, toxicity assays are an important part of research on medicinal plants (Harchaoui et al., 2018).

# MATERIALS AND METHODS

# 1. Materials

## 1.1. Collection and authentication of plant material

Fresh stems and roots of *Pituranthos scoparius* were obtained from the mountain Djebel Zdimm (located in the north-eastern of Algeria between 5º23'–5º29' east longitude and between 36º–36º27' north) at an altitude of 1258 m above sea level in February–April 2017 (Figure 8). Prof. Laouer H., a botanist at Laboratory of Valorization of Natural Biological Resources, University Setif 1, Algeria, identified and authenticated the plant. A voucher specimen (013/DBEV/UFA/18) was deposited at the herbarium located at the Department of Vegetal Biology and Ecology, University Setif 1, Algeria. After harvesting, the collected material was cleaned, washed, air-dried in the shadow to preserve its properties, crushed and then separately powdered using an electric grinder.

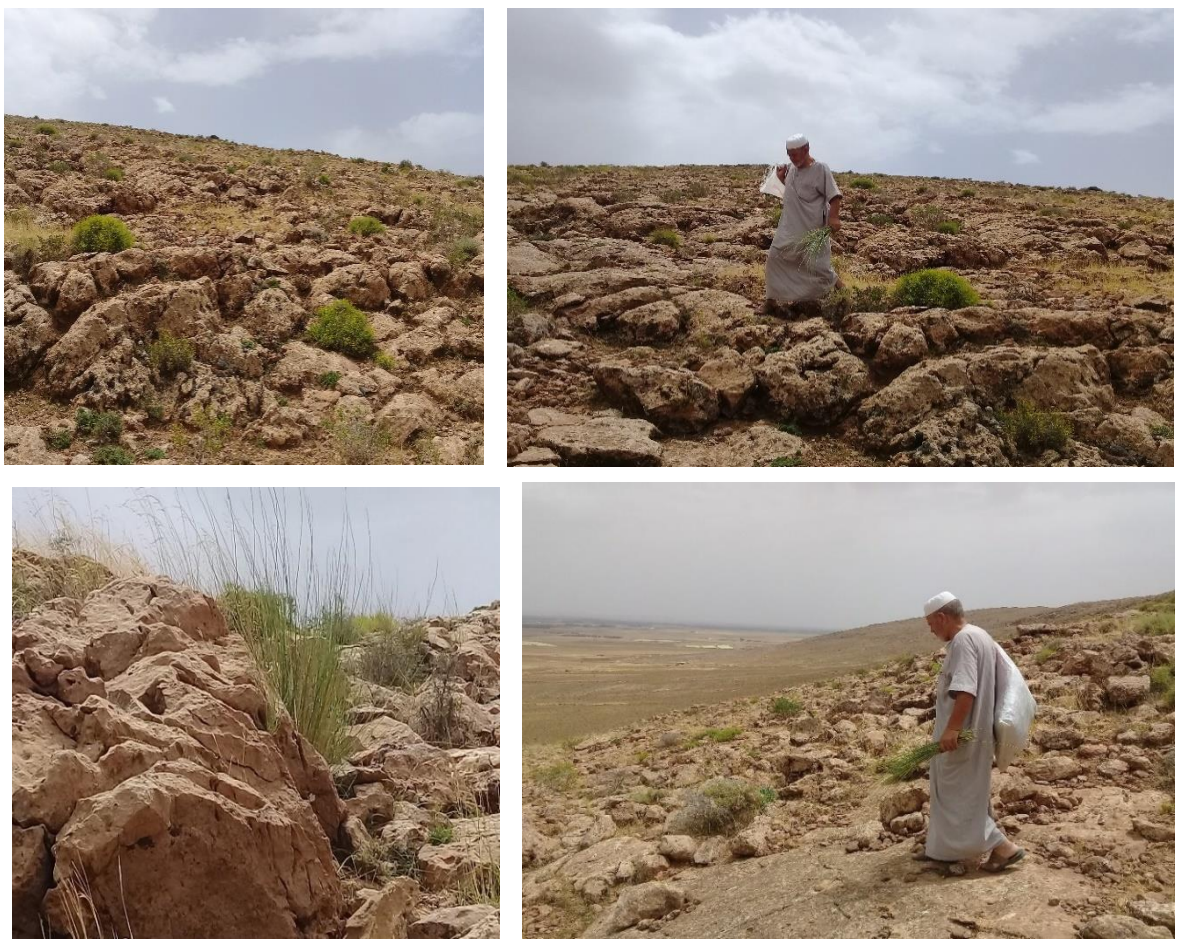

**Figure 8:** Collection of *Pituranthos scoparius* from the mountain Djebel Zdimm (Setif-Algeria).

## 1.2. Animals

Adult female Wistar rats (150–180 g) and albino mice (25–30 g) were used. These animals were purchased from 'Institut Pasteur d'Algérie', Algiers. Rats and mice were housed in cages under standard conditions of 12:12 h light/dark cycle and 25 ± 1°C for seven days before the experiments. They were given free access to water and standard diet (*ad libitum*), and kept under standard conditions mentioned in the Animals By-Laws N° 425 (OECD, 2008). The experimental assays were approved by the Committee of the 'Association Algerienne des Sciences en Experimentation Animale' (http://aasea.asso.dz/ articles) under law No. 88-08/1988, associated with veterinary medical activities and animal health protection (N° JORA: 004/1988).

## 1.3. Human red blood cells

Blood samples were collected from healthy human volunteers from our Laboratory of Applied Biochemistry (BAL), Setif-Algeria, and were clearly informed about the methods and objectives of this study. Blood was collected according to the International Federation of Blood Donor Organizations (IFBDO) with a standard operating procedures, and approved by the Research Ethics Committee of «Hospitalo-Universitary Centre», CHU of Setif, Algeria, which adopted the directive 2001/20/EC concerning implementation of good clinical practice in the conduct of clinical trials on medicinal products for human use.

## 1.4. Chemicals and reagents

All used chemicals and reagents were obtained from Merck, Sigma (Germany), Fluka, Prolab and Biochem. Purified solvents were used for thin layer chromatography, while commercial available solvents were used for column chromatography. All other reagents were of analytical grade. Silica gel S (70–230 mesh, Fluka, Steinheim, Germany) was subjected in column chromatography (CC). Purification of compounds was performed by thin layer chromatography (TLC) using TLC glass plates which were prepared manually using silica gel G-UV254 (Macherey-Nagel). In some cases, final purification was performed on commercial glass plates G-UV 254 (0.5 mm, Macherey–Nagel, Easton, PA, USA). Routine TLC was performed on silica gel 60 $F_{254}$ (Merck, Darmstadt, Germany) pre-coated glass plates (0.25 or 0.50 mm in thickness) and on silica gel G-UV 254 glass plates (0.25 mm, Macherey–Nagel). Sephadex LH-20 was used in purification of some fractions.

### 1.5. Instrumentation

$^1$H- and $^{13}$C-NMR spectra were recorded on a Bruker DPX-500 MHz spectrometer (Bruker TopSpin, Berlin, Germany) with DMSO-$d_6$ as solvent and TMS as an internal standard. Chemical shifts are expressed in $\delta$ units, whereas coupling constants ($J$-values) are given in Hertz. High-resolution mass spectra (HRMS) were obtained with electrospray ionization (ESI, negative mode) technique by collision-induced dissociation with the aid of a Bruker APEX-4 instrument (Bruker, Bremen, Germany).

## 2. Methods

### 2.1. Ethnopharmacological survey

In order to identify and establish a list of plants used in traditional medicine due to their pharmacological properties, an ethnobotanical survey was conducted according to Miara et al. (2018) with few modifications. The current study was carried out for information about different parts, treated diseases and route of applications of *Pituranthos scoparius* according to the traditional medicine due to their constituents compounds used in pharmacological properties (Figure 9).

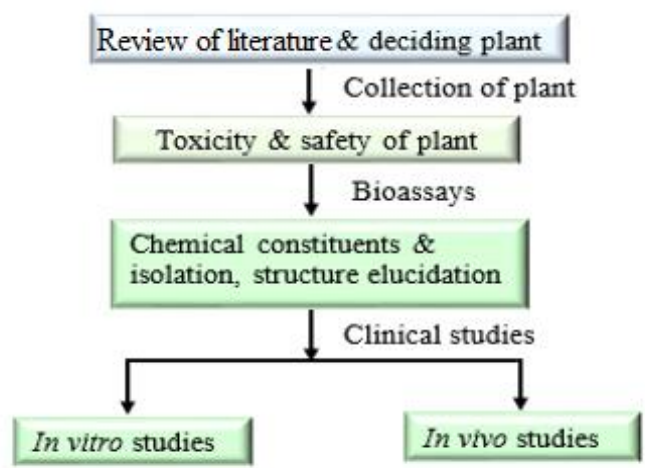

**Figure 9:** Structure of ethno-pharmacological study plan

### 2.1.1. Study region

The area of djebel Zdimm is located in the north-eastern part of Algeria between 5º23'–5º29' east longitude and between 36º–36º27' north, at an altitude of 1258 m above sea level. This area offers a floristic variety and a substantial traditional therapeutic knowledge. Three urban areas along and around of djebel Zimm's were studied (Figure 10).

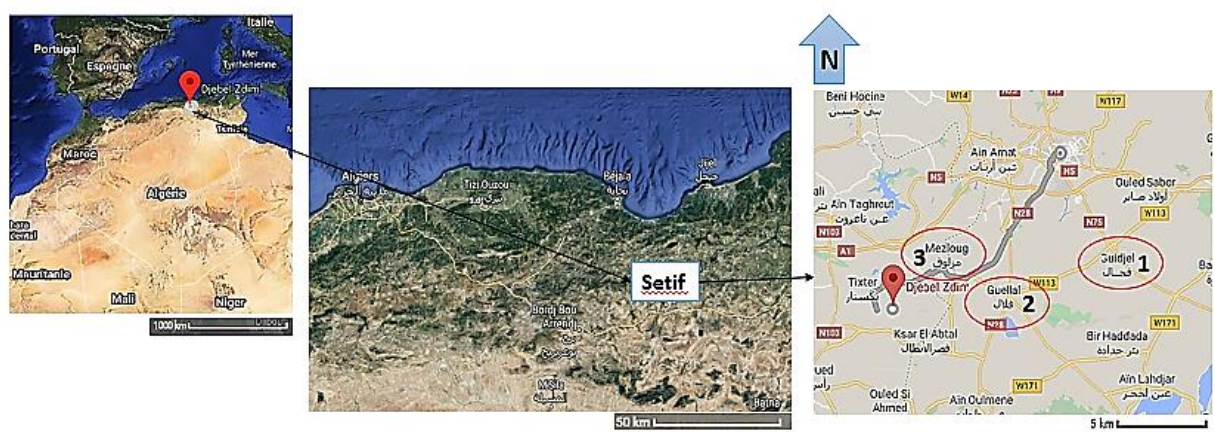

**Figure 10:** Geographical location of the area of study. Urban districts: **1** (Guedjel), **2** (Guellal), **3** ( Mezloug) along and around of djebel Zimm's (Setif).

### 2.1.2. Questionnaire and sampling

The present work aims to organize such overall information on medicinal plants along to the differents parts *Pituranthos scoparius* and its relation with therapeutic usefulness in traditional medicine. All the participants (villagers, traditional healers and herbalists) were native to the study area. A total of 192 informants (113 males and 79 female) of different ages which ranged between 30 to 65 years old participated in this survey. In addition, all citizens in urban and farming areas were questioned about their knowledge on *Pituranthos scoparius.*

## 2.2. Extraction procedures profile

### 2.1.1. Preparation of decoction extracts

The preparation of the plant extracts were carried out according to the method of Chanda et al. (2013). Decoction extracts were prepared by boiling 100g of dried plant materials (stems and roots) in 1L of distilled water for 20 min. Then the solution was filtered through muslin cloth and centrifugated at 4000 rpm for 20 min. The dried extracts thus obtained were stored at 4°C then, screened for their pharmacological properties.

### 2.1.2. Preparation of crude methanolic extracts and sub-fractions

The extraction was done by fractionation using liquid- liquid method (Baghiani et al., 2012). Approximately 100 g of powdered plant materials (stems and roots) was soaked separately in a methanol-water mixture (85 %, v/v) for 24 h at room temperature with occasional stirring. The mixtures were filtered separately and residues were extracted with two additional 1L portions of methanol: water (1:1) for 4 h. Then the solutions were filtered through muslin cloth, and the solvents were evaporated under reduced pressure to produce an initial crude extract (CrE). CrE was successively extracted with different solvents of increasing polarity: hexane for defatting, chloroform for aglycone flavonoids and ethyl acetate for glycoside flavonoids, each fraction was evaporated to dryness under reduced pressure to produce chloroform extract (ChE) and ethyle acetate extract (EaE), respectively. The remaining aqueous extracts was labeled (AqE). All the solvents were eliminated by evaporation under reduced pressure and the resulting sub-fractions were stored at 4°C for further pharmacological studies (Figure 11).

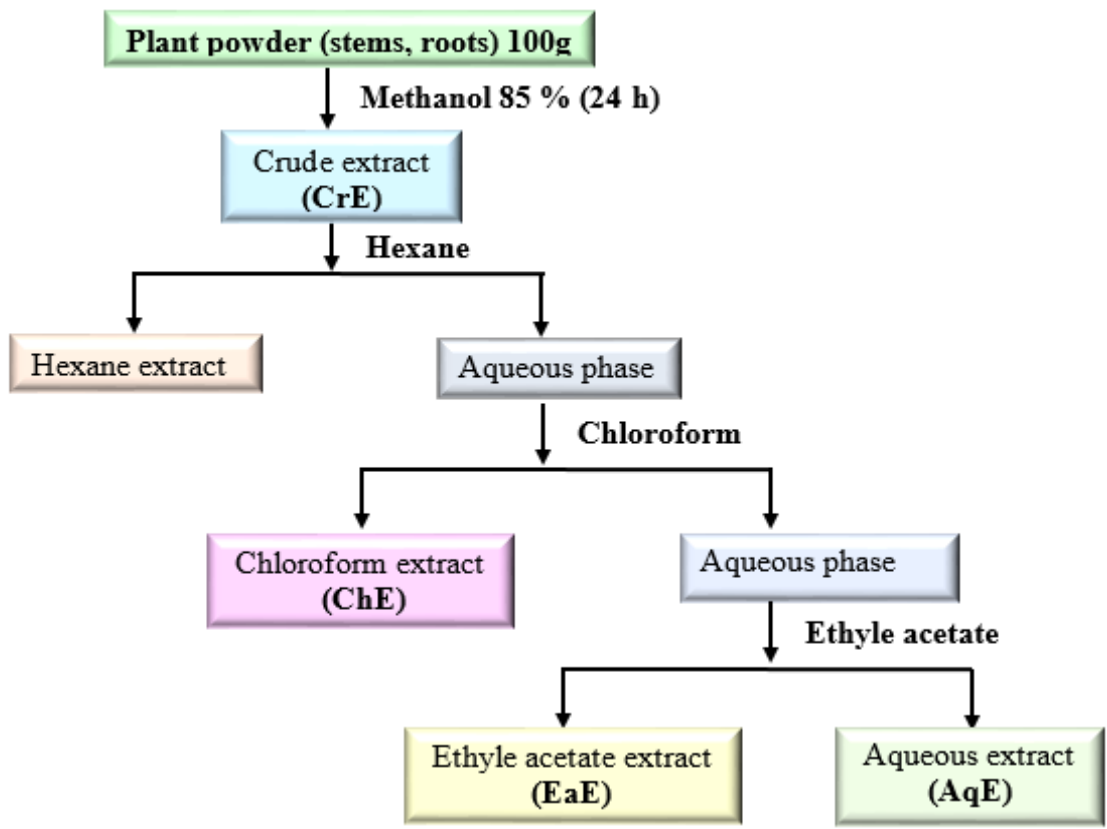

**Figure 11:** A sequential extraction procedure to prepare the sub-fractions.

**2.1.3. Preparation and fractionation of the crude ethanolic stems extract**

Approximately 7 kg of powdered stem materials was defatted by extraction with *n*-hexane (25 L) at room temperature for 4 days. Then the defatted material was air-dried for about 30 min and extensively extracted with absolute ethanol (25-30 L, four times, four days each) at room temperature. The combined crude ethanolic extracts were evaporated under reduced pressure and the resulting concentrated ethanolic extract (282.0 g) was partitioned between chloroform and water (1: 1, 2L). The chloroform layer was evaporated under vacuum to give a gummy residue (94 g), which was afterwards partitioned between *n*-hexane and 10% aqueous methanol (1:1, 2L) to provide the aqueous methanol extract (43g) and hexane extract. Polar organic compounds were extracted from water layer with *n*-butanol to afford the water and the *n*-butanol extract (52g). Fractionation of the crude ethanolic extract from stems of *P. scoparius* is depicted in figure 12.

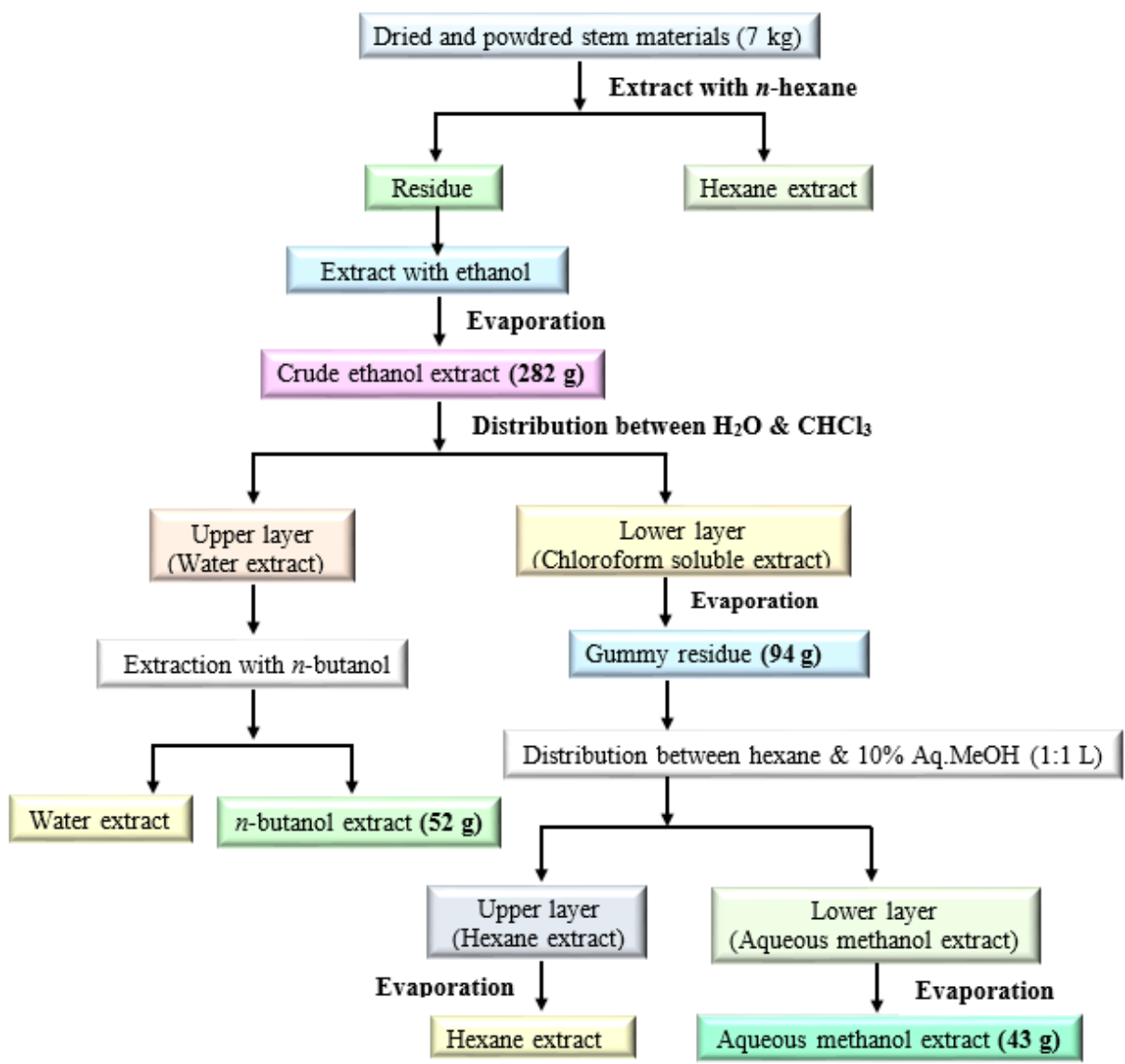

**Figure 12:** Fractionation of the crude ethanol stems extract of *Pituranthos scoparius.*

## 2.3. Phytochemical evaluation

### 2.3.1. Qualitative phytochemical analysis

The stem extracts (SE), roots extracts (RE) and *n*-butanol extract (ButE) of *P. scoparius* were subjected to screening to detect the presence of potential phytochemical constituents such as alkaloids, flavonoids, polyphenols, tannins, quinones, anthraquinones, coumarins, and reducing sugars according to published procedures (Bekro et al., 2007; karumi et al., 2004; Oloyede et al., 2005). These are qualitative analyses based on coloring and/or precipitation reactions.

***2.3.1.1. Detection of alkaloids:*** fifty mg of residue is taken up in 3 mL of ethanol-water (60:40 v/v). After stirring, 2 drops of Dragendorff reagent were added. The appearance of an orange-red or reddish-brown precipitate indicates a positive test.

***2.3.1.2. Detection of polyphenols:*** a few drops of 2% alcoholic solution of ferric chloride were added to 2 mL of the extract. The appearance of a darkish blue or green color indicates the presence of polyphenols.

***2.3.1.3. Detection of flavonoids:*** five mL of the extract was treated with few drops of concentrated HCl. After adding little quantity of magnesium turnings, the appearance of a red or orange color indicates the presence of flavonoids.

***2.3.1.4. Detection of tannins:*** two mL of extract was reacted with little sodium acetate ($CH_3CO_2Na$), then, a few drops of a 2% $FeCl_3$ aqueous solution were added. The reaction is positive if a blue-black color appears.

***2.3.1.5. Detection of free quinones:*** few drops of sodium hydroxide (1% in $H_2O$) were added to 5 mL of extract. Appearance of a color that turns yellow, red or purple indicated the presence of free quinones.

***2.3.1.6. Detection of anthraquinones:*** ten mL of the extract was treated with 5 mL of ammonium hydroxide (10%). Formation of a red ring indicated the presence of anthraquinones.

***2.3.1.7. Detection of coumarins:*** two mL of alcoholic stock solution obtained from stock solution (dissolving 500 mg of each extract in 100 mL of ethanol) are introduced in 2 test tubes. In one of the test tubes is added 0.5 mL of 10% NaOH, and the test tubes are heated in a water bath to boiling. After cooling, 4 mL of distilled water are added to each test tube.

The reaction is positive; if the test tube in which the alkaline solution has been added is transparent compared to the control test tube. After acidifying the alkaline solution with a few drops of concentrated HCl, it loses its yellow color and becomes cloudy or a precipitate forms.

*2.3.1.8. Detection of reducing sugar:* five mL of the extract was treated with 5 mL of Fehling's reagent, and heated in a water-bath at 70 °C. Formation of a brick red precipitate indicated the presence of sugar.

**2.3.2. Quantitative phytochemical analysis**

**2.3.2.1. Determination of total polyphenol**

We estimated the total phenolic content in the extract using the method of Folin-Ciocalteu's (Arrar et al., 2013). In brief, we mixed an aliquot of 100 µL of the stem extract with 500 µL of Folin-Ciocalteu's reagent for 4 min, followed by the addition of 400 µL of a 7.5% aqueous $Na_2CO_3$. Then, absorbance of solution was determined at 765 nm after 2h of incubation. Polyphenolic content was expressed as µg gallic acid equivalent (GE)/mg dried extract (DE). The amount of total polyphenols in different extracts was determined from a standard curve of gallic acid ranging from 0.00 to 160 µg/mL (Figure 13).

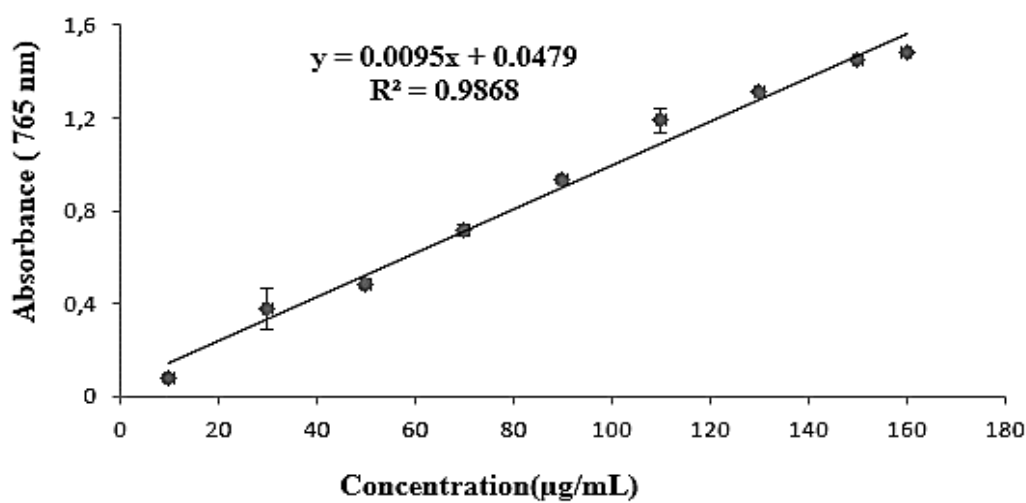

**Figure 13:** Standard curve of gallic acid for determination of total polyphenols in *Pituranthos scoparius* extracts. Each value represent mean ± SD (n = 3).

**2.3.2.2. Determination of total flavonoids**

The total flavonoids content was evaluated by the method of aluminum chloride (Arrar et al., 2013). According to this method, 1 mL of extract was added to 1 mL of aluminum chloride solution (2%). This was followed by measuring the absorbance of the mixture at 430 nm versus a methanol blank, after 10 min of incubation. Total flavonoids were reported as µg of quercetin equivalent (QE)/mg dried extract (DE). The amount of total flavonoids in different extracts was determined from a standard curve of quercetine ranging from 0.00 to 40 µg/mL (Figure 14).

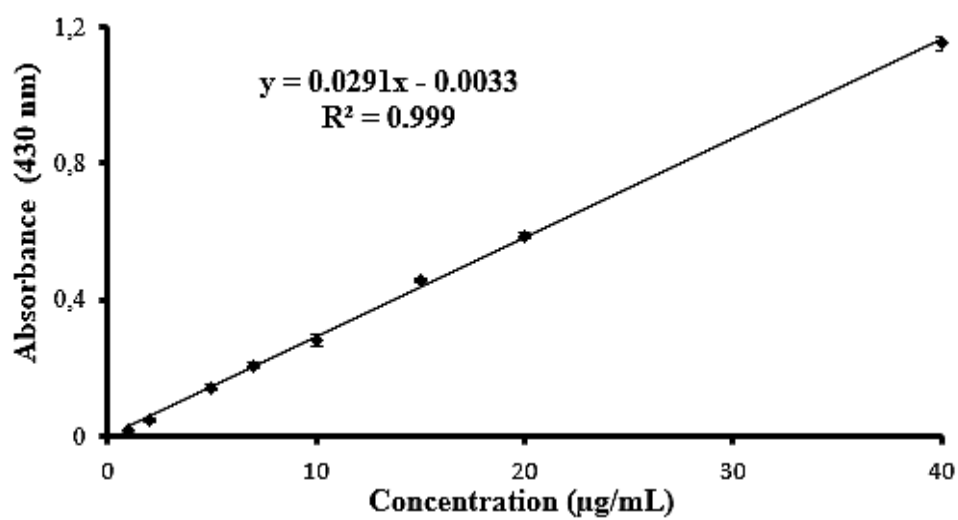

**Figure 14:** Standards curve of quercetine for determination of total flavonoids in *Pituranthos scoparius* extracts. Each value represent mean ± SD (n=3)

**2.3.2.3. Determination of tannins**

We employed the procedure bellow to measure the precipitation of hemoglobin by tannins (Khennouf et al., 2013). Briefly, a 500 µL aliquot of different concentrations of extracts was mixed with 500 µL of hemolyzed sheep blood. This mixture was centrifuged for 10 min after 20 min of incubation. Absorbance of the supernatant was then measured at 576 nm, and effectiveness of the precipitation of solutions tested was expressed as µg tannic acid equivalent (TE)/mg dried extract; a mixture of equal volumes of tannic acid (100–600 µg/mL) and hemolyzed blood was employed to obtain the calibration curve (Figure 15).

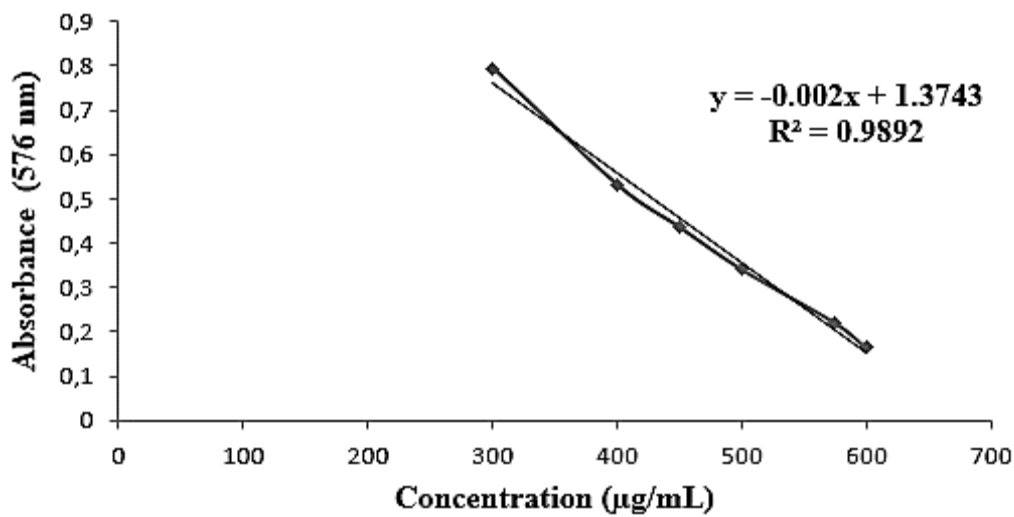

**Figure 15:** Standards curve of tannic acid for determination of tannins in *Pituranthos scoparius* extracts. Each value represent mean ± SD (n=3)

## 2.4. Isolation, detection and identification of compounds

### 2.4.1. Gel chromatography column analysis of *n*-Butanol extract (PSB)

The crude *n*-butanol extract (51.9 g) was adsorbed on 50 g of silica gel S (70–230 mesh, Fluka, Steinheim, Germany) and was introduced into a proper column (119×5.0 cm) using about 300 g of the same adsorbent packed with chloroform and eluted with the same solvent. The polarity of the solvent was increased with methanol until 100% methanol was used. A total of 126 fractions (500 mL each) were collected and grouped into 4 major fractions according to their TLC behaviour. Further purification was carried not for these groups by a combination of gel filtration (Sephadex$^R$ LH-20, Fluka, Steinheim, Germany) and thin layer chromatography (TLC) using suitable solvent systems.

**Fraction I** was loaded on a sephadex column eluted with methanol and dichloromethane (1:1*v/v*). This fraction afforded an impure solid. Further washing with methanol yielded a light yellow solid (600 mg).

**Fraction II** was loaded on a sephadex column eluted with methanol and dichloromethane (*v/v*). A total of 24 fractions were collected and grouped into 4 groups depending on their TLC behaviour, then TLC was used for further purification. The subfraction 2 yielded an amorphous yellow solid (220 mg). The mother liquor of this fraction did not show any important spots other than the solid so it was not investigated any further.

**Fraction III** was loaded on a sephadex column eluted with methanol and dichloromethane (*v/v*). This fraction gave a white solid (100 mg). The mother liquor of this fraction did not show any important spots other than the solid so it was not investigated any further.

**Fraction IV** was loaded on a sephadex column eluted with methanol and dichloromethane (30:70, *v/v*). This fraction afforded an impure solid. Further washing with methanol yielded a yellow solid (950 mg). The columns used for separation of the *n*-butanol extract are shown in table 2, while the compounds isolated from this fraction are shown in table 3.

**Table 2:** Columns used for chromatography of *n*-butanol extract frations of *Pituranthos scoparius*

| Fractions | Weight (g) | Length× Diameter (cm) | No. of fractions | No. of Group | Fraction volume (mL) | Mobile phase | Type of adsorbent |
|---|---|---|---|---|---|---|---|
| FI | 5.5 | 74×3.5 | 39 | 5 | 10 | Dich/MeOH | Sephadex LH-20 |
| FII | 2.5 | 74×3.5 | 48 | 4 | 10 | Dich/MeOH | Sephadex LH-20 |
| FIII | 3 | 74×3.5 | 40 | 3 | 10 | Dich/MeOH | Sephadex LH-20 |
| FIV | 4 | 74×3.5 | 38 | 4 | 10 | Dich/MeOH | Sephadex LH-20 |

**MeOH:** Methanol, **Dich:** Dichloromethane

**Table 3:** Compounds isolated from the *n*-butanol extract of *Pituranthos scoparius*

| Compounds | Compounds names | Approx. weight (mg) |
|---|---|---|
| **PSB1** | Isorhamnetin-3-*O*-β-glucoside | 600 |
| **PSB2** | Isorhamnetin-3-O-β-apiofuranosyl (1→2)-β-glucopyranoside | 220 |
| **PSB3** | D-mannitol | 100 |
| **PSB4** | Isorhamnetin-3-β-glucosyl-(1→6)-β-glucoside | 950 |

**2.4.2. Detection of compounds**

Detection of flavonoids was achieved using short UV radiation, while detection of terpenes was achieved by anisaldehyde-sulfuric acid spray reagent followed by heating on a hot plate. The spraying reagent was prepared by the addition of 16 mL of concentrated sulphuric acid to 400 mL ethanol followed by the addition of 8 mL of *p*-anisaldehyde and then swirling.

**2.4.3. Determination of the ultraviolet absorption spectra of flavonoids**

*-Preparation of reagent stock solutions*

1. A stock solution of the flavonoid was prepared by dissolving a small amount of compound (0.1 mg) in 10 L spectroscopic methanol.

2. Sodium methoxide (NaOMe) solution was prepared by the addition 2.5 g of freshly cut metallic sodium (cautiously in small portions) to 100 mL of dry spectroscopic methanol.

3. Aluminium chloride solution ($AlCl_3$) was prepared by the addition of 5.0 g of anhydrous $AlCl_3$ cautiously to 100 mL spectroscopic methanol.

4. Hydrochloric acid solution was prepared by the addition of 50 mL concentrated HCl solution to 100 mL of distilled water.

*-Measurement of UV-absorption spectrum of flavonoids*

1. The UV-absorption spectrum of the methanolic solution was measured firstly.

2. The NaOMe spectrum was measured immediately after the addition of 3 drops of NaOMe reagent to the methanolic solution of the flavonoid.

3. The $AlCl_3$ spectrum was measured immediately after the addition of 6 drops of $AlCl_3$ reagent to the methanolic solution of the compound.

4. The $AlCl_3$/HCl spectrum was measured immediately after the addition of 3 drops of HCl solution to the solution containing $AlCl_3$ reagent (solution of the previous step).

## 2.5. Pharmacological investigations

### 2.5.1. Screening of *in vitro* antioxidant activity

#### 2.5.1.1. DPPH radical scavenging assay

Free radical scavenging capacity of the extracts was assessed using the 2,2'-diphenyl-1-picrylhydrazyl (DPPH) assay by measuring the decrease in the DPPH maximum absorbance at 517 nm (Charef et al., 2015). In this method, 50 µL of various concentrations of the extracts/standard were mixed with 1.25 mL of DPPH solution in methanol (0.004%). Absorbance of the sample was measured at 517 nm after a 30 min of incubation in the dark at room temperature. Butylated hydroxytoluene (BHT) was used as positive control. The scavenging capacity was calculated according to the following equation (Equation (1):

$$I\% = (A_{blank} - A_{test} / A_{blank}) \times 100. \qquad (1)$$

Where $A_{blank}$ = absorbance of the solution except the tested sample, and $A_{test}$ = absorbance of the tested sample.

#### 2.5.1.2. ABTS radical scavenging assay

$ABTS^+$ radical scavenging assay was assessed by colorimetric analysis (Bouaziz et al., 2015) with slight modifications. The $ABTS^+$ solution was formed by the reaction of 7 mM of ABTS solution in 2.45 mM potassium persulfate. The mixture was saved in the dark at room temperature for 16 h before use. The solution was diluted with absolute ethanol and equilibrated at room temperature to give an absorbance of 0.7 at 734 nm. Then, 20 µL of the extracts/standard dilutions was mixed with 2 mL of $ABTS^+$ solution and kept for six min at room temperature. Vitamin C (VitC) was used as positive control. The absorbance was measured at 734 nm. The scavenging capability of $ABTS^+$ radical was calculated according to the same formula of DPPH assay.

**2.5.1.3. β-Carotene bleaching by linoleic acid assay**

Inhibition of oxidative discoloration of β-carotene by the products of oxidation of linoleic acid can be used to determine the antioxidant capacity of the extracts (Charef et al., 2015) according to the following procedure: An amount of 0.5 mg of β-carotene was dissolved in 1 mL of chloroform. To this solution, 25 μl of linoleic acid and 200 mg of Tween 40 were added. After evaporation of the chloroform by means of a rotary evaporator, 100 mL of distilled water saturated with $O_2$ was added, and the solution was vigorously shaken to form a stable emulsion. Then, 350 μL of the extracts/standard (BHT) prepared at a concentration of 2 mg/mL was added to 2.5 mL of this mixture, followed by incubation for 24 h in the dark at room temperature. Kinetics of discoloration of the emulsion in the presence and absence of the antioxidant was measured at 490 nm at intervals during 24h (0, 1, 2, 3, 4, 6 and 24) of incubation at room temperature and in the dark. The antioxidant activity was calculated as follows (Equation (2):

$$\text{Antioxidant activity (\%)} = (A_s/A_{BHT}) \times 100 \qquad (2)$$

Where $A_s$ = absorbance in the presence of sample; $A_{BHT}$ = Absorbance in the presence of BHT.

**2.5.1.4. Reducing power assay**

The reducing power of the extracts from *Pituranthos scoparius* was determined by the capacity to reduce $Fe^{+3}$ to $Fe^{+2}$ ions assay (Bouaziz et al., 2015), with some modifications. According to this procedure, an aliquot of 400 μL of extract was mixed with an identical volume of both phosphate buffer (0.2 M, pH = 6.6) and potassium ferricyanide (1%). This mixture was then incubated for 20 min at 50 °C in a water bath. The reaction was terminated by adding 400 μL of trichloroacetic acid (TCA) (10%), and the mixture was centrifuged at 3000 rpm for 10 min. The supernatant (400 μL) was added to distilled water (400 μL) and 80 μL of 0.1% ferric chloride. The colour intensity of the mixture was measured at 700 nm after 10 min of incubation. In this context, a high absorbance of the solution means a high reducing power.

**2.5.1.5. Hydroxyl radical scavenging assay**

Hydroxyl radical scavenging ability of the extracts was via a spectrometric method (Ates et al., 2008). Briefly, a mixture containing 1 mL of $FeSO_4$ (1.5 mM), 700 μL of $H_2O_2$ (6 mM) was mixed with varying concentrations of extract or ascorbic acid as a positive control. Then, 300 μL of sodium salicylate (20 mM) was added, followed by incubation for 20 min at 37°C. Absorbance of the obtained mixture was measured at 562 nm. Scavenging capacity of the extract was evaluated using the following equation (Equation (3):

$$I(\%) = [1-(A_e-A_c)/ A_0]\times100. \qquad (3)$$

Where $A_0$ = absorbance of control, $A_e$ = absorbance in the presence of sample

$A_c$ = absorbance without sodium salicylate.

**2.5.1.6. Iron chelating assay**

Ferrous iron-chelating capacity of the extracts was evaluated by spectrometric assay (Baguiani et al., 2012). According to this method, a mixture containing 250 μL of extract or EDTA as a positive control, 50 μL $FeCl_2$ (0.6 mM in distilled water) and 450 μL methanol was made. The control included all reaction reagents except the test samples. This mixture was mixed well and allowed to react at room temperature for 5 min, followed by addition of 50 μL ferrozine. Absorbance at 562 nm was measured after 10 min of incubation. An $IC_{50}$ value was defined as the effective concentration of test material which produces 50 % of the maximal scavenging effect.

## 2.5.2. Screening of anti-inflammatory activity

**2.5.2.1. The *in vitro* evaluation of the anti-inflammatory activity**

The anti-inflammatory activity of the CrE from the stems (S), roots (R) along with ButE and isolated compounds was evaluated by their effect on protein denaturation and was conducted according to the method described by Akkouche et al. (2012), with some modifications. Eggs were washed and cleaned, and then broken. The white and yolk separated, taking the necessary measures of removing chalazae. The volume of egg white was measured using a test-tube and then adjusted with the buffer solution Tris-HCl (20 mM, pH 6.87) to obtain a dilute solution of 1:100. The mixture was stirred for 10 min, followed by filtration with a strip agase. A mixture (4 mL) of 2 mL of egg white suspension (1%), and 2 mL of CrE/ButE and isolated compounds or aspirin as a standard (0.5, 1 and 2 mg) was mixed, and was incubated in a water bath at 70 °C for 15 min. After cooling, absorbance was measured at 650 nm. The percentage inhibition of protein denaturation was calculated as follows (Equation (4)):

$$\% \text{ inhibition of denaturation} = 100 \times (A_C - A_S)/A_C \qquad (4)$$

where $A_C$ = absorption of the control sample, and $A_S$ = absorption of the test sample.

**2.5.2.2. The *in vivo* investigations of the anti-inflammatory activity**

**2.5.2.2.1. Xylene-induced ear edema**

The anti-inflammatory activity of the crude extracts of stems and roots from *P. scoparius* was investigated in xylene-induced ear edema in mice, following the procedure outlined by Delaporte and coworkers (2004). Mice were divided into eight groups of 6 animals each (n = **6**). Group n°=1 (Positive control): received orally with indomethacin used as a standards drugs (50 mg/Kg), group n°=2: (Negative control), received orally distilled water, groups 3-5 received orally CrE from stems (100, 300, and 600 mg/kg in 0.5 mL H$_2$O, respectively), groups 6-8 received orally CrE from roots by the same doses of groups 3-5. After 60 min of treatment, edema was then topically induced in each mouse using 30 μL/ear of xylene. The thickness of the ear measured with a digital caliper before and 2 h after the xylene application. The inhibition percentage of ear edema was computed as follows (Equation (5)):

$$\text{Inhibition percentage (\%)} = 100 \times (D_n - D)/ D_n, \qquad (5)$$

where D = the difference of ear edema thickness in the treated group and $D_n$ = the difference of ear edema thickness in the negative group.

### 2.5.2.2.2. Croton oil-induced ear edema

Anti-inflammatory activity was carried out according to the method of Dulcetti Junior et al. (2004) with slight modifications. Mice were randomized into eight groups of 6 mice each. Group 1 (Positive control), received orally indomethacin used as a standard drug (50 mg/Kg), group 2 (Negative control), received orally distilled water, whereas groups 3–5 received orally CrE from stems (100, 300, and 600 mg/kg in 0.5 mL $H_2O$, respectively), groups 6–8 received orally CrE from roots by the same doses of groups 3–5. After 1 h, induction of inflammation with 15 μL croton oil solution (80 μg in 50% water-acetone v/v) was locally applied in the inner surface of the right ear of mice. The thickness of the ear was measured by means of a digital caliper before treatment and 6 h after the induction of inflammation. The inhibition percentage of ear edema was computed as in the equation of xylene-induced ear edema activity.

### 2.5.2.2.3. Topical anti-inflammatory

The topical anti-inflammatory activity of ButE and isolated compounds was evaluated using the croton oil-induced ear edema method in mouse model, according to the method of Manga et al., (2004). Mice were randomly divided into *thirteen* groups of six mice each. Group 1: treated locally with indomethacin (used as a standards drug) (0.5 and 1 mg/ear) (positive control). Group 3: treated locally with croton oil (80 mg/kg) dissolved in 15 μL of acetone-water solution (1:1, *v/v*) (negative control). Group 4 and 5: treated with ButE (0.5 and 1 mg/ear) dissolved in 15 µL of acetone-water solution (1:1, *v/v*). Groups 6–13: treated locally with isolated compounds **PSB1**, **PSB2**, **PSB3**, and **PSB4**, respectively (0.5 and 1 mg/ear), dissolved in 15 µL of acetone-water solution (1:1, *v/v*). First, 15 µL of acetone-water solution (1:1, *v/v*) containing 80 µg of croton oil was locally applied at the inner surface of the right ear of each mouse. Simultaneously, 15 µL of acetone-water solution (1:1, *v/v*) containing 0.5 mg of different products and indomethacin (used as a standard drug) were topically applied at the same place of the ear; mice of the control group topically received only the solution of croton oil. The thickness of the ear was measured by means of a digital caliper before treatment and 6 h after the induction of inflammation. The inhibition percentage of ear edema was computed as in the equation of xylene-induced ear edema activity.

### 2.5.2.2.4. Carrageenan-induced rat paw inflammation

The carrageenin-induced edema assay was assessed according to the method of Igbe and Inarumen (2013). Tested Wistar rats (150–180 g) were randomly divided into nine groups of five animals each (n = 5). Groups 1 and 2 (Positive controls): treated with diclofenac sodium and indomethacin used as standard drugs (20 and 50 mg/Kg, respectively), group 3: (Negative control), received distilled water, whereas groups 4–6 received orally CrE from stems (100, 250, and 500 mg/kg in 1 mL $H_2O$, respectively), groups 7-9 received orally CrE from roots by the same doses of groups 4–6. After 1 h, edema was induced in the right hind paw of each rat by sub plantar injection of 100 µL freshly prepared carrageenan suspension (1% in NaCl 0.9 %). Paw thickness was measured using a vernier caliper at hourly intervals for 5 h.

### 2.5.2.2.5. Analgesic activity by acetic acid induced writhing in mice

Writhing study was performed by following the method of Ishola et al. (2011) to investigate the analgesic effect of extracts. In this protocol, female mice were divided into eight groups of five animals each (n=5). Group 1 (Positive control): treated orally with aspirin, used as a standard drug (100 mg/Kg). Group 2: (Negative control), treated orally with distilled water. Groups 3–5: treated orally with CrE from stems (100, 250, and 500 mg/kg in 0.5 mL $H_2O$, respectively), groups 6-8 received orally CrE from roots by the same doses of groups 3-5.. After 60 min, writhes was induced in mice through intra-peritoneal injection with 0.1% (v/v) acetic acid. After 5 min of injection, the number of abdominal contractions was counted over a period of 30 min. The percentage inhibition of writhing reflex was calculated using the formula (Equation (6)):

$$\text{Inhibition (\%)} = 100 \times (C_n - C_t)/C_n, \tag{6}$$

where $C_n$ = mean of contractions' count in animals in the negative control, and $C_t$ = Mean of contractions' count in animals treated with different concentrations of extracts or aspirin.

**2.5.3. Screening of toxicity study**

**2.5.3.1. *In vitro* cytotoxicity against red blood cells**

Hemolysis test was performed using human blood of a healthy donor according to a procedure outlined by Malagoli et al. (2008). Whole blood was obtained from healthy young human donors and was collected according to International Federation of Blood Donor Organizations (IFBDO) standard operating procedures. According to this procedure, human erythrocytes from blood were isolated by centrifugation at 3000 rpm for 10 min and washed three to four times with NaCl (0.9%) solution until the supernatant became colorless. After the centrifugation, the blood volume was measured and reconstituted as a 2% (*v/v*) suspension with isotonic buffer solution (10 mM sodium phosphate buffer pH 7.4). Then, a mixture (1 mL) containing 500 µL of ButE/isolated compounds (0.5 and 1 mg/mL) or Triton X-100 (0.1%) and 500 µL of erythrocyte suspension (2%) was incubated at 37 °C for 1 h. After the incubation periode, the suspension was centrifuged at 3000 rpm for 10 min.

Absorbance of supernatant was determined by spectrophotometrically at 540 nm and hemolysis was calculated by the following formula (Equation (7)):

$$\text{Hemolysis \%} = (\text{Abs}_{\text{test}} / \text{Abs}_{\text{Triton X-100}}) \times 100 \qquad (7)$$

where the $\text{Abs}_{\text{test}}$ is the absorbance in the presence of tested compound, and $\text{Abs}_{\text{Triton X-100}}$ is the absorbance in the presence of Triton X-100 as a positive control. The positive control induces 100% hemolysis.

### 2.5.3.2. Acute oral toxicity study

**2.5.3.2.1. Observations**

The acute toxicity of the crude extracts of stems and roots is evaluated in rats, using the internationally accepted guidelines (OECD, 2008). Rats were divided into three groups (n = 5) and subjected to fasting for 12 h before the experiment. Doses of CrE from stems and roots (2 and 5 g/kg in 1 mL $H_2O$, respectively) were orally given to each rat in the first and second groups, respectively, whereas rats in the third group (negative control) were given distilled water. Body weight and observations of symptoms of toxicity were recorded systematically for 48 h and then once a day for 14 days. Surviving if in animals were sacrificed then blood and organs were subjected to biochemical and histological analyses, respectively.

**2.5.3.2.2. Plasma collection and biochemical Analysis**

Fresh blood was quickly collected in heparin-tubes from sacrificed animals and the serum was separated by centrifuging at 3000 rpm for 15 min. The serum was then subjected to analysis to determine the biochemical parameters: urea, creatinine, protein, albumin, cholesterol, triglyceride, uric acid, total bilirubin, LDH, AST, ALT, and ALP using Biolysis 100 analytic and commercial kits (Spinreact, Spain, and Chroma test react).

**2.5.3.2.3. Relative Organs weight**

After sacrificing the animal, various organs such as the liver, kidneys, heart, spleen, stomach, and lungs were weighed. The relative organ weights (ROW) of each animal were estimated using the following formula:

ROW = organ weight (g)/ body weight of animal on sacrifice day (g) x 100.

**2.5.3.2.4. Histological investigation of kidney and liver**

Fresh organ portions (4–5 µm$^2$) were cut, fixed in 10% formalin, and inserted in a cubical block of paraffin. We then employed haematoxylin and eosin to stain sections according to the procedure of Jarrar and Taib (2012). The sections were examined under a microscope (Leica DM-1000 microscope) at 10, 25, and 40× magnification. Digital images of both control and treated groups were obtained with the aid of a Leica camera (Vision DTC 495) linked with the microscope. Images were investigated using image processing software.

**2.6. Statistical analysis**

*In vitro*, data were subjected to Student's *t*-test for significance. Determinations were conducted in triplicate and results were expressed as the mean ± standard deviation (SD). *In vivo*, data were subjected to one-way analysis of variance (ANOVA) and "Dunnett's multiple comparison test" was used to compare with control group followed by "Tukey's multiple comparisons test" were used to determine significant differences between groups, for significance. We employed GraphPad Prism-6 (GraphPad Software, San Diego, California USA, http://www.graphpad.com) to analyze data obtained from this investigation. Differences were considered significant at $p \leq 0.05$.

# RESULTS AND DISCUSSION

## 1.1. Ethnopharmacological survey

The medicinal plant is any vegetal containing substances that can be used therapeutically. These plants are extensively employed by alternative medicine. The interaction between humans and medicinal plants has been long described as one of the factors affecting human civilization, particularly in medicinal fields. Documentation of the medicinal application of plants through ethnobotanical investigations allows the development of conventional drugs and treatments as well as for plant conservation (Jadid et al., 2020). Ethnopharmacology is an interdisciplinary and multidisciplinary approach in drug discovery, including observation, description, and biological activity investigation of folk medicines (Süntar, 2019).

The main fieldwork of this study was to assemble information along with therapies on the use of *Pituranthos scoparius* by native people of the study area. Shown in Table 4 are results of the ethnobotanical survey. Botanical identification showed that 13 families were recorded among which Apiaceae with 02 species: *Pituranthos scoparius* Coss. & Dur. (47 citations) and *Thapsia garganica* L (07 citations), and 24.47% confirmed the use of *Pituranthos scoparius* in traditional folk medicine.

For a long time, *Pituranthos scoparius* has been classified among the most used medicinal plant in this area. In a similar fashion, among 47 persons who utilized this plant, another survey was conducted for the uses of different parts of this medicinal plant in folk medicine; results are depicted in Figure 16.

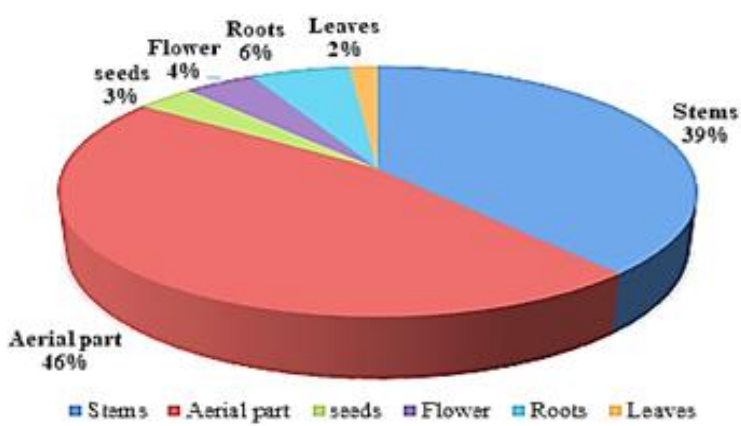

**Figure 16:** Percentage of uses of different parts of *Pituranthos scoparius* in Setif

**Table 4:** Ethno-medicinal survey data of some medicinal plants used in folk medicine (Djebel Zimm's-Setif, North of Algeria).

| Plant spices with botanical families | Local name | Parts used | Preparation made | Citations |
|---|---|---|---|---|
| *Thapsia garganica* L. **Apiaceae** | Derias | Rhizome | Infusion, decoction | 07 |
| *Pituranthos scoparius* Coss. & Dur. **Apiaceae** | Guezzah | Flower, leaves, stems, roots, and aerial part | Infusion, decoction, and powder | 47 |
| *Pistacia atlantica* Desf. **Anacardiaceae** | Qezwan | Resin, seeds, and leaves | Masticatory | 01 |
| *Helichrysum staechas* (L.) DC. **Asteraceae** | Chernessa, Chegara | Flowers | Infusion, decoction | 07 |
| *Hertia cheirifolia* (L.) O.K. **Asteraceae** | Khertchoun | Seeds and leaves | Infusion, extract | 01 |
| *Artemisia herba-alba* Asso. **Asteraceae** | Chih | Leaves | Infusion, tisane | 11 |
| *Ormenis africana* L **Asteraceae** | Gartofa | Leaves | Poudre, freshly | 06 |
| *Pallenis spinosa* (L.)Coss **Asteraceae** | Nouged or rebian | Flowers and seeds | Infusion, decoction | 04 |
| *Borago officinalis* L. **Boraginaceae** | Borage, lcen thour | Leaves, flowers, stems, and aerial parts | Freshly, decoction | 30 |
| *Juniperus oxycedrus* L. **Cupressacea** | Ttaga | Resin | Tisane, decoction | 05 |
| *Argyrolobium saharae* Pomel. **Fabaceae** | Tea | Leaves and seeds | Infusion, tisane, decoction | 03 |
| *Globularia alypum* L. **Globulariaceae** | Tasselgha | Leaves and seeds | Tizane | 12 |
| *Thymus ciliatus* Desf. ssp. *eu-ciliatus* Maire. **Lamiaceae** | Zaitra or djertil | Leaves and flowers | Decoction, tisane | 05 |
| *Asphodelus microcarpus* Salzm.et Viv. **Liliaceae** | Berouag | Tubercle | Poudre | 18 |
| *Rhamnus alaternus* L. **Rhamnaceae** | Amlilece, Mlila | Fruit and leaves | Tizane | 02 |
| *Ruta chalepensis* L **Rutaceae** | Fidjel | Leaves and seeds | Tizane, decoction | 13 |
| *Daphne gnidium* L **Thymeleaceae** | Lazzaz | Leaves and seeds | Freshly, Masticatory | 05 |
| *Peganum harmala* L. **Zygophylacea** | Alharmal | Leaves and seeds | Tizane, decoction | 15 |

In these citations, different diseases treated with *Pituranthos scoparius* were surveyed; results are presented in Table 5.

**Table 5:** Percentage of the diseases treated by *Pituranthos scoparius* in Setif, Algeria.

| Diseases | Asthma | Jaundice | Rheumatic pains | Fever | Dermatosis | Others |
|---|---|---|---|---|---|---|
| Citations | 20% | 10% | 20% | 13% | 30% | 7% |

Table 6 presents the demographic data of 47 informants, whereas table 7 provided detailed information of *Pituranthos scoparius* parts and their preparation routes used for therapeutic purposes.

**Table 6:** Demographic data of the respondents

| Variables | Demographic classes | Citations |
|---|---|---|
| **Gender** | Male | 32 |
| | Female | 15 |
| **Age groups** | 30-45 | 17 |
| | 46-65 | 30 |
| **Education attainment** | Illiterate | 20 |
| | Primary | 9 |
| | Secondary | 11 |
| | Graduate | 7 |
| **Occupation** | Housewives | 9 |
| | Villagers | 23 |
| | Traditional healthers | 10 |
| | Herbalists | 5 |

**Table 7:** Plant parts and their preparation routes used for therapeutic purposes

| Part used | Preparation route | Traditional uses |
|---|---|---|
| **Stems** | Decoction and infusion | Dermatose, migraine, asthma, nervous breakdown, scorpion-bites and snake-bites |
| **Leaves** | Powder and decoction | |
| **Flowers** | Decoction and infusion | Relieve indigestion, stomach aches and lower abdomen |
| **Aerial part** | -Decoction<br>-Infusion | -Digestive disorders, diarrhea and eczema<br>- Postpartum care, diabetes, hepatitis, digestive disorders and urinary tract infections |

The documentation of this rich traditional ethnomedicinal knowledge has provided us with novel information of this undocumented knowledge about the methods of preparation and routes of application of *Pituranthos scoparius* (Figure 17).

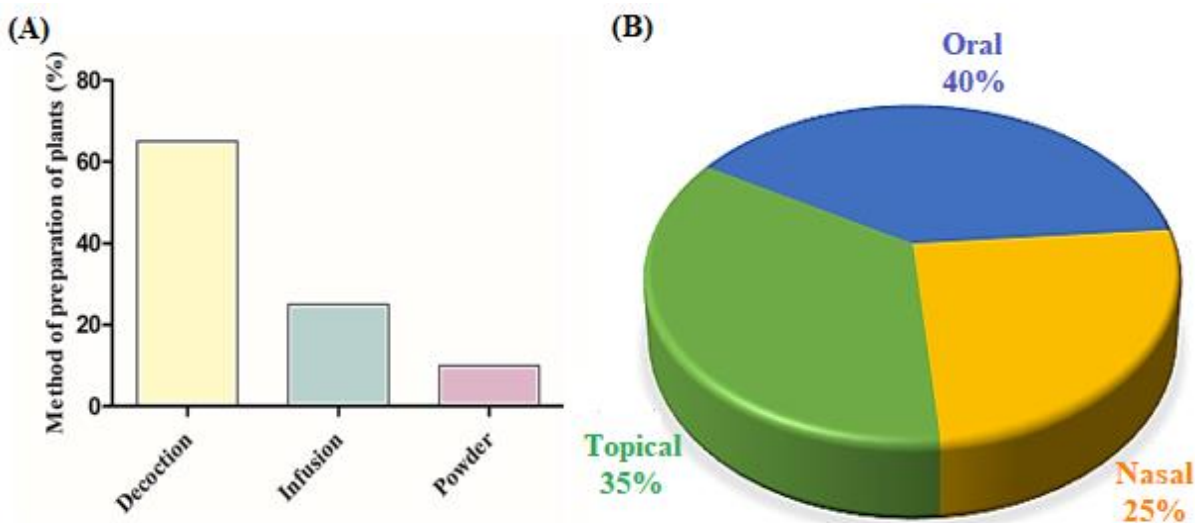

**Figure 17:** Methods of preparation (**A**) and routes of application (**B**) of *Pituranthos scoparius*

Results from ethnopharmacological survey revealed that most of the interviewers were males. In relationship to the age of the interviewed people, the predominance of older individuals demonstrated that erudition about *P. scoparius* is transferred from generation to another. The present study provides useful information about the traditional uses of *Pituranthos scoparius* used by local populations in various diseases. The future spread of this information helps the popular acceptance of uses phytoterapy.

## 2. Extractions and yields profile

Extracts yield of *P. scoparius* roots and stems were prepared by two methods: decoction method using water and by fractionation method using various polar and non-polar solvents. The estimate of yields in relation to the total powder weight (100 g) of plants, shows that the decoction extract (DecE), crude extract (CrE) and aqueous extract (AqE) of stems (S) from *P. scoparius* has the high yield which range from $8.15 \pm 0.71$ to $11.46 \pm 2.89$ % followed by the roots extracts with the yield which range from $4.86 \pm 0.50$ to $6.88 \pm 0.39$ % while the chloroform extracts (ChE) and ethyle acetate extracts (EaE) of stems and roots parts displayed a lower yields with range from $0.45 \pm 0.16$ to $0.76 \pm 0.19$ %. The yields of different extracts are summarized in Table 8.

**Table 8:** The extracts yields of stems (S) and roots (R) from *Pituranthos scoparius*

| % yield of extracts | S | R |
|---|---|---|
| **DecE** | 9.62 ± 0.20 | 6.88 ± 0.39 |
| **CrE** | 11.46 ± 2.89 | 6.36 ± 0.54 |
| **ChE** | 0.51 ± 0.24 | 0.76 ± 0.19 |
| **EaE** | 0.45 ± 0.16 | 0.71 ± 0.48 |
| **AqE** | 8.15 ± 0.71 | 4.86 ± 0.50 |

The results of the extraction of the plant material obtained show that the crude extract from the stems is recovered with a better yield compared to the roots extract. For this plant, our results indicate that extraction yield was substantially larger than that obtained by Adida et al. (2015); this could be ascribed to the solvent and extraction procedure employed. Extraction yield was used as an indicator of the effects of the extraction conditions. According to the study of Dhanani et al. (2017), the difference in polarities of extracting solvents might affect the solubility of chemical components in a sample and its extraction yield.

## 3. Phytochemical evaluation

### 3.1. Qualitative phytochemical analysis

Phytochemicals found in the *Pituranthos scoparius* stem extract include polyphenolic compounds such as tannins, flavonoids, phenolic acids, and others. These phytochemicals have attracted the attention of scientists and medicinal chemists due to their many biological activities and their impact on human health (Pereira et al., 2009). In this context, De Lima et al. (2018) reported that the use of crude phytochemical extracts from medicinal plants could be preferable and acceptable to the scientific community. This can be attributed to several factors including the cost of production, availability, and accessibility, and to lower toxicity in most cases compared to synthetic drugs.

Qualitative phytochemical studies were performed on extracts using suitable chemicals and reagents to confirm the presence of alkaloids, flavonoids, polyphenols, quinones, tannins, coumarins, anthraquinones, and reducing compounds. Our results revealed the presence of polyphenols, flavonoids, tannins and free quinones in both S and R extracts. However, alkaloids, anthraquinones, coumarins and reducing sugars were not found. On the other hand, the results revealed the presence of alkaloids, polyphenols, flavonoids, tannins, free quinones, and coumarins; nevertheless, reducing sugars and anthraquinones were not found in the *n*-butanol extract (ButE). The phytochemical screening results of the stems and roots extracts are reported in table 9.

**Table 9:** Phytochemical screening of different stems and roots extracts from *P. scoparius*

| Phytochemical compound | S and R | | | | | S |
|---|---|---|---|---|---|---|
| | DecE | CrE | ChE | EaE | AqE | ButE |
| Alkaloids | - | - | - | - | - | + |
| Polyphenols | + | + | + | + | + | + |
| Flavonoids | + | + | + | + | + | + |
| Tannins | + | + | + | + | + | + |
| Free quinones | + | + | + | + | + | + |
| Anthraquinones | - | - | - | - | - | - |
| Coumarines | - | - | - | - | - | + |
| Reducing sugars | - | - | - | - | - | - |

+/-: Presence/absence of the compound.

However, the absence of some phytochemicals in plant extracts does not mean that these compounds are not present. The extraction methods used in the studies might not be efficient enough to extract desired metabolites. For example, phytochemical analysis reported by some researchers (Adida et al., 2014; Adida et al., 2015) on the *P. scoparius* extracts revealed the absence of anthraquinones, which does agree with our results. Similarly, Adida et al. (2015) reported the absence of alkaloids, and Benalia et al., (2016) reported the absence of coumarins in *P. scoparius* extracts, which does not corroborate with our results for *n*-butanol extract (ButE).

## 3.2. Quantitative phytochemical analysis

### 3.2.1. Determination of total polyphenol, flavonoids and tannins

Quantitative tests were performed to determine the concentration of the phytochemicals in the extracts. Polyphenols, flavonoids and tannins were quantified using spectrometric methods. The total phenolic contents (TPC), total flavonoids contents (TFC) and tannins contents (TC) of different *P. scoparius* stems and roots extracts were evaluated employing the Folin-Ciocalteu reagent, aluminum chloride and haemoglobin precipitation methods, respectively. Results are shown in Table 10 (**A**), that TPC in the stems extracts has the high amounts which ranging from $76.82 \pm 1.24$ to $434.34 \pm 2.75$ µg gallic acid equivalent (GE)/ mg dried extract (DE) compared to the roots extracts which range from $30.62 \pm 0.97$ to $109.28 \pm 1.46$ µg GAE/ mg DE. Whereas the TFC in the SE varied from $0.81 \pm 0.38$ to $207.49 \pm 1.03$ µg quercetin equivalent (QE)/ mg DE, while in the RE displayed a lower TFC with rat varied from $0.77 \pm 0.12$ to $9.96 \pm 0.58$ µg QE/ mg DE. Table 10 (**B**) also presents the relative contents of tannins in these extracts. In general TPC, TFC and TC levels exhibited a great difference among various stems and roots extracts. The highest level of TPC, TFC and TC levels were recorded in the SE especially in EaE followed by CrE which exhibited a value higher than those of RE.

From these results, it is clear to conclude that the highest total phenolics, flavonoids and tannins contents were found in ethyl acetate ($434.34 \pm 2.75$ µg GE; $207.49 \pm 1.03$ µg QE and $126.32 \pm 1.32$ µg TE/ mg DE, respectively).

Therefore, the selection of an appropriate solvent system is one of the most relevant steps in optimizing the recovery of total phenolic contents and other bioactive compounds from extracts.

**Table 10:** Total polyphenols, flavonoids (**A**) and tannins (**B**) contents of stems and roots extracts from *P. scoparius.* The values are expressed as mean ± SD (*n*=3).

| (A) | *Total phenolic content (µg gallic acid equivalent (GAE) / mg dried extract (DE))* | | *Total flavonoids content (µg quercetin equivalent (QE) / mg dried extract (DE))* | |
|---|---|---|---|---|
| *Extracts* | S | R | S | R |
| **DecE** | 150.89 ± 0.68 | 50.04 ± 0.73 | 0.81 ± 0.38 | 1.31 ± 0.07 |
| **CrE** | 125.37 ± 0.97 | 43.60 ± 1.70 | 24.68 ± 0.2.19 | 0.77 ± 0.12 |
| **ChE** | 233.95 ± 0.19 | 79.86 ± 0.97 | 4.21 ± 0.51 | 6.10 ± 0.02 |
| **EaE** | 434.34 ± 2.75 | 109.28 ± 1.46 | 207.49 ± 1.03 | 1.22 ± 0.20 |
| **AqE** | 76.82 ± 1.24 | 30.62 ± 0.97 | 6.41 ± 0.32 | 9.96 ± 0.58 |
| **ButE** | 143.86 ± 0.97 | / | 60.11 ± 0.51 | / |

| (B) *Extracts* | *Tannins content (µg tannic acid equivalent (TE) / mg dried extract(DE))* | |
|---|---|---|
| | S | R |
| **DecE** | 71.24 ± 0.08 | 40.64 ± 0.04 |
| **CrE** | 133.37 ± 0.97 | 37.51 ± 0.04 |
| **ChE** | 57.70 ± 0.26 | 67.85 ± 1.23 |
| **EaE** | 126.32 ± 1.32 | 72.60 ± 0.30 |
| **AqE** | 87.87 ± 0.97 | 43.64 ± 1.19 |

## 4. Structure elucidation of compounds

### 4.1. Chemical constituents of the *n*-butanol extract of *P. scoparius*

The *n*-butanol extract (**PSB**) was chromatographed on a silica gel S column, and eluted with $CHCl_3$: MeOH mixtures of increasing polarity. 126 Fractions were collected (450 to 500 mL each) and divided into 4 groups according to their TLC behavior. The butanol extract of this plant gave upon Column Chromatography (CC) and Thin Layer Chromatogrphy (TLC) four compounds. These compounds are: isorhamnetin-3-*O*-β-glucoside (**PSB1**) (Yuan et al., 2006; Wang and Zeng, 2019), isorhamnetin-3-O-β-apiofuranosyl (1→2)-β glucopyranoside (**PSB2**), D-mannitol (**PSB3**) (Branco et al., 2010), and isorhamnetin-3-O-β-glucopyranosyl-(1→6)-β-glucopyranoside (**PSB4**) (Hsieh et al., 2004) were isolated from *n*-butanol extract (ButE). All four compounds were isolated in pure form, as indicated by their TLC profiles and NMR spectra, for the first time from *P. scoparius*.

## 4.2. Identification and spectral data of isolated constituents

### 4.2.1. Isorhamnetin-3-*O*-β-glucoside (PSB1)

After loading fraction **I** on sephadex column, an impure yellow solid was obtained. Several washings with distilled methanol achieved further purification. The resulting bright yellow solid was identified as isorhamnetin-3-*O*-β-glucoside (Figure 18). This flavonoid was isolated for the first time from stems of *Pituranthos scoparius*.

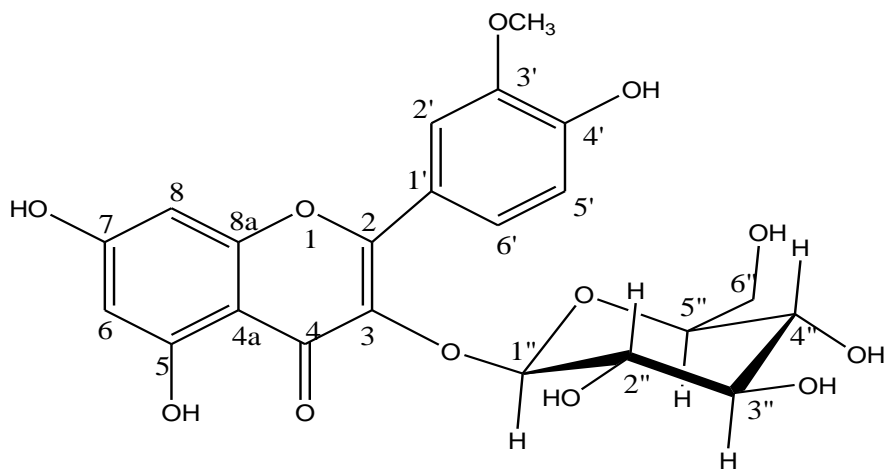

**Figure 18:** Chemical structure of isorhamnetin-3-*O*-β-glucoside.

$^1$H-NMR (DMSO-$d_6$, 500 MHz): 3.80 (3H, s, OCH$_3$), 3.55–5.33 (m, sugar protons), 5.53 (1H, d, *J* = 7.1 Hz, H-1″), 6.17 (1H, s, H-6), 6.41 (1H, s, H8), 6.88 (1H, d, *J* = 7.9 Hz, H-5′), 7.66 (1H, bd, *J* = 7.9 Hz, H-6′), 8.06 (1H, bs, H-2′), 12.54 (1H, bs, 5-OH) as presented in Figure 19.

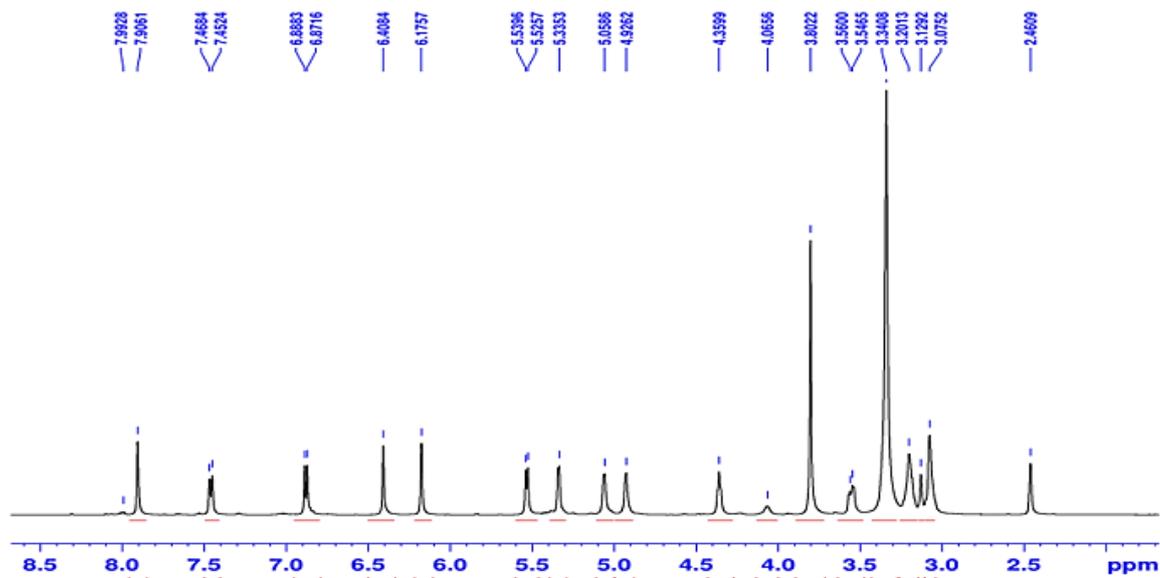

**Figure 19:** $^1$H-NMR (DMSO-$d_6$) spectrum of isorhamnetin-3-$O$-β-glucoside

$^{13}$C-NMR (DMSO-$d_6$, 125 MHz): 56.1 (OCH$_3$), 66.7 (C-6″), 70.0 (C-3″), 70.2 (C-4″), 74.2 (C-2″), 75.8 (C-5″), 94.2 (C-6), 99.2 (C-8), 101.2 (C-1″: anomeric carbon), 113.9 (C-5′), 115.7 (C-2′), 121.5 (C-1′), 122.5 (C-6′), 133.4 (C-3), 147.4 (C-4′), 149.9 (C-3′), 156.8 and 156.9 (C-2 and C-8a), 161.7 (C-5), 104.5 (C-4a), 164.6 (C-7), 177.9 (C-4) ppm as presented in Figure 20.

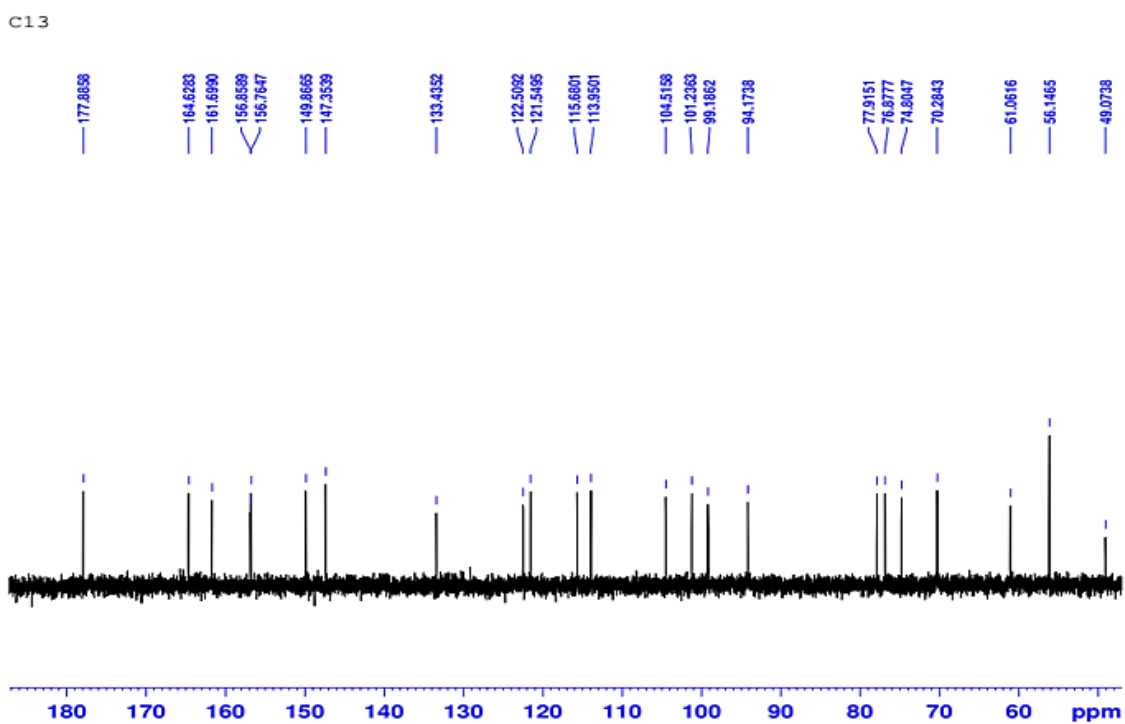

**Figure 20:** $^{13}$C-NMR (DMSO-$d_6$) spectrum of isorhamnetin-3-$O$-β-glucoside

HRMS (ESI) *m/z*: calcd for $C_{22}H_{21}O_{12}$ [M − H]⁻ 477.10330, found 477.10565 (corresponding to the molecular formula: $C_{22}H_{22}O_{12}$). NMR spectral data are in agreement with the literature (Touil et al., 2006).

UV λ$_{max}$ (MeOH) nm: 213, 256 (band II), 358 (band I); + MeO⁻ Na⁺: 214, 271, 330, 413; + $AlCl_3$: 215, 267, 363, 399, + $AlCl_3$ + HCl: 215, 268, 399.

### 4.2.2. Isorhamnetin-3-*O*-β-apiofuranosyl (1→2)-β-glucopyranoside (PSB2)

After loading fraction **II** on sephadex column, the mother liquor of fraction **II** gave a major UV active spot. This spot was purified using TLC and $CHCl_3$: MeOH: $H_2O$ (70:28:2) as an eluent. Then, was collected, washed with 30% MeOH/$CH_2Cl_2$ mixture, and the solvent was then evaporated under vacuum. The resulting amorphous compound was identified as the rare isorhamnetin-3-*O*-β-apiofuranosyl (1→2)-β-glucopyranoside (Figure 21). This glucoside flavonoid was isolated for the first time from *Pituranthos scoparius*.

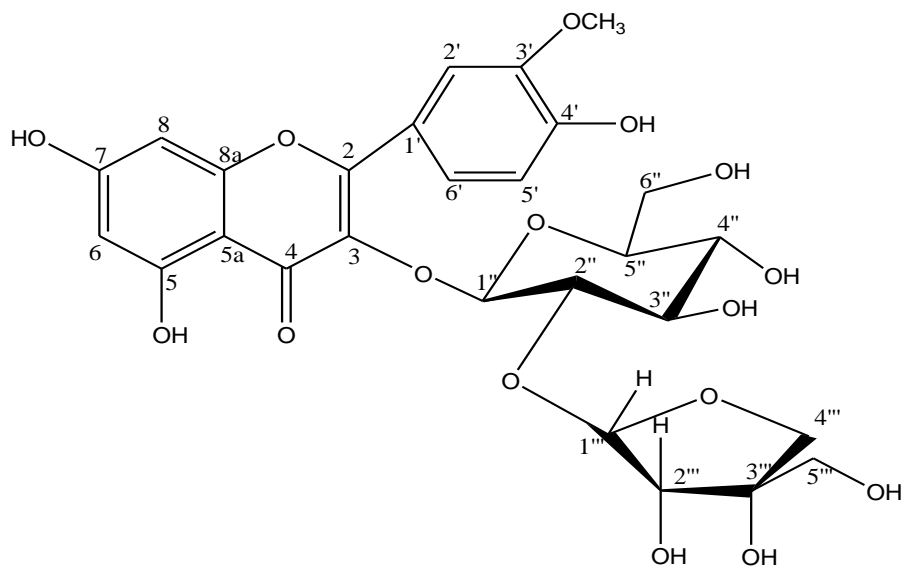

**Figure 21:** Chemical structure of isorhamnetin-3-O- β-apiofuranosyl (1→2)-β-glucopyranoside,

¹H-NMR (DMSO-*d₆*, 500 MHz): 3.80 (3H, s, $OCH_3$), 3.00–3.77 (m, sugar protons), 5.28 (1H, bs, H-1‴), 5.63 (1H, d, *J* = 7.9 Hz, H-1″), 6.15 (1H, s, H-8), 6.39 (1H, s, H-6), 6.89

(1H, d, *J* = 8.5 Hz, H-5'), 7.50 (1H, bd, *J* = 8.5 Hz, H-6′), 7.90 (1H, bs, H-2′), 12.58 (1H, bs, 5-OH) as presented in Figure 22.

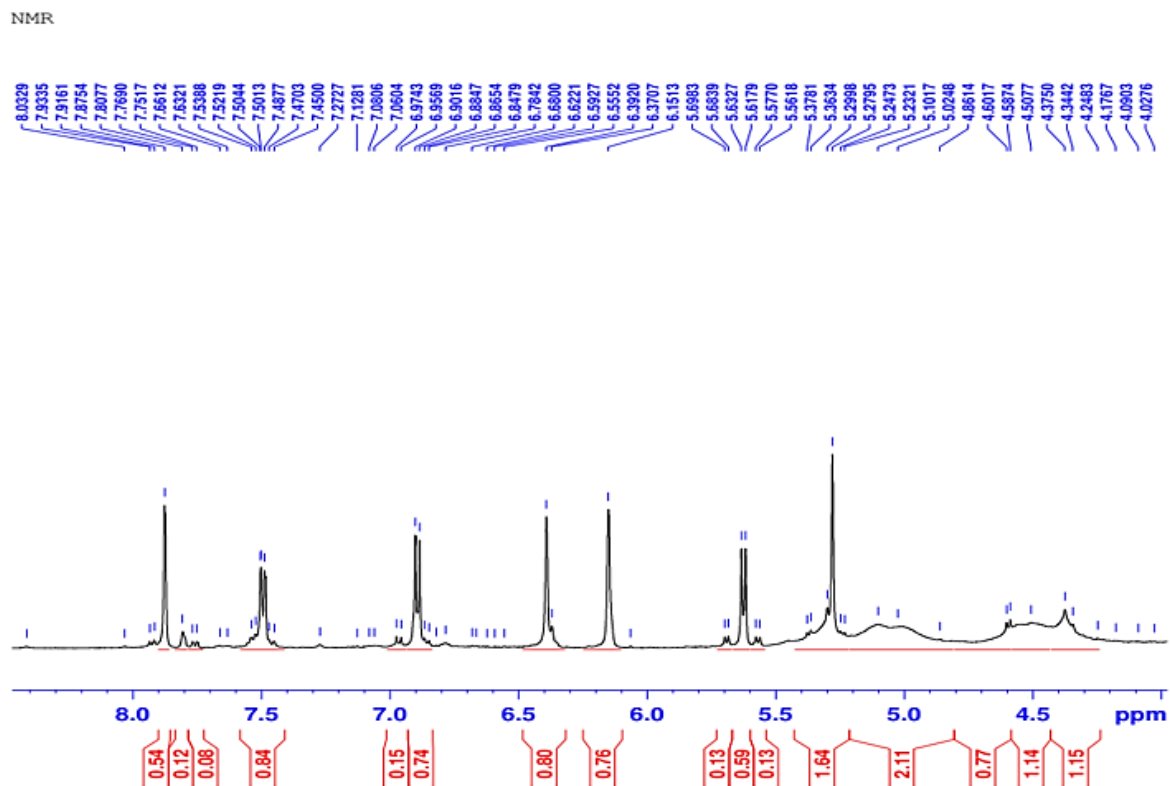

**Figure 22:** $^1$H-NMR (DMSO-$d_6$) spectrum of isorhamnetin-3-*O*-β-apiofuranosyl (1→2)-β-glucopyranoside

$^{13}$C-NMR (DMSO-$d_6$, 125 MHz): 56.1 (OCH$_3$), 60.9 (C-6″), 64.8 (C- 5‴), 70.5 (C-4″), 74.5 (C-4‴), 76.5 (C-2‴), 77.4 (C-5″), 77.5 (C-3″), 77.9 (C-2″), 79.8 (C-3‴), 94.3 (C-8), 99.1 (C-6), 99.5 (C-1″), 104.0 (C-5a), 109.0 (C-1‴), 113.7 (C-2′), 115.7 (C-5′), 121.5 (C-1′), 122.5 (C-6′), 133.2 (C-3), 147.4 (C-4′), 150.0 (C-3'), 156.1 (C-8a), 156.8 (C-2), 161.6 (C-5), 165.0 (C-7), 177.5 (C-4) ppm as presented in Figure 23. HRMS (ESI) *m/z*: calcd for C$_{27}$H$_{29}$O$_{16}$ [M – H]$^−$ 609.14556, found 609.12873 (corresponding to the molecular formula C$_{27}$H$_{30}$O$_{16}$).

UV λ$_{max}$ (MeOH) nm: 212, 256, 357; + MeO$^−$Na$^+$: 214, 272, 330, 406, + AlCl$_3$: 215, 269, 360, 400; + AlCl$_3$ + HCl: 213, 268, 357, 400.

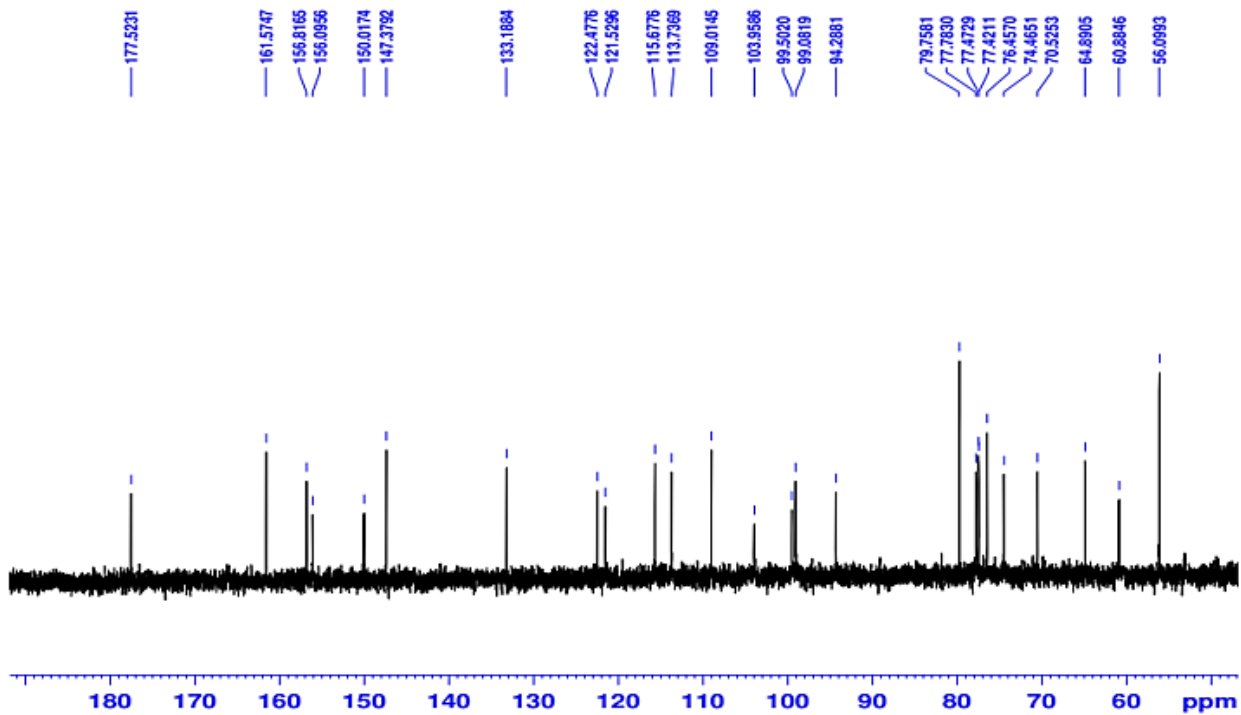

**Figure 23:** $^{13}$C-NMR (DMSO-d$_6$) spectrum of isorhamnetin-3-*O*-β-apiofuranosyl (1→2)-β-glucopyranoside

### 4.2.3. D-Mannitol (PSB3)

A large amount of an impure solid precipitated from the *n*-butanol extract (Mass = 3 g). Also, after loading fraction **III** on sephadex column, a solid was obtained. The solid was washed with distilled methanol to give a white solid (100 mg) identified as D-Mannitol (Figure 24). This compound was isolated for the first time from *Pituranthos scoparius*.

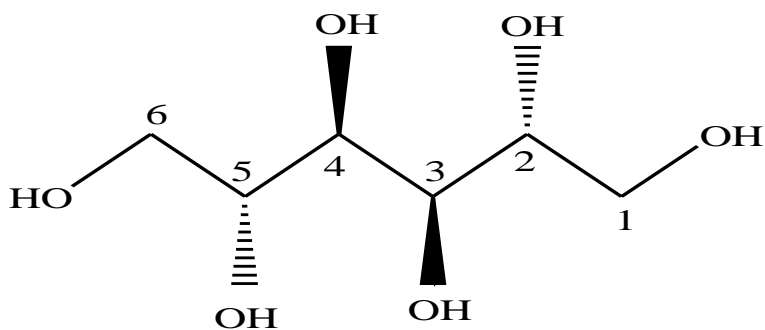

**Figure 24:** Chemical structure of D-mannitol

$^1$H-NMR (DMSO-$d_6$, 500 MHz): 3.31–3.60 (m, 8 protons), 4.10 (1H, d, $J$ = 7.0 Hz, H-2,5), 4.30 (4H, t, $J$= 5.5 Hz, H-1,6), 4.33 (2H, t, $J$= 5.3 Hz, H-3,4) as presented in Figure 25.

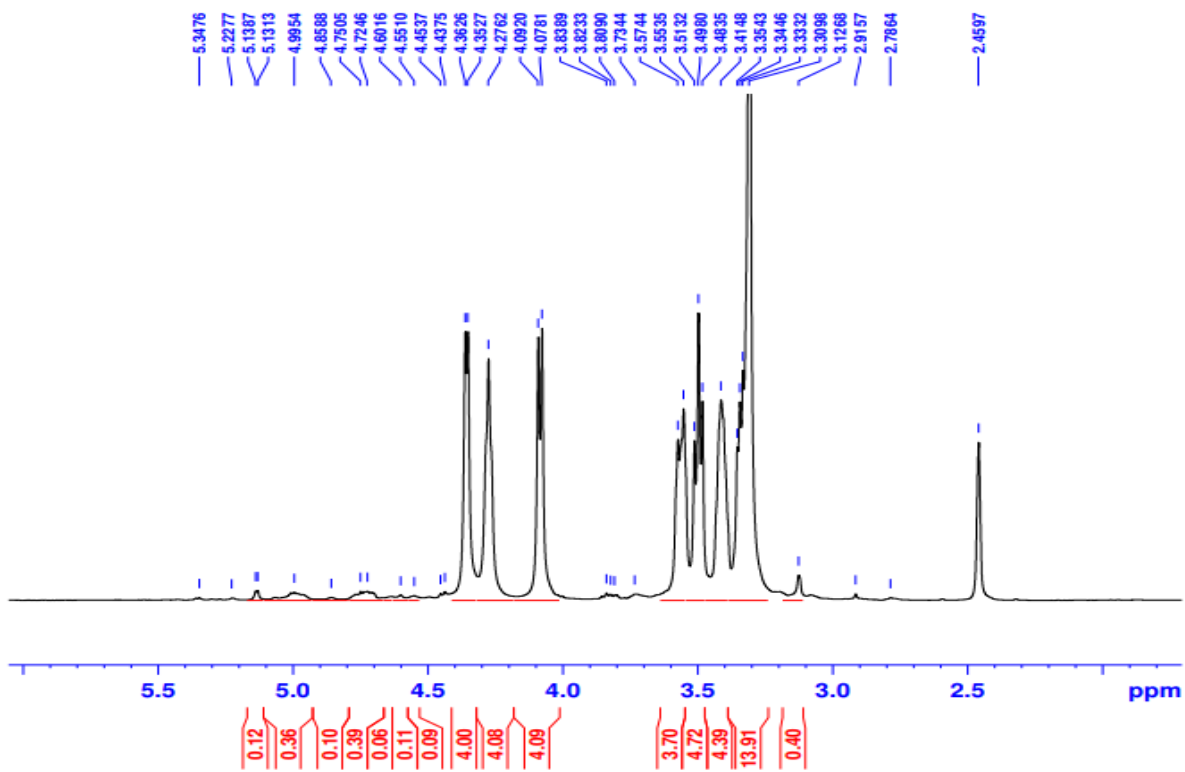

**Figure 25:** $^1$H-NMR (DMSO-$d_6$) spectrum of D-Mannitol

$^{13}$C-NMR (DMSO-$d_6$, 125 MHz): 64.3 (C-1,6), 70.1 (C-2,5), 71.8 (C-3,4) as presented in Figure 26. HRMS (ESI) *m/z*: calcd for $C_6H_{13}O_6$ [M – H]$^-$ 181.07121, found 181.07176 (corresponding to the molecular formula $C_6H_{14}O_6$). NMR spectral data are in agreement with the literature (Branco et al., 2010).

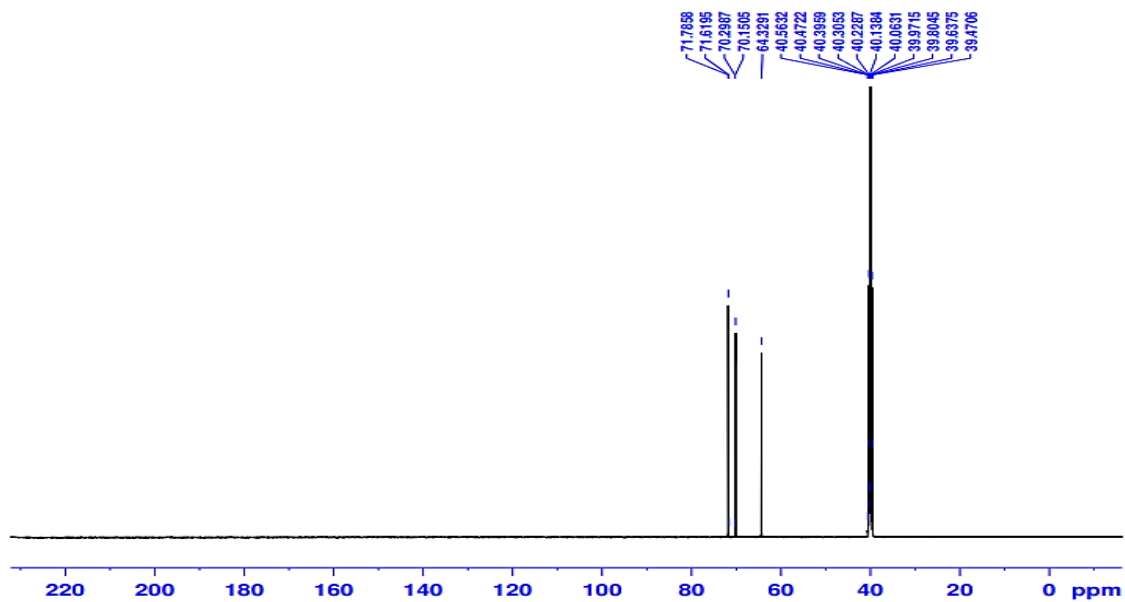

**Figure 26:** $^{13}$C-NMR (DMSO-$d_6$) spectrum of D-Mannitol

### 4.2.4. Isorhamnetin-3-O-β-glucopyranosyl-(1→6)-β-glucopyranoside (PSB4)

After loading fraction IV on sephadex column, an impure yellow solid was obtained upon treatment of fraction IV with methanol. Further washing with methanol gave a pure yellow solid, identified as isorhamnetin-3-O-β-glucopyranosyl-(1→6)-β-glucopyranoside (Figure 27). This di-glucoside flavonoid was isolated for the first time from *Pituranthos scoparius*.

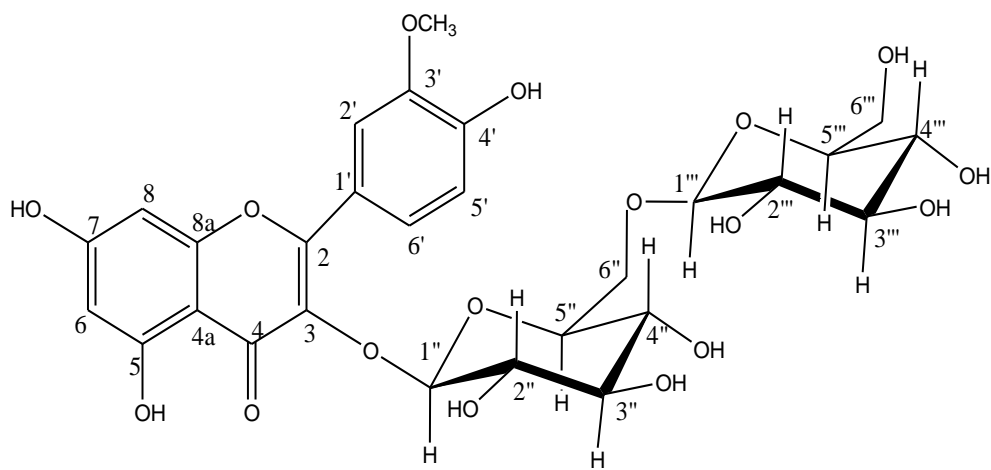

**Figure 27:** Chemical structure of isorhamnetin-3-O-β-glucopyranosyl-(1→6)-β-glucopyranoside.

$^1$H-NMR (DMSO-$d_6$, 500 MHz): 3.80 (3H, s, OCH$_3$), 3.55–5.33 (m, sugar protons), 5.47 (1H, d, *J* = 6.7 Hz, H-1″), 6.17 (1H, s, H-6), 6.41 (1H, s, H-8), 6.88 (1H, d, *J* = 7.9 Hz, H-

5′), 7.46 (1H, bd, *J* = 7.9 Hz, H-6'), 7.89 (1H, bs, H-2′), 12.54 (1H, bs, 5-OH) as presentd in Figure 28.

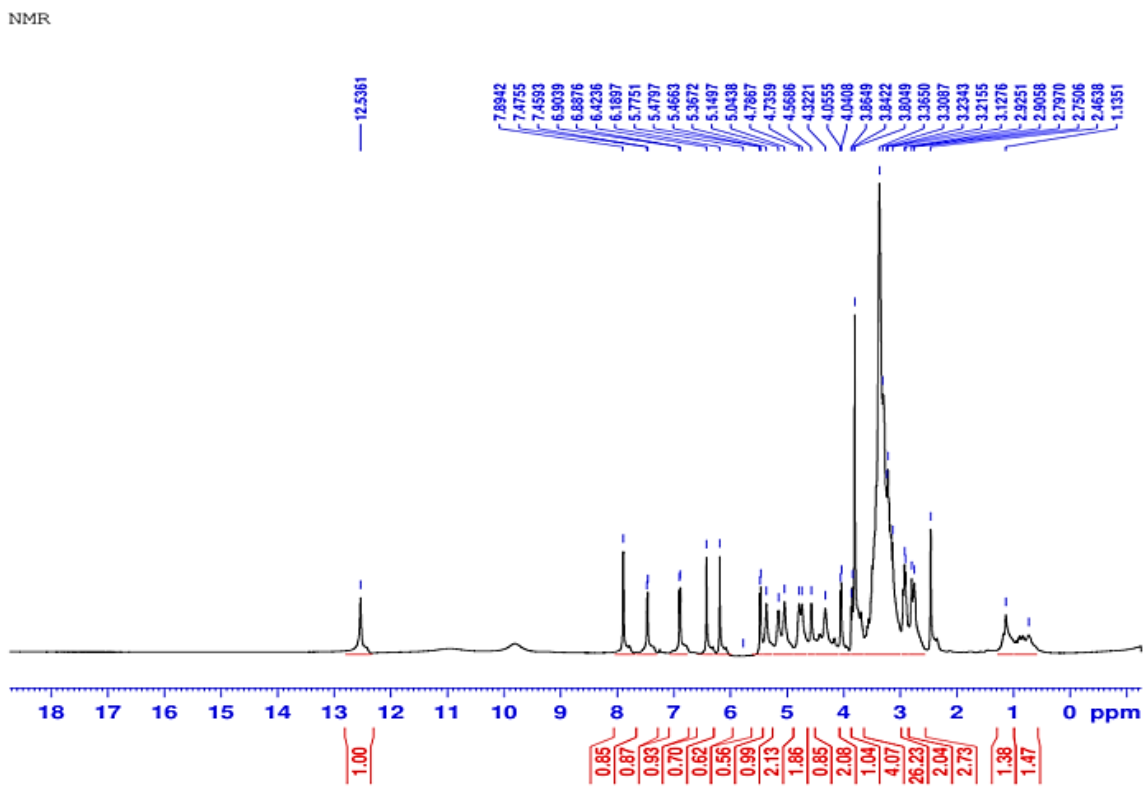

**Figure 28:** ¹H-NMR (DMSO-$d_6$) spectrum of isorhamnetin-3-O-β-glucopyranosyl-(1→6)-β-glucopyranoside

¹³C-NMR (DMSO-$d_6$, 125 MHz): 56.1 (OCH$_3$), 61.5 (C-6‴), 66.7 (C-6″), 68.2 (C-5‴), 70.0 (C-3″), 70.2 (C-4″), 70.5 (C-2‴), 71.7 (C-4‴), 74.2 (C-2″), 75.8 (C-5″), 76.3 (C-3‴), 94.2 (C-6), 99.2 (C-8), 103.6 (C-1‴), 101.2 (C-1″), 113.9 (C-5′), 115.7 (C-2′), 121.5 (C-1′), 122.5 (C-6′),133.4 (C-3), 147.4 (C-4′), 149.9 (C-3'), 156.8 (C-2), 156.9 (C-8a), 161.7 (C-5), 104.5 (C-4a), 164.6 (C-7), 177.9 (C-4) as presented in Figure 29.

HRMS (ESI) *m/z*: calcd. For C$_{28}$H$_{32}$NaO$_{17}$ [M + Na]$^+$ 663.15372, found 663.15317 (corresponding to the molecular formula C$_{28}$H$_{32}$O$_{17}$). NMR spectral data are in agreement with the literature (Yen et al., 2009).

UV λ$_{max}$ (MeOH) nm: 213, 255, 359; + MeO⁻ Na⁺: 216, 271, 330, 415; + AlCl$_3$: 215, 268, 364, 399, + AlCl$_3$ + HCl, 215, 268, 365, 402.

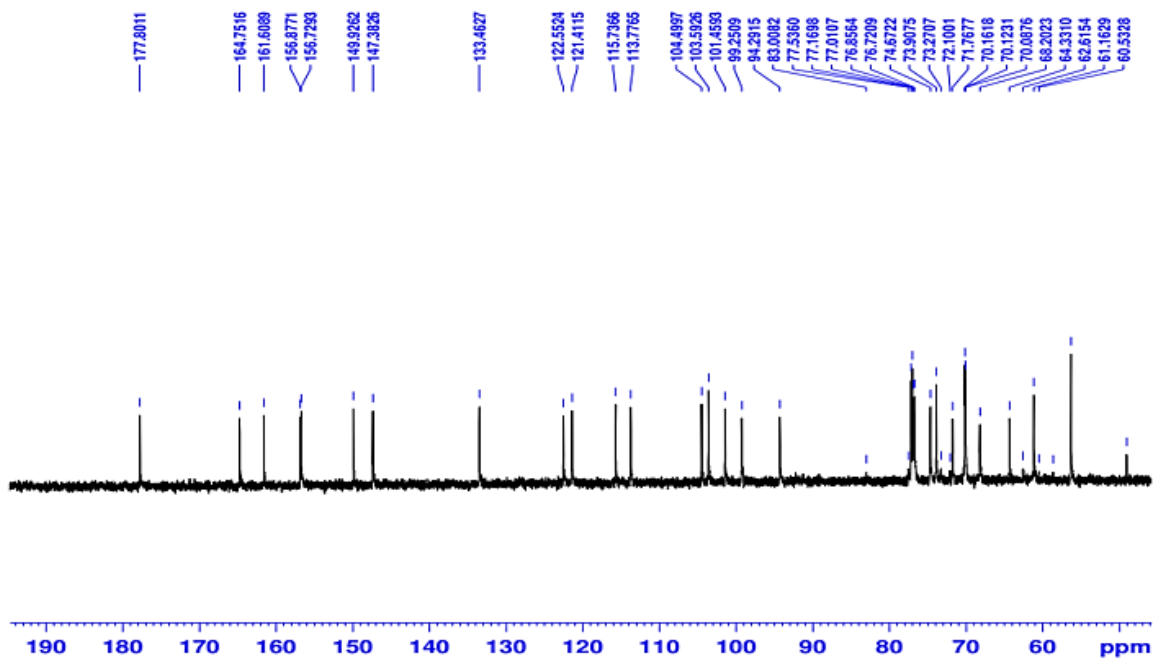

**Figure 29:** $^{13}$C-NMR (DMSO-d$_6$) spectrum of isorhamnetin-3-O-β-glucopyranosyl-(1→6)-β-glucopyranoside

Structures of the isolated compounds **PSB1**, **PSB3**, and **PSB4** were confirmed by mass spectrometry (MS) and NMR spectral data. They were identified by comparison of their spectroscopic properties with literature data. These data, detailed in the experimental part, are consistent with the suggested structures. Thus, the mass spectra displayed the correct molecular ion peaks for which the measured high resolution (HRMS) data are in good agreement with the calculated values. In addition, structures were verified by comparing their NMR data with those reported in the literature. These known compounds were isolated for the first time from the *n*-butanol stem extract of this species.

Compound **PSB2** was isolated as an amorphous yellow solid. HRMS showed a molecular ion peak at *m/z*: 609.12873 [M – H]⁻, corresponding to the correct molecular formula $C_{27}H_{30}O_{16}$. $^1$H-NMR and $^{13}$C-NMR spectra showed signals for an isorhamnetin and a glucosylapioside moieties. The anomeric proton of the glucoside appeared as a doublet at δ 5.63 (*J* = 7.4 Hz) (δ C = 99.5) indicating a β-glucoside, whereas that of the apiose moiety appeared as a broad singlet at δ 5.28 (δ C = 109.0) confirming the β-stereochemistry of the apioside. The linkage of apiose at C-2" of glucose was evident from the shift of the δ value of this carbon which normally appears around 74 ppm to 77.9 ppm in compound **PSB2**. $^1$H- and $^{13}$C-NMR chemical shifts of the compound are comparable to those reported for isorhamnitine-3-glucosyl apioside (Aspers et al., 2002). The location of the methoxyl group at C-3' and the sugar unit at C-3 was evident from the UV-Vis spectrum of compound **PSB2**. The spectrum in methanol (MeOH) showed absorption bands at 256 (band II) and 357 nm (band I). Upon addition of MeO$^-$ Na$^+$, band I was shifted to 406 nm indicating a free OH group at C-4' and a new band appeared at 330 nm indicating a free OH at C-7. In addition, the spectrum of the compound in MeOH + AlCl$_3$ was not affected by the addition of HCl confirming the abscence of two ortho hydroxyl groups in ring B. These results confirm the location of the methoyl group and the sugar unit at C-3' and C-3, respectively. Further confirmation of the location of the methoxyl group was evident from the HMBC correlation of the methoxyl $^1$H-NMR signal (δ = 3.80) with the $^{13}$C-NMR signal of C-3' (δ = 150.0).

## 5. Pharmacological investigations

### 5.1. Screening of *in vitro* antioxidant activity

#### 5.1.1. DPPH Radical-scavenging assay

Many plants extracts exhibit efficient antioxidant properties due to their phyto-constituents, including phenolic acids and flavonoids (Barriada-Bernal et al., 2013). In the present study, reduction of DPPH• radicals can be significantly observed at 517 nm by the stems (S) and roots (R) extracts of *P. scoparius*. Measured by DPPH• method (Figure 30) the free radical scavenging activity (IC$_{50}$) of the stems extracts (SE) and the roots extracts (RE) of *P. scoparius* are ranged from 42.16 to 126.39 µg/mL and from 176.49 to 356.44 µg/mL, respectively compared to BHT as standard (87.26 µg/mL). The free radical scavenging activity of the EaE and ChE from the stems of the plant were the best showing an IC$_{50}$= 42.16 and 56.72 µg/mL, respectively. This is due in part to the presence of phenolic compounds and flavonoids. This radical scavenging activity of extract could be related to the nature of phenolics (Hammami et al., 2011), thus contributing to their electron transfer or hydrogen donating ability.

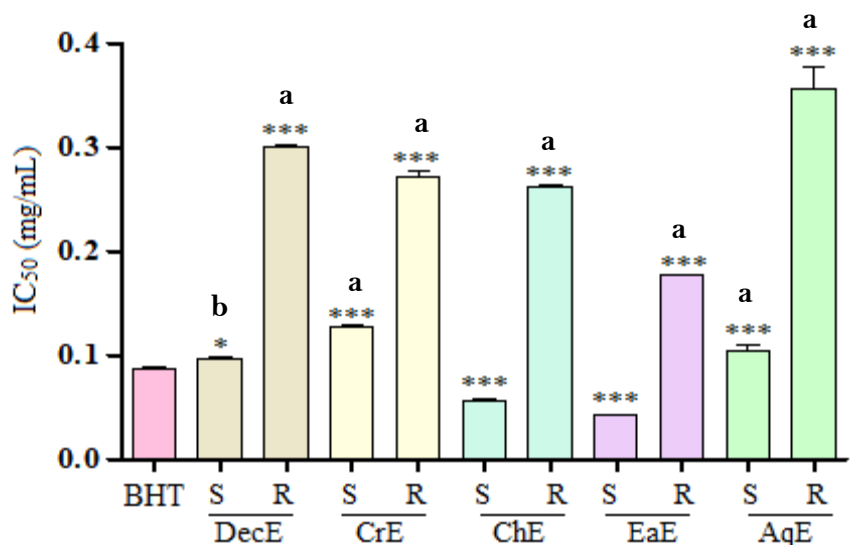

**Figure 30:** Free radical scavenging activity of stems (S) and roots (R) extracts of *Pituranthos scoparius* using DPPH assay. Results represent mean ± SD ($n = 3$). *: $p < 0.1$, ***: $p < 0.001$ compared to the control.

Recapitulated results revealed that SE carry the antioxidative effect for free radical scavenging (DPPH•), these property may be due to the combined activity of the above mentioned bioactive components with other components, most specifically, polyphenolic compounds that have been previously reported to be responsible for the antioxidant and are proficient of donating hydrogen to a free radical (DPPH•) to convert them to n on-reactive species. Numerous phenolic compounds such as flavonoids, phenolic acids, tannins and other compounds are widely distributed in plants and have gained much attention due to their antioxidant activities and free radical scavenging abilities, which potentially have benefit for human health (Pereira et al., 2009). Strong DPPH radical scavenging activity of EaE of SE can be due to higher content of total phenolic and tannin compounds.

**5.1.2. ABTS radical scavengig assay**

ABTS assay is considered as an important method for estimating antioxidant activity. This method is based on the spectrophotometric measurement of ABTS cation radical (ABTS$^{·+}$) concentration changes resulting from the ABTS$^{·+}$ reaction with antioxidants (Dawidowicz and Olsowy, 2011). In the present study, the ABTS radical scavenging activity was determined for all extracts of *P. scoparius* (Figure 31), the ABTS radical scavenging activity of the SE is ranged from values of 9.60 to 27.83 µg/mL, whereas, the IC$_{50}$ of RE is ranged from values 17.84 to 69.30 µg/mL. The highest ABTS radical scavenging activity was found in EaE and ChE with values of 9.60 and 10.04 µg/mL, respectively.

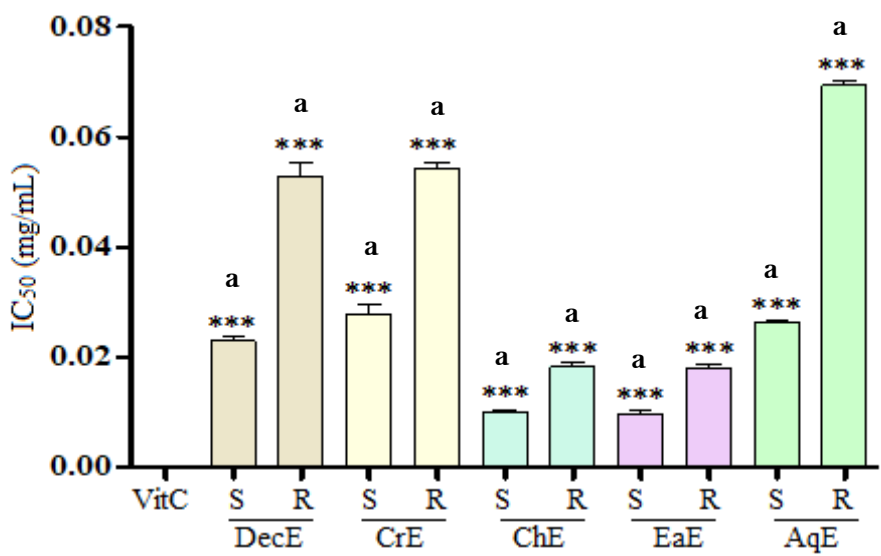

**Figure 31:** Free radical scavenging activity of stems (S) and roots (R) extracts of *P.scoparius* using ABTS assay. Data were presented as means ± SD (*n*=3). \*\*\*: *p* <0.001 compared to the control. VitC: Vitamin C

The scavenging of the ABTS$^+$ radical by the extracts was found to be much higher than that of DPPH radical. Factors like stereoselectivity of the radicals or the solubility of the extract in different testing systems have been reported to affect the capacity of extracts to react and quench different radicals (Adedapo et al., 2009).

**5.1.3. β-Carotene bleaching by linoleic acid assay**

β-carotene/linoleic acid assay determines the inhibition ratios of the oxidation of linoleic acid as a method to study the ability of SE and RE to inhibit lipid peroxidation. β-carotene in this method undergoes rapid discoloration in the absence of antioxidants. The addition of the extracts of *P. scoparius* or BHT used prevented the bleaching of β-carotene. High inhibition percentage indicates higher antioxidant activity. In this test, all extracts presented significant ability and exhibited varying degrees of antioxidant capacity in the ß-carotene/linoleic acid test (Figure 32).

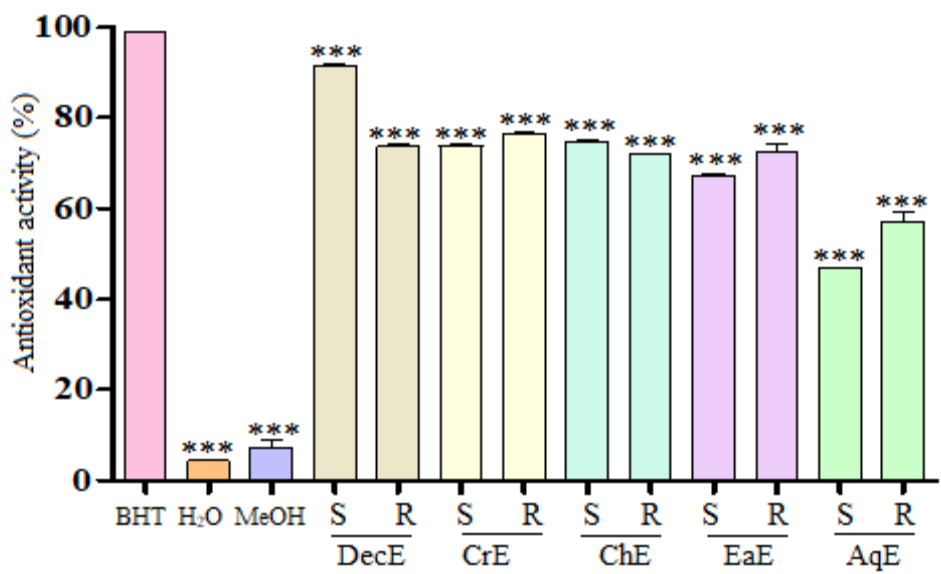

**Figure 32:** Antioxidant capacity of stems (S) and roots (R) extracts of *P.scoparius* (2 mg/mL at 24 h of incubation) measured by β-carotene bleaching method. Values are represent as means ± SD ($n=3$). ***: $p<0.001$ compared to BHT as standard and ($H_2O$ and methanol) as blanks.

Results showed that the SE inhibiting the percentage of β-carotene bleaching ranged from values of 46.86 ± 0.22 % to 91.53 ± 0.98 %, whereas, RE present an inhibition percentage of β-carotene bleaching ranged from 55.24 ± 0.73 % to 76.73 ± 0.65 %). The relatively highest antioxidant activity was obtained in DecE of stems (91.53 ± 0.98 %) followed by CrE of roots (76.73 ± 0.65 %). The results obtained from extracts of *P. scoparius* root were all significantly different ($p < 0.05$), compared to BHT as standard (99.13% ± 0.08%). Results obtained from experimental data revealed that there might be a correlation between total phenolic and antioxidant capacity of different extracts of *P. scoparius*. However, some literature demonstrated that antioxidant was not solely dependent on phenolic content, but it may be due to other phytoconstituents as tannins, triterpenoid or combined effect of them.

Different types of phenolic compounds have different antioxidant activity, which mainly depends on their structure as extract contains different types of phenolic compounds, which have different antioxidant capacities (Tatiya et al., 2011).

**5.1.4. Reducing power assay**

Figure 33 shows the antioxidant activity curves of the reducing powers of SE and RE. In this assay, all the extracts showed a degree of electron donation ability to reduce $Fe^{3+}$ to $Fe^{2+}$ form by donating an electron. Increased absorbance of the final reaction mixture at 700 nm indicates a greater reducing ability of the compounds (Ates et al., 2008). In the current test, the reducing power activity was determined for all extracts of *P. scoparius* (Figure 25), the reducing power of the SE is ranged from values of 38.61 to 131.27 µg/mL, whereas, the $IC_{50}$ of RE is ranged from values of 93.14 to 338.86 µg/mL. EaE and ChE of stems (S) showed significantly the highest reducing potential than other extracts with an $IC_{50} = 38.61$ µg/mL and 61.16 µg/mL, respectively (Figure 33).

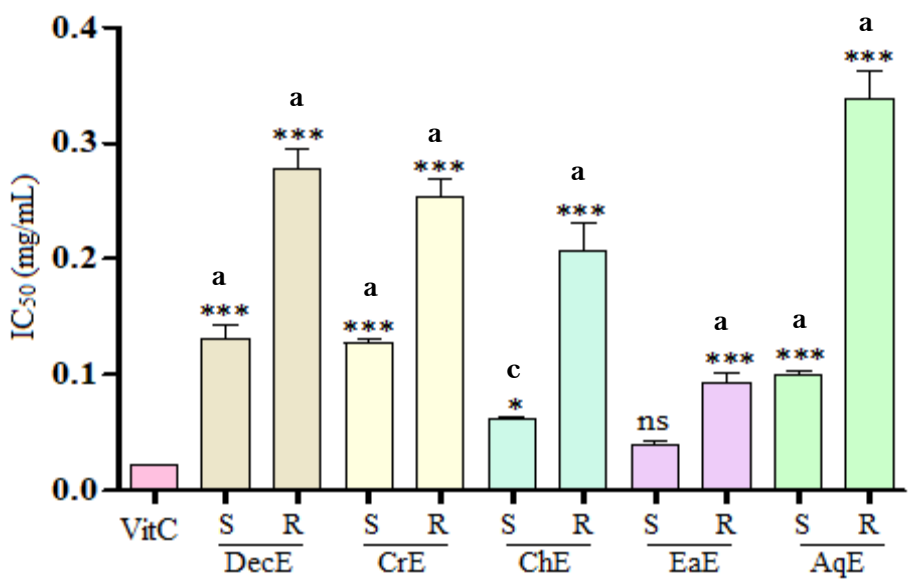

**Figure 33:** Reducing power activity of stems (S) and roots (R) extracts of *P. scoparius* extracts. Values are expressed as means±SD (*n*=3). ns: no significant difference, *: $p < 0.05$, ***: $p < 0.001$ as compared with the control.

The reducing power of the extracts may provide a significant indication of the potential antioxidant capacity of the plant. The reducing properties linked with the presence of reductones, which have been shown to exhibit antioxidant action by breaking the chain reactions by donating a hydrogen atom. Reductones are also reported to react with certain precursors of peroxide, thus preventing peroxide formation (Ates et al. 2008).

**5.1.5. Hydroxyl radical scavenging assay**

Hydroxyl radicals were generated by the Fenton reaction could oxidize $Fe^{2+}$ to $Fe^{3+}$ which is reflected by the degree of decolorization of the reaction solution. In this assay, OH·radicals were generated using a system containing $FeSO_4$ and $H_2O_2$ and detected by their ability to hydroxylate salicylate, and the absorption was measured at 562 nm. Vitamin C was used as a standard antioxidant for comparison. The hydroxyl radical scavenging activity of SE ranged from values of 375.15 to 1185.71 µg/mL, whereas, the $IC_{50}$ of RE is ranged from values of 14.00 to 458.43 µg/mL. AqE, CrE and DecE of roots (R) showed significantly the highest reducing potential than other extracts with an $IC_{50}$ = 14.00 µg/mL, 14.97 and 48.19 µg/mL, respectively. The results are summarized in Figure 34.

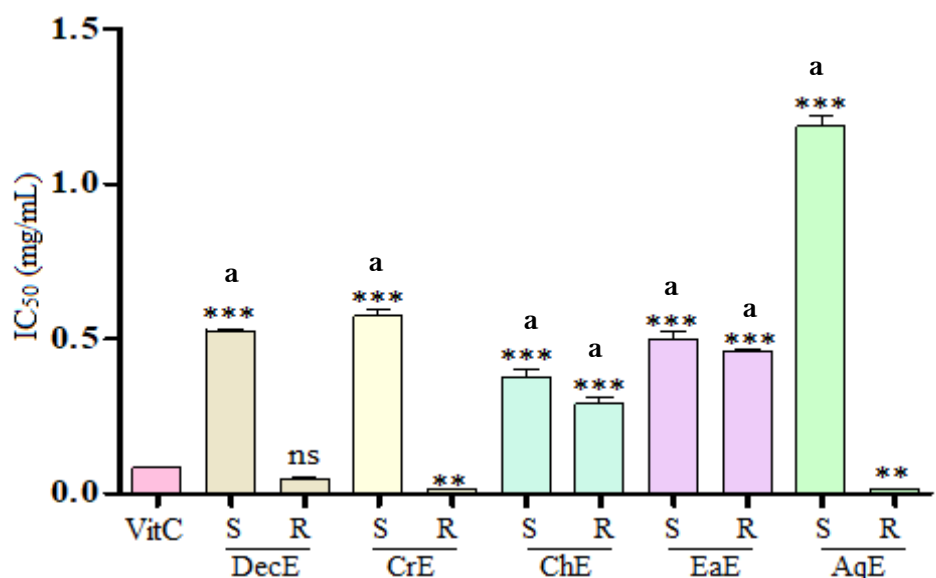

**Figure 34:** Hydroxyl radical scavenging activity of stems (S) and roots (R) extracts of *P. scoparius*. Data were presented as means ± SD (*n*=3). ns: no significant difference, **: $p < 0.01$, ***: $p < 0.001$ as compared with the control.

In general, RE showed potent antioxidant activity with an effective degree for AqE, CrE and DecE mainly due to the interactions of different phenolic compounds. Therefore, the higher potency of the scavenging hydroxyl radicals may be attributable to the presence of the hydrogen donating ability of phenolic compounds in the extracts; which is highly related to the presence of hydroxyl groups (Pavithra and Vadivukkrasi, 2015). Hydroxyl radical is a potent cytotoxic factor able to attack almost every molecule in the body resulting in peroxidation of cell membrane lipids and in the formation of malondialdehyde, which is mutagenic and carcinogenic (Salar et al., 2015). Therefore, the scavenging of hydroxyl radical by extracts may provide significant protection to biomolecules by their ability to remove hydroxyl and superoxide free radicals due to inhibition of respective mechanisms involved in the formation of radicals (Kumar et al., 2013).

**5.1.6. Iron chelating assay**

The metal chelating assay is based on the ability of extract to chelate transition metals by binding them to the ferrous ($Fe^{2+}$) ion catalyzing oxidation and inhibition the formation of $Fe^{2+}$-ferrozine complex. As seen in Figure 35, all extracts interfered with the formation of ferrous and ferrozine complex, suggesting that the SE and RE exhibited appreciable chelating activity. The metal chelating ability of SE ranged from values of 36.30 to 1397.73 µg/mL, by comparing to RE, the chelating activity ranged from 17.81 to 235.88 µg/mL. The values of $IC_{50}$ of the various tested extracts show a big variation; the AqE of roots (R) expresses most powerful effect with an $IC_{50}$=17.81 ± 0.002 µg/mL followed by DecE of stems (S) and DecE of roots (R) with an $IC_{50}$= 36.30 ± 0.003 µg/mL and 38.82 ± 0.005 003 µg/mL, respectively ( Figure 35).

Iron can stimulate lipid peroxidation by the Fenton reaction, and also quickens peroxidation by decomposing lipid hydroperoxides into peroxyl and alkoxyl radicals that can themselves abstract hydrogen and perpetuate the chain reaction of lipid peroxidation (Elmastaş et al., 2006). Baghiani et al. (2012) showed that phenolic compounds proved to be chelators of free metal ions. Nevertheless, in this study, a small and no significant correlation was observed between the chelating extracts of SE and RE and their contents of phenolic compounds.

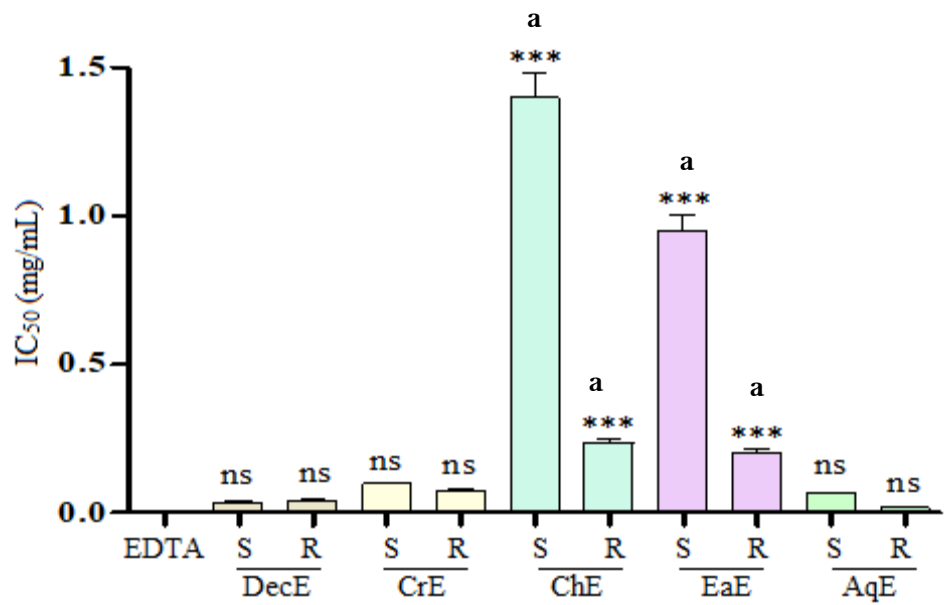

**Figure 35:** Metal chelating capacity of stems (S) and roots (R) extracts of *P. scoparius*. Data were presented as means ± SD (*n*=3). ns: no significant difference ns: no significant difference, ***: $p < 0.001$ as compared with the control.

Similar results have been reported in other recent work, for instance in a study of extracts of some medicinal plants from Iran, Ebrahimzadeh et al. (2008) have found that no correlation was found between polyphenols and flavonoids content of an extract and its chelating activity. Besides, the scavenging potential and the metal chelating ability of the antioxidants are dependent upon their unique phenolic structure and the number of hydroxyl groups (Sowndhararajan and Kang, 2013). Therefore, some extracts with high polyphenols and flavonoids contents showed feeble chelating activity, but others with little polyphenol and flavonoid contents showed good chelating activity (Ebrahimzadeh et al., 2008).

## 5.2. Screening of anti-inflammatory activity

### 5.2.1. The *in vitro* evaluation of the anti-inflammatory activity

The in vitro anti-inflammatory activity was evaluated using the protein denaturation inhibition test. Results showed that CrE of stems (S) and roots (R) are able to significantly inhibit protein denaturation and their inhibitory effect at different concentrations (0.5, 1, and 2 mg/mL) on protein denaturation is depicted in Figure 36. Inhibition % of protein denaturation of CrE of stems ranged from 61.93 to 73.17 % and CrE of roots ranged from 71.67 to 74.70 % at the concentration of 0.5, 1, and 2 mg/mL, compared with aspirin as standard. In general, CrE of roots exhibited significant ($p < 0.05$) level of inhibition compared to CrE of stems (Figure 36).

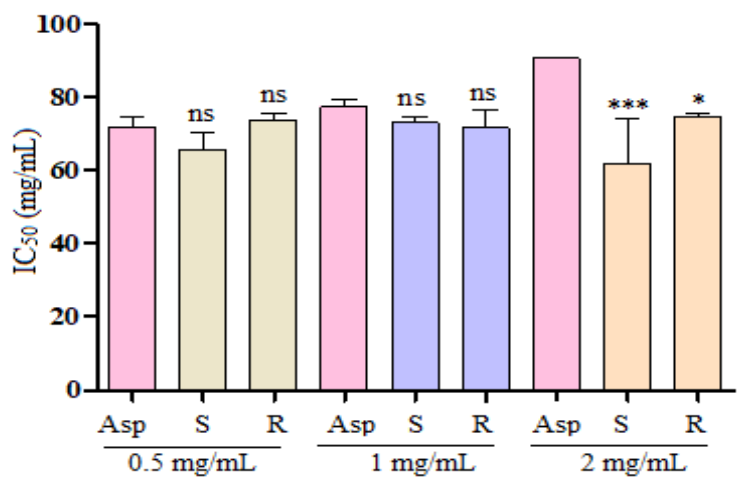

**Figure 36:** Effects of CrE of stems (S) and roots (R) on protein denaturation at concentrations of 0.5, 1, and 2 mg/mL. Data are presented as the mean ± SD (n = 3). ns: no significant difference, *: $p < 0.05$, ***: $p < 0.001$ as compared with the control.

Whereas, results showed that ButE and isolated compounds are able to significantly inhibit protein denaturation, and their inhibitory effect at different concentrations (0.5, 1, and 2 mg/mL) on protein denaturation is depicted in Figure 37.

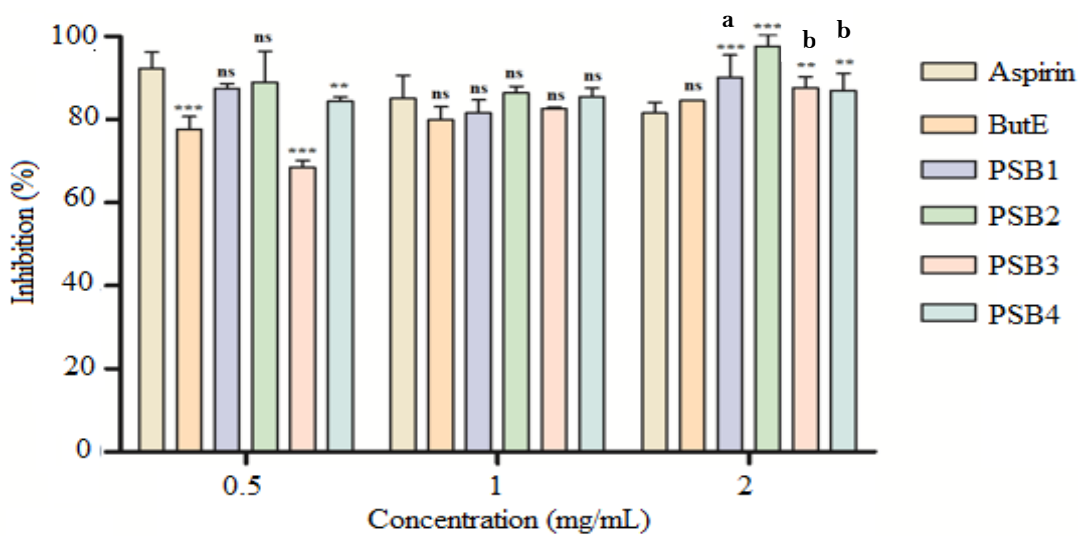

**Figure 37:** Effects of ButE and isolated compounds on protein denaturation at concentrations of 0.5, 1, and 2 mg/mL. Data are presented as the mean ± SD (n = 3). ns: no significant difference, *: $p < 0.05$, **: $p < 0.01$, ***: $p < 0.001$ as compared with the control.

Inhibition % of protein denaturation of *n*-butanol extract (ButE) and the four compounds ranged from 68.50% to 97.60% at the concentration of 0.5, 1, and 2 mg/mL, compared with aspirin as standard. At the concentration of 2 mg/mL, compounds **PSB 2**, **PSB 1**, **PSB 3**, and **PSB 4,** and ButE exhibited significant ($p < 0.05$) level of inhibition with values of 97.60 ± 2.12%, 90.13 ± 1.35%, 87.60 ± 1.76%, 86.97 ± 2.67%, and 84.52 ± 1.87% respectively, compared to aspirin with an inhibition of 81.58 ± 2.06% (Figure 37). The present investigation showed that ButE and the four compounds exhibit potent anti-inflammatory activity. Consequently, by inhibition of protein denaturation, the inflammatory action can similarly be inhibited (Dharmadeva et al., 2018; Ben Nasr et al., 2020). Ben Nasr and coworkers (2020) reported that the aqueous extract from *Pituranthos chloranthus* inhibits lipoxygenase which might be related to phytochemical constituents present in the plant. Thermal denaturation of protein results in loss of biological properties of protein molecules. It is one of the causes of the inflammatory reaction through production of auto-antigens in autoimmune diseases, such as arthritic diseases, diabetes and cancer (Dharmadeva et al., 2018). Inhibition of protein denaturation may play an important role in the anti-rheumatic activity of non-steroidal anti-inflammatory drugs (NSAIDs). Moreover, denatured protein expresses antigens, associated with type III hyper-sensitive reaction, which are related to diseases such as glomerulo-nephritis and serum sickness (Ahmad et al., 1992). Therefore, the ability of substances to prevent the protein denaturation may also help to prevent inflammatory disorders.

### 5.2.2. The *in vivo* investigations of the anti-inflammatory activity
### 5.2.2.1. Xylene-induced ear edema assay

Results from this study revealed that the crude extract (CrE) from stems and roots at doses of 100–600 mg/kg exhibits substantial anti-inflammatory activity in the xylene-induced ear edema test as depicted in Figure 38. The two extracts exerted significant dose-dependent activity (100, 300 and 600 mg/kg) against edematous response caused by xylene and lowered the edematous response to a certain degree. Additionally, our findings indicated that CrE from stems at a dose of 600 mg/kg reduces the edematous response by 70.16 % compared to CrE from roots. On the other hand, the non-steroidal anti-inflammatory drug indomethacin lowered the edematous response by 79.01 % at the dose of 50 mg/kg.

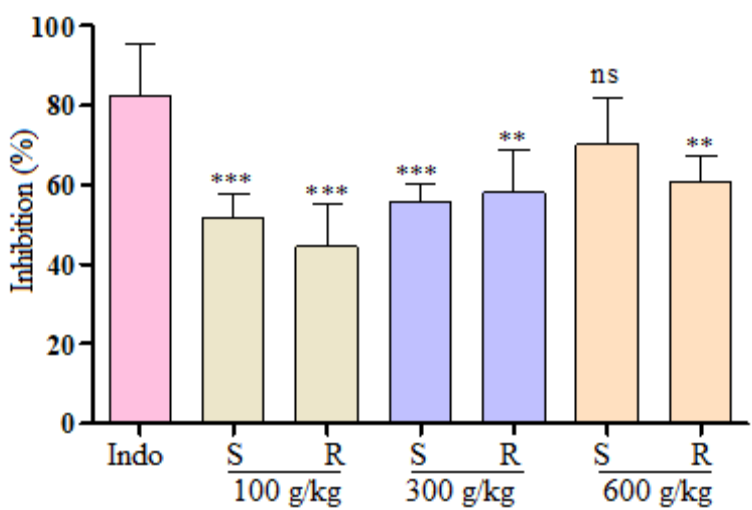

**Figure 38:** Anti-inflammatory effect of CrE of stems (S) and roots (R) on xylene-induced ear edema in mice. Data are presented as the mean ± SEM (n = 6), ns: no significant difference, **: $p < 0.01$, ***: $p < 0.001$. Indo: Indomethacin

Xylene-induced ear edema in mice is a commonly used acute inflammation model (Zhao et al., 2018), it is a reproducible experimental model and provides good predictive values for screening anti-inflammatory agents (Lu et al., 2006). Xylene can stimulate vasodilation and raise blood vessel permeability and then induce edema. The molecular and cellular mechanism by which xylene induces inflammation involves sensory neurons sensitive to capsaicin which, following stimulation, releases a number of mediators that can initiate the inflammatory response. This phenomenon is known as neurogenic inflammation (Richardson and Vasko, 2002). Substance P and peptide linked to the calcitonin gene are the main initiators of neurogenic inflammation. They induce vasodilation and plasma oxidation by acting on the smooth muscles of blood vessels and endothelial cells (Rotelli et al., 2003) as they can directly activate mast and other immune cells. It is also recognized that sensory neurons have cyclooxygenase which can synthesize pro-inflammatory prostaglandins (Richardson and Vasko, 2002). Along this line, the effect of indomethacin on inflammation can be explained by the inhibition of pro-inflammatory prostaglandin synthesis. Similarly, pretreatment of mice with CrE from the two parts of *P. scoparius* caused significant inhibition of the development of edema. This could suggest that the extracts reduces the release of substance P or antagonizes its action (Zhou et al., 2008), which may be due to suppression of phospholipase $A_2$ that is associated with the pathophysiology of inflammation triggered by xylene (Rahman et al., 2019). In this regard, Kim et al. (2006) showed that gallic acid and its derivatives are responsible for the inhibition of Kappa-B nuclear factor (NF-κB) binding which is necessary for pro-inflammatory cytokines expression. Flavonoids can also affect the expression of pro-inflammatory cytokines by inhibition of NF-κB transcription (González-Gallego et al., 2007).

**5.2.2.2. Croton oil-induced ear edema**

The effect of CrE from stems and roots on croton-oil induced ear edema in mice is shown in Figure 39. The reduction of edema by croton-oil ear in mice of CrE of stems ranged from 42.56 to 61.02 % and CrE of roots ranged from 36.00 to 40.92 %, compared with indomethacin as standard. In the test, CrE from stems at doses of 300 and 600 mg/kg, showed significantly reduction of edema by croton-oil ear in mice at 60.00 % and 63.07 %, respectively. The standard drug, indomethacin (50 mg/kg), produced a clear anti-edematous effect in mice (63.08% inhibition).

The anti-inflammatory activity of CrE from stems and roots was further evaluated by the inhibition of croton oil-induced ear edema in mouse model. Croton oil is a highly irritating agent that contains 12-*O*-tetradecanoyl-13-phorbol acetate (TPA), which can stimulate an inflammatory response and then induces edema (Da Silva et al., 2015). The molecular and cellular mechanism by which croton oil induces inflammation may be linked to the activation of numerous protein kinases C (Passos et al., 2013) by the secretion of high levels of intracellular factors such as calcium and diacylglycerol. Activation of the receptors coupled to G protein results in the production of these factors.

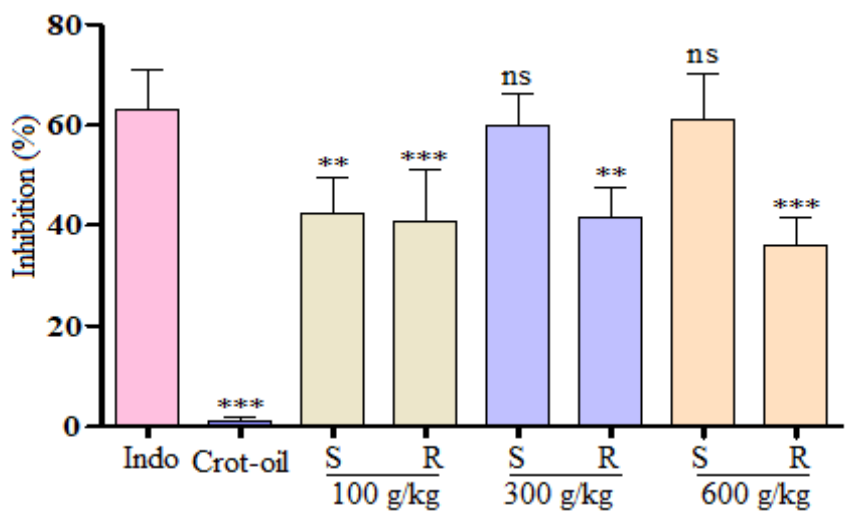

**Figure 39:** Anti-inflammatory effect of CrE from stems (S) and roots (R) of *P. scoparius* on croton oil-induced ear edema in mice. Data are presented as the mean ± SEM (n = 6), ns: no significant difference, **: $p < 0.01$, ***: $p < 0.001$.

Diverse intracellular signal transduction pathways mediated by protein kinases C (PKC) such as phospholipase $A_2$ associated with the release of arachidonic acid and eicosanoid production, influence the pathogenesis of inflammation. The mechanism of action of indomethacin on inflammation is based on the inhibition of pro-inflammatory prostaglandin synthesis. This inhibition may be due to reduction of the release of PKC or its inhibition, and to the suppression of the exudative phase in acute inflammation.

**5.2.2.3. *In vivo* topical anti-inflammatory activity**

The topical anti-inflammatory activity of ButE and isolated compounds was evaluated using the croton oil-induced ear edema method in mouse model. Displayed in Figure 40 are results of our study of the anti-inflammatory activity of ButE and the four isolated compounds in mice, using the croton oil-induced ear edema method.

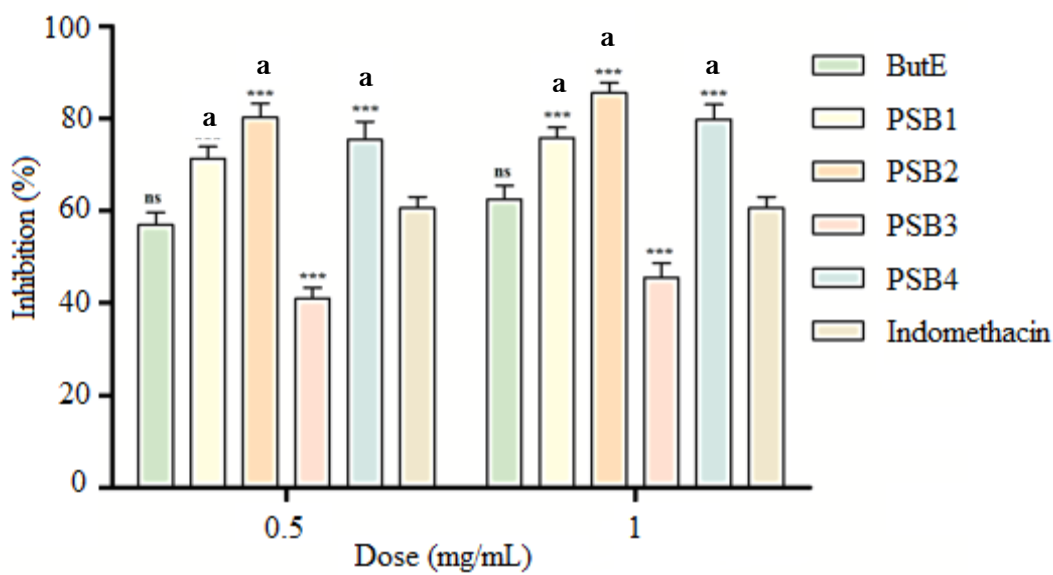

**Figure 40:** Percentage inhibition of the ear edema of *n*-butanol extract (ButE) and isolated compounds in mice after 6 h, at a dose of 0.5 and 1 mg/ ear. Data are presented as the mean ± SEM (n = 6). ns: no significant difference, ***: $p < 0.001$, as compared with the control.

Our findings revealed that ButE and compounds **PSB1**, **PSB2**, and **PSB4,** at a dose of 0.5 mg/mL caused significant inhibition of edema induced by croton oil in mice (at 6 h) with percent inhibition ranging from 40.92 ± 3.98% to 81.53 ± 5.90%. On the other hand, the non-steroidal anti-inflammatory drug indomethacin caused less edematous reduction by 62.02 ± 3.42% at the same dose (Figure 40). Similarly, at a dose of 1 mg/mL, compounds **PSB2**, **PSB4**, and **PSB1** exhibited significantly higher ($p < 0.05$) activity with an inhibition of 85.50 ± 2.78%, 79.78 ± 4.68%, and 75.78 ± 2.98%, respectively, whereas ButE and **PSB3** showed the lower activity with an inhibition of 62.42 ± 4.68% and 45.52 ± 4.68%, respectively.

Croton oil-induced dermatitis represents a model of acute inflammatory response when the response is maximal. This test is widely used for identifying potential anti-inflammatory drugs for the therapy of skin diseases (Mo et al, 2013). The thickness of the ear edema reaches its maximum 6 h after the application of croton oil (Tubaro et al, 1985). The phlogenic effect of croton oil is due to the presence of the 12-*O*-tetradecanoyl-13-phorbol acetate (TPA) moiety.

TPA induces an inflammatory response characterized by significant production of pro-inflammatory mediators, increased vascular permeability, edema, and neutrophil infiltration (Delaporte et al., 2004). The high anti-inflammatory effect exhibited by compounds **PSB1**, **PSB2**, and **PSB4** can be explained by the inhibition of protein kinases C (PKC) whose involvement in the inflammatory process is clearly evident in this experimental model (Delaporte et al., 2004).

A wide range of flavonoids with diverse chemical structures have been linked with various anti-inflammatory mechanistic effects (Howes, 2018). Moreover, structure-activity relationship (SAR) of flavonoid compounds and natural product investigations on anti-inflammation have also been reported (Gautam and Jachak, 2009). In this context, the presence of the four isolated compounds in ButE may be responsible for suppressing the exudative phase of acute anti-inflammatory properties and then pain.

**5.2.2.3. Carrageenan-induced rat paw inflammation**

The carrageenan-induced paw edema in rat model was utilized to evaluate the anti-inflammatory activity of CrE of stems and roots of *P. scoparius*. In this model, the two extracts of *P. scoparius* significantly inhibited the edema at 100, 250 and 500 mg/kg within 3 h (Table 11). Diclofenac and indomethacin showed an inhibition of the inflammation induced by carrageenan compared to control group ($p < 0.05$). Harchaoui and coworkers (2018) reported that *P. scoparius* exhibit high anti-inflammatory effect in formalin and carrageenan-induced rat paw edema methods. These results suggest that the presence of high content of polyphenols in CrE of both stems and roots might partly contribute to its potency.

**Table 11:** Effect of CrE of stems (S) and roots (R) on carrageenan-induced paw edema in rats. Results are presented as mean ± SEM (n =5 for each group), ns: no significant difference, *: $p < 0.05$, **: $p < 0.01$, ***: $p < 0.001$.

| Treatment (mg/kg) | | Paw thickness (mm) | | | | |
|---|---|---|---|---|---|---|
| | | 0h | 1h | 2h | 3h | 5h |
| Control | | 2.76 ± 0.13 | 5.22 ± 0.85 | 4.37 ± 0.69 | 6.05 ± 0.84 | 5.39 ± 0.69 |
| Dicl | (20) | 2.97 ± 0.29 ns | 3.76 ± 0.35 ns | 2.74 ± 0.05 *** | 2.48 ± 0.26 *** | 2.59 ± 0.60 *** |
| Indo | (50) | 3.04 ± 0.41 ns | 4.56 ± 0.35 ns | 4.31 ± 0.10 ns | 4.17 ± 0.24 *** | 4.18 ± 0.25 * |
| CrE (S) | (100) | 2.23 ± 0.14 ns | 4.62 ± 0.39 ns | 3.54 ± 0.34 ns | 3.91 ± 0.42 *** | 4.68 ± 0.19 ns |
| | (250) | 2.32 ± 0.17 ns | 5.33 ± 0.75 ns | 4.52 ± 0.51 ns | 4.61 ± 0.45 ** | 5.37 ± 0.42 ns |
| | (500) | 2.27 ± 0.13 ns | 5.03 ± 0.55 ns | 4.26 ± 0.58 ns | 4.24 ± 0.60 *** | 5.32 ± 0.74 ns |
| CrE (R) | (100) | 2.51 ± 0.13 ns | 4.39 ± 0.52 ns | 3.67 ± 0.20 ns | 3.94 ± 0.36 *** | 5.06 ± 0.43 ns |
| | (250) | 3.05 ± 0.94 ns | 4.97 ± 0.73 ns | 3.73 ± 0.43 ns | 3.89 ± 0.20 *** | 4.68 ± 0.39 ns |
| | (500) | 2.30 ± 0.09 ns | 4.45 ± 0.80 ns | 3.94 ± 0.45 ns | 3.74 ± 0.24 *** | 4.69 ± 0.56 ns |

Carrageenan-induced paw edema is a widely used primary assay for investigating new anti-inflammatory agents and is believed to be biphasic (Igbe and Inarumen, 2013). The initial phase occurs within 1–2 h after carrageenan injection, due to the release of serotonin and the increase of prostaglandin, histamine, and bradykinin. Whereas, the $2^{nd}$ phase occurs 3–5 h after carrageenan injection, which is correlated with the production and release of kinins and prostaglandins (Zhou et al, 2008). Our results show that the extracts exhibits inhibitory activity against carrageenan-induced paw edema over a period of 3 and 5 h, comparable to the standards (diclofenac and indomethacin). The anti-inflammatory activity of the extract might act through a mechanism that involves inhibition of COX associated with the inflammatory cascade induced by carrageenan (Reanmongkol and Songkram, 2013).

### 5.2.2.4. Analgesic activity

Our findings showed that CrE of stems and roots exhibits analgesic activity against acetic acid-induced abdominal constriction. Results in Figure 41 reveal that this extract exhibits a considerable antinociceptive effect against acetic acid-induced writhing in mice in a dose-dependent fashion. The reduction of abdominal constriction induced by acetic acid in mice of CrE of stems ranged from 26.31 to 70.68 % and CrE of roots ranged from 56.89 to 66.91 %, compared with indomethacin as standard. In the test, CrE from roots at doses of 100 and 250 mg/kg, showed significant reduction of abdominal constriction at 60.00 % and 63.07 %, respectively. At a dose of 500 mg/kg, CrE of stems showed significantly reduction of abdominal constriction at 70.68 %. The standard drug, aspirin (100 mg/kg), produced a clear analgesic activity against acetic acid-induced abdominal constriction in mice (79.32 % inhibition).

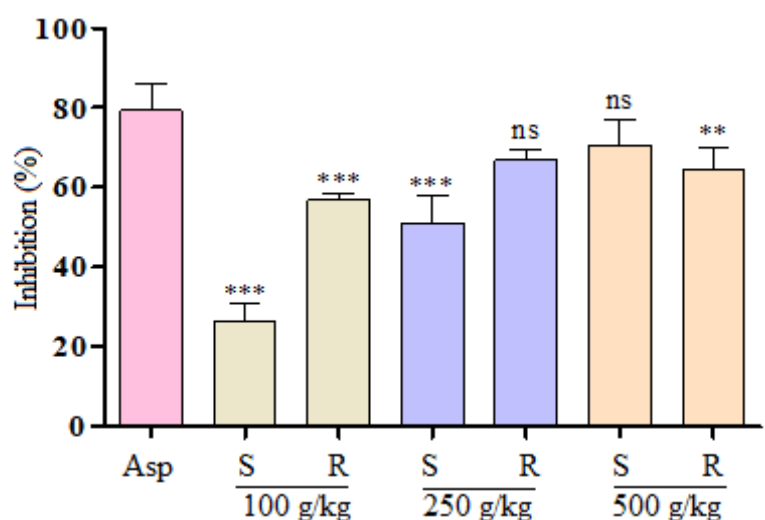

**Figure 41:** Analgesic effect of CrE of stems (S) and roots (R) against acetic acid-induced abdominal contractions in mice. Data are given as the mean ± SEM (n = 5), ns: no significant difference, \*\*: $p < 0.01$, \*\*\*: $p < 0.001$. Asp: Aspirin

Analgesic effect is normally screened by the acetic acid test which is widely recognized for its ability to identify both peripheral and central molecules. This activity was highlighted by the abdominal cramps test (Haider et al., 2011). In this context, acetic acid is known to cause inflammation through the release of mediators including serotonin, histamine, and prostaglandins ($PGE_2$, $PGE_2\alpha$ and $PGF_2$) in peritoneal fluids, along with the release of lipooxygenase products which excite the nociceptive neurons (Bagepalli et al., 2009). This leads to dilatation of arterioles and venules (Carey et al., 2008) with contraction and division of endothelial cells, leading to amplified vascular permeability, and therefore, extravasation of plasma proteins to the peritoneal cavity (Okoli et al., 2007). Along this line, the anti-inflammatory and analgesic activities have analogous underlying mechanisms (Ettebong et al., 2014). Numerous non-steroidal anti-inflammatory drugs (NSAIDS) such as indomethacin exhibit analgesic as well as anti-inflammatory activities. In this investigation, pretreatment of mice with CrE of stems and roots caused a significant inhibition of writhing. This may suggest the ability of the extract to inhibit vascular permeability and may modulate the amplitude of the inflammatory response, which may explain its anti-edematous and analgesic effects. These results suggest that the two extracts exerts both anti-inflammatory and analgesic activities as in the case of aspirin, probably by inhibiting cyclooxygenase (Carey et al., 2008), which is responsible for the production of pro-inflammatory prostaglandins. These findings suggest that the extracts can effectively inhibit the acute inflammation and then pain.

Aspirin is an NSAIDS which is widely used clinically as an anti-inflammatory and analgesic agent. It exerts its effect through inhibition of cyclooxygenase-mediated of $PGI_2$ and $PGE_2$. These two substances display pro-inflammatory and pro-nociceptive activity and increase the effect of other mediators such as histamine and 5-hydroxytryptamine at the site of inflammation (Al Swayeh et al., 2000). Results from this investigation show that CrE of stems and roots significantly inhibited abdominal contractions in a dose-dependent fashion as compared to the control (aspirin). This suggests that the two extracts exerts peripheral analgesic activity. Research findings showed that any agent that reduces the number of abdominal contortion exhibits analgesic effect by suppression of postaglandin synthesis, a peripheral mechanism of pain inhibition (Alam et al., 2009). In our case, the analgesic effect of these extracts may be due to the presence of significant amounts of polyphenols, flavonoids, and tannins, in this plant. These compounds in CrE of stems and roots could explain the pharmacological properties of these extracts. These results confirm the use of *P. scoparius* as an anti-inflammatory agent.

## 5.3. Screening of toxicity study

### 5.3.1. *In vitro* cytotoxicity against red blood cells

The cytotoxicity of ButE and isolated compounds was studied by examining the hemolytic activity against human red blood cells (RBCs) using Triton X-100 as a positive control. Percentage hemolysis was evaluated by comparing the absorbance of sample and that of Triton X-100; the positive control showed about 100% hemolysis. Results reveal that the percentage of hemolysis of RBCs caused by the isolated compounds and extract ranged from 1.85% to 2.45% at the concentration of 0.5 and 1 mg/mL respectively, compared to Triton x-100. At the concentration of 1 mg/mL, compounds **PSB1**, **PSB2**, **PSB4**, **PSB3**, and ButE, exhibited significantly low ($p < 0.05$) level of hemolysis with percentages of 2.45 ± 1.20%, 2.24 ± 0.63%, 2.19 ± 0.24%, 1.85 ± 0.39%, and 2.03 ± 0.52 %, respectively; these results are shown in Figure 42.

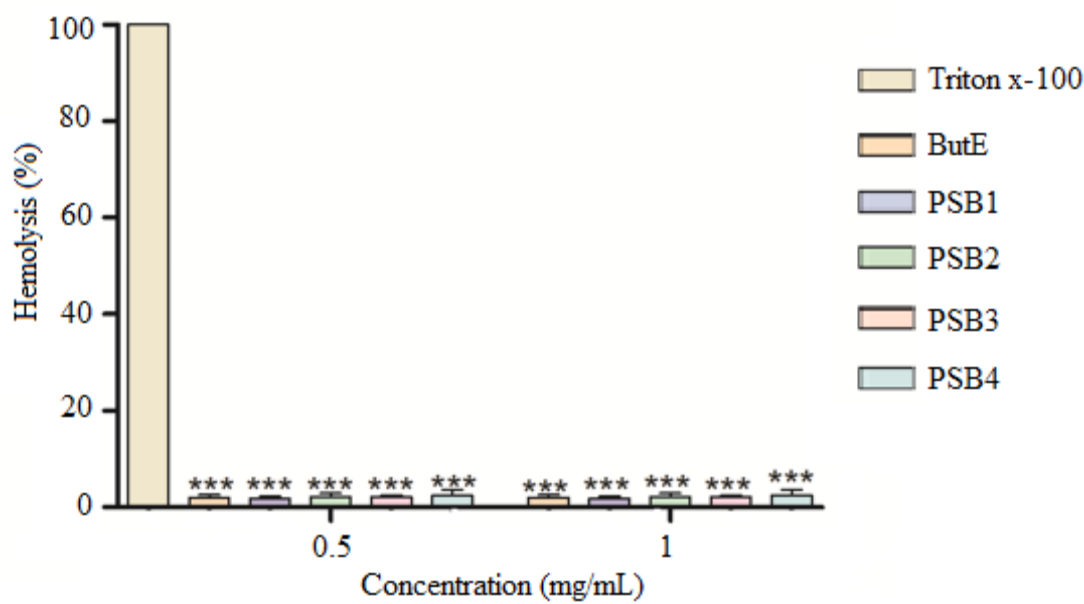

**Figure 42:** Hemolysis percentage of *n*-butanol extract (ButE) and isolated compounds at dose (0.5 and 1 mg/mL) against red blood cells. Data are presented as the mean ± SD (n = 3), ***$p < 0.001$ compared with control (Triton x-100).

Our results indicate that ButE and isolated compounds showed a minor percentage of lysis as compared to the positive control. Our findings are supported by those of Haddouchi et al. (2016) which showed that the methanol extract of *P. scoparius* exhibits a very weak hemolytic effect against red blood cells, characterized by the absence of risk of cytotoxicity. Moreover,

Asgary et al. (2005) reported on the protective effect of flavonoids against red blood cell hemolysis, whereas Kalita et al. (2011) reported on the effect of *Lantana camara* leaves extracts on the hemolytic assay, and highlighted the importance of phytochemicals in medicinal plants.

Riaz et al. (2012) report that the good sign to estimate in vitro the influences of various compounds is the mechanical stability of the membrane of red blood cells (RBCs) when screening for cytotoxicity. Furthermore, it is well documented that the hemolytic activity of a compound is an indication of its general cytotoxicity towards normal cells (Signoretto et al., 2016). On the other hand, the mechanical stability of the RBCs membranes is a good indicator to evaluate, in vitro, the effects of different compounds for cytotoxicity (Baillie et al., 2009). In this context, the majority of healthy humans may suffer at adequately high concentrations of hemolytic drugs. Therefore, it is very important to investigate the effect of toxic substances on the hemolytic activity of RBCs (Lin and Haynes, 2010). Treating cells with a cytotoxic compound may cause various health problems, where cells may undergo a loss of membrane integrity and die rapidly as a result of cell hemolysis (Tiwari et al, 2011). Inhibition of red blood cell hemolysis may provide insights into the inflammatory process. Therefore, substances that contribute significant protection of cell membrane against harmful substances are important in inhibiting the progression of inflammation.

**5.3.2. Acute oral toxicity study**

**5.3.2.1. Observations**

Our findings indicated that CrE of both S and R did not exhibit any behavioral changes in treated rats throughout the examination period. Moreover, no death was recorded and no symptoms or signs of toxicity noticed in any of the treated animals during fourteen days of experiments; $LD_{50}$ was higher than 5 g/kg BW for rats. Similarly, Harchaoui et al. (2018) demonstrated that the aqueous extract of *P. scoparius* is non-toxic even at a dose of 16 g/kg in an acute toxicity study.

### 5.3.2.2. Relative organs weight

Results from this investigation revealed that oral administration of CrE of both stems (S) and roots (R) did not cause any variation in relative weight of the different organs in the treated female rats in comparison with the control group as shown in Table 12.

**Table 12:** Relative organs weight of female rats treated with CrE of stems (S) and roots (R) of *P. scoparius* and control. Results are expressed as mean ± SEM (n = 5), ns: no significant difference.

| Treated groups | Control group | (S) | | (R) | |
|---|---|---|---|---|---|
| | | 2 g/kg BW | 5 g/kg BW | 2 g/kg BW | 5 g/kg BW |
| Liver | 4.60 ± 0.60 | 4.82 ± 0.68 $^{ns}$ | 5.10 ± 0.39 $^{ns}$ | 4.51 ± 0.69 $^{ns}$ | 4.41 ± 0.60 $^{ns}$ |
| Kidneys | 0.78 ± 0.06 | 0.79 ± 0.08 $^{ns}$ | 0.78 ± 0.09 $^{ns}$ | 0.77 ± 0.10 $^{ns}$ | 0.76 ± 0.06 $^{ns}$ |
| Spleen | 0.40 ± 0.04 | 0.41 ± 0.04 $^{ns}$ | 0.40 ± 0.09 $^{ns}$ | 0.39 ± 0.09 $^{ns}$ | 0.35 ± 0.03 $^{ns}$ |
| Heart | 0.39 ± 0.02 | 0.40 ± 0.03 $^{ns}$ | 0.40 ± 0.04 $^{ns}$ | 0.36 ± 0.03 $^{ns}$ | 0.41 ± 0.07 $^{ns}$ |
| Lungs | 0.78 ± 0.09 | 0.80 ± 0.15 $^{ns}$ | 0.68 ± 0.09 $^{ns}$ | 0.70 ± 0.06 $^{ns}$ | 0.70 ± 0.11 $^{ns}$ |
| Stomach | 0.87 ± 0.12 | 0.93 ± 0.28 $^{ns}$ | 1.02 ± 0.24 $^{ns}$ | 0.90 ± 0.14 $^{ns}$ | 0.96 ± 0.05 $^{ns}$ |

### 5.3.2.3. Progression of body weight

Our results show no significant changes in body weight of treated groups of CrE of both stems (S) and roots (R) as compared to the control as shown in Figure 43.

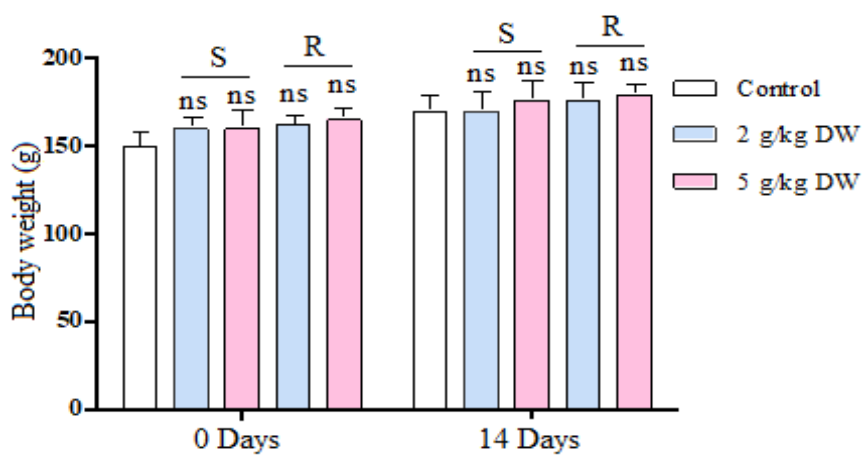

**Figure 43:** Body weight of rats administered with CrE of stems (S) and roots (R). Values are presented as the mean of three measurements ± SD (n=5), ns: no significant difference.

### 5.3.2.4. Biochemical analysis

Our findings indicate that there were no significant variations in most biochemical parameters among treated rats by CrE of both stems (S) and roots (R) as compared to the control, whereas a significant ($p < 0.05$) dose-dependent decrease was observed in cholesterol, total bilirubin, triglyceride, and uric acid in the treated female rats. In addition, a significant drop in albumin was seen in the treated rats (Figure 44 and 45). In the toxicity study, various biochemical parameters such as the enzymes ALT, AST, and ALP could be utilized as markers of hepatocellular effects (Brandt et al., 2009). Results from our investigation revealed that CrE of stems and roots at 2 and 5 g/kg doses did not cause significant changes in the levels of enzymes such as AST, ALT and ALP in treated rats when compared to the control group, and did not cause death or physical changes in rats in fourteen days of the investigation. These results indicate the non-hepatotoxic potential of CrE of stems and roots. Furthermore, urea and creatinine are regarded as biomarkers of renal failure (Lameire et al., 2005).Our findings also revealed a significant ($p < 0.05$) dose-independent reduction in albumin in rats. This reduction in albumin level in CrE of stems and roots-administered rats could associated with increased C reactive protein (Gotsman et al., 2019).

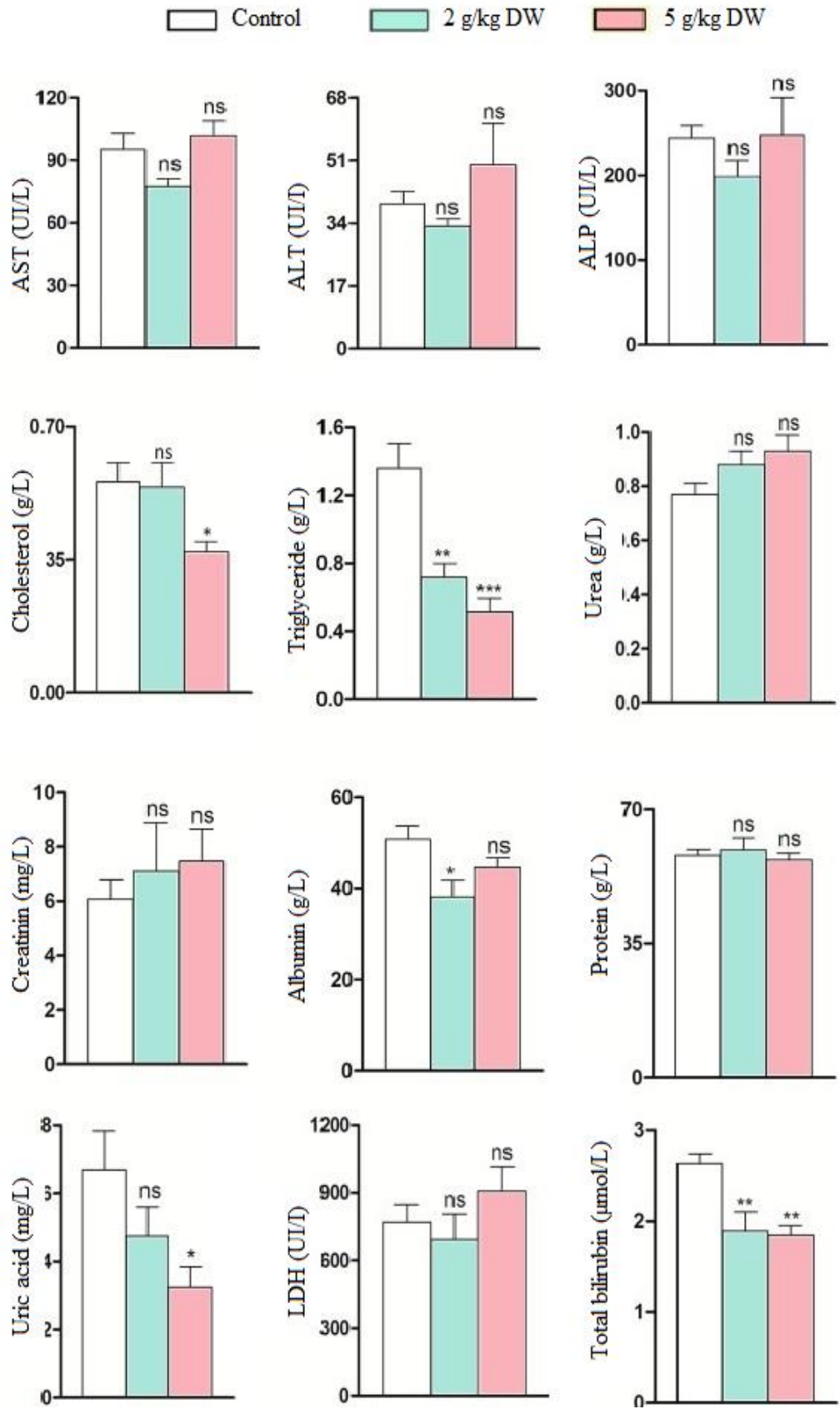

**Figure 44:** Biochemical parameters of rats administered with CrE of stems measured throughout the acute toxicity experiment. Values are expressed as mean ± SEM (n = 5), ns: no significant difference, *: $p < 0.05$, **: $p < 0.01$, ***: $p < 0.001$.

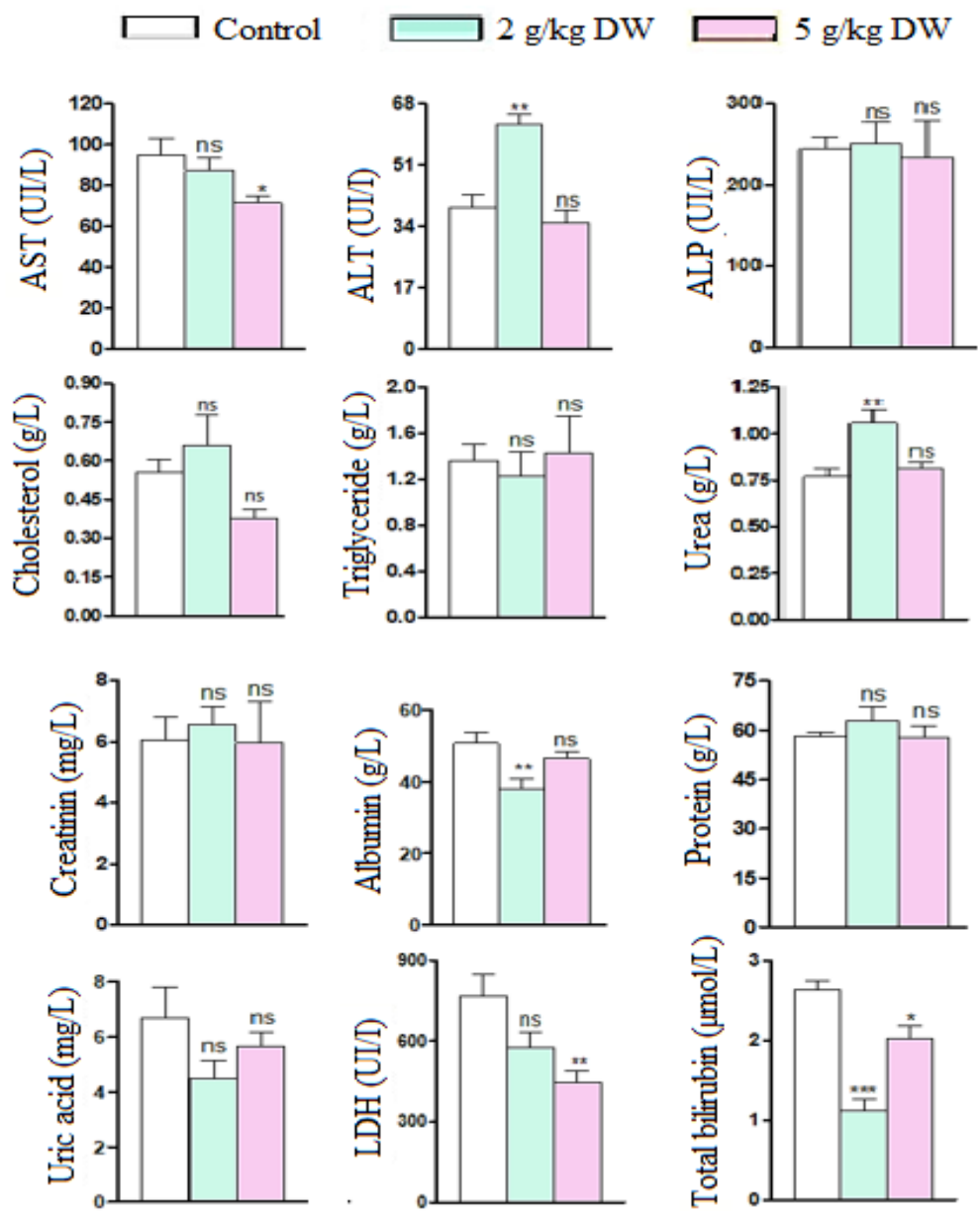

**Figure 45:** Biochemical parameters of rats administered with CrE of roots measured throughout the acute toxicity experiment. Values are expressed as mean ± SEM (n = 5), ns: no significant difference, *: $p < 0.05$, **: $p < 0.01$, ***: $p < 0.001$.

Furthermore, rats treated with these extracts showed a significant dose-dependent decrease in blood uric acid and total bilirubin level. Uric acid plays a key role in the development of stones as it promotes CaOx deposition by absorbing glutamic acid and other organic compounds (Kaushik et al., 2019).

Our findings showed that the concentration of uric acid was reduced after the treatment, which reiterated the antilithic efficacy of CrE of stems and roots. Then, the presence of various phytochemicals in CrE of stems and roots may be play role as antioxidants thereby stabilizing the cellular membrane of hepatocytes exhibiting bilirubin lowering property (Patil et al., 2015). In addition, we detected a considerable lowering in triglycerides and cholesterol levels in rats treated with either 2 or 5 g/kg BW. This may suggest that the extract assists in the oxidation of triglycerides by lipoprotein lipase to fatty acids and/or facilitates their elimination by the bile after their conversion to bile acids (Alnouti, 2009).On the other hand, histological analysis revealed the presence of vascular congestion in liver of rats treated with 5 g/kg BW. This can be attributed to the vasoconstriction effect of CrE of stems and roots on the wall of blood vessels (Gonzalez et al., 2015).

**5.3.2.5. Histological investigation of kidney and liver**

Investigation of histological parts (liver and kidneys) of treated rats indicated a conservation of the cellular structure of the pair organs compared to the controls. At a dose of 2 g/kg, the liver histological parts did not show any structural changes in the treated rats CrE of both stems (S) and roots (R) compared to the controls. However, a vascular congestion was observed in few hepatic tissues of rats administered with 5 g/kg by CrE of stems (Figure 38). This can be attributed to the vasoconstriction effect of CrE of stems on the wall of blood vessels (Gonzalez et al., 2015). However, a peliosis showed in histological cutting of liver of rats treated with CrE of roots. Moreover, kidneys' histological section did not show any structural alterations for rats treated with 2 and 5 g/kg doses of CrE of both stems (S) and roots (R) compared to the control as depicted in Figure 46.

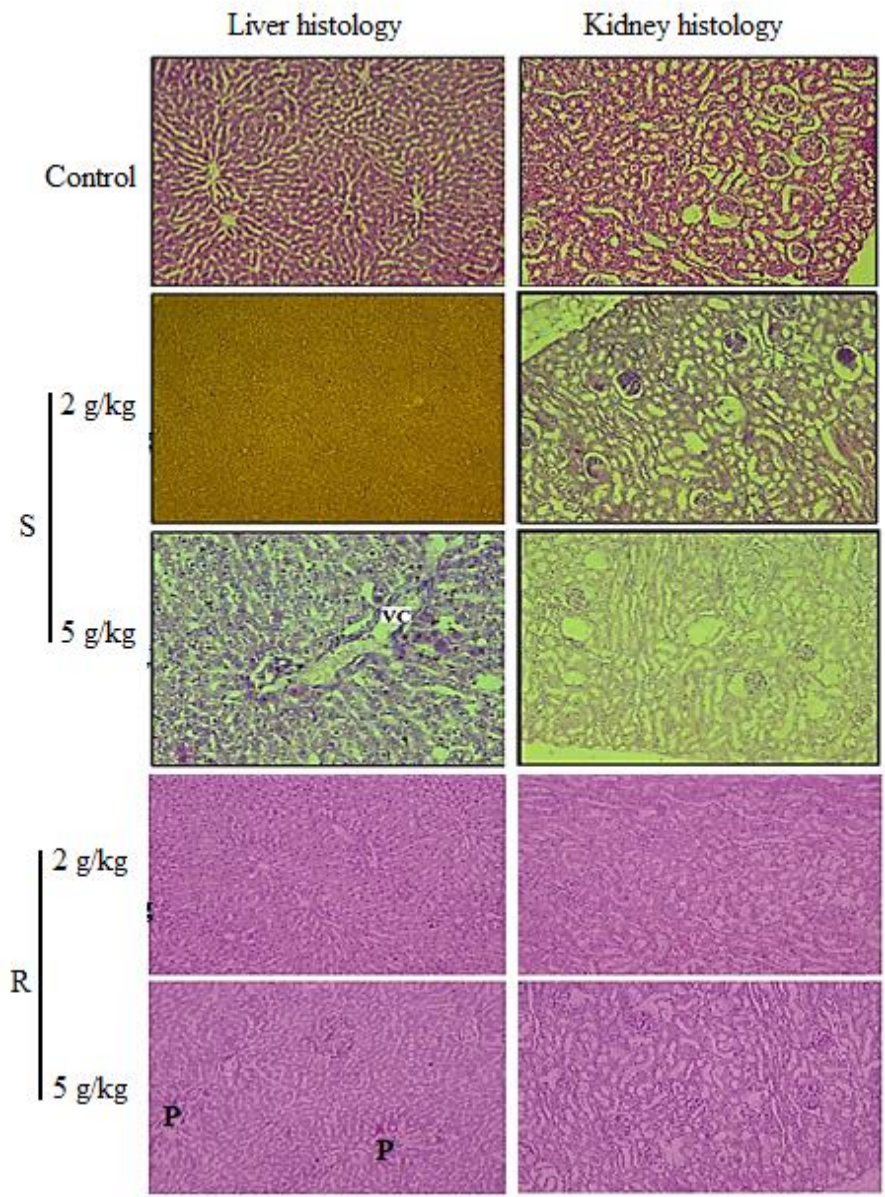

**Figure 46:** Histological study of rats' liver and kidney administered with CrE of stems (S) and roots (R) in the acute toxicity (x10). VC: vascular congestion, P: peliosis.

# CONCLUSION AND PERSPECTIVE

# CONCLUSION AND PERSPECTIVES

In Folk medicine, medicinal plants have been used for years as an alternative source of medications, such as anti-inflammatory and antioxidants agents. However, lack of experimental records concerning the safety-catch and effectiveness of medicinal plants is one of the main concerns about the use of these plants in medicine. *Pituranthos scoparius* is a medicinal plant belonging to the Apiaceae family, grows in North Africa, and is widely distributed in the high plateau of most parts of the Sahara in Algeria. The local Algerian population applies this plant to treat numerous infectious diseases, dermatoses, nervous breakdowns, digestive disorders, and calm abdominal pain.

In the present work, we present experimental results pertaining to ethnopharmacological study, *in vitro* antioxidant, the acute oral toxicity, anti-inflammatory, and analgesic effect of *Pituranthos scoparius* stems and roots extracts. Furthermore, extraction, fractionation, isolation and structure elucidation of constituents were conducted from the *n*-butanol stems extract (ButE).

In the first point, results from an ethnomedicinal questionnaire revealed that the most people employed this plant in traditional medicine in both oral and topical applications.

Additionally, the extraction of polyphenols was carried out in different methods include decoction, extraction and fractionation with various solvent increasing polarities. Different conditions of include time of collection the plant, nature of phytochemical compounds and solvents used could affects on the extraction yields. Results obtained from the qualitative phytochemical analysis of stems extracts (SE), roots extracts (RE) and *n*-butanol extract (ButE) of stems of *P. scoparius* revealed the presence of polyphenols, flavonoids, tannins and free quinones in both SE, RE and ButE, weareas, the alkaloids and coumarins were found in the ButE. In quantitative phytochemical analysis, the most important highest amount of total polyphenols, flavonoids and tannins were found in EaE of stems part with values of 434.34 ± 2.75 µg gallic acid equivalent; 207.49 ± 1.03 µg quercetin equivalent and 126.32 ± 1.32 µg tannic acid equivalent/ mg dried extract, respectively.

In a similar fashion, we have succeeded in isolating four known compounds by column chromatography analysis on silica gel. Using $^1$H-NMR, and $^{13}$C-NMR, the compound was identified as isorhamnetin-3-O-β-apiofuranosyl (1→2)-β-glucopyranoside, D-mannitol, isorhamnetin-3-O-β-glucoside and isorhamnetin-3-O-β-glucopyranosyl-(1→6)-β-glucopyranoside, for the first time, from the *n*-butanol stem extract of *P. scoparius*.

Various assays including DPPH and ABTS scavenging assays, lipid peroxidation, hydroxyl scavenging ability, iron-chelation and reducing power activities were employed to evaluate the *in vitro* antioxidant properties. Results from the *in vitro* antioxidant activity of stems and roots extracts of *P. scoparius* revealed that the EaE and ChE of stems presented a high scavenging activity against DPPH scavenging assay, whereas, EaE and ChE of both stems and roots showed an excellent scavenging activity against ABTS scavenging test.

The AqE and CrE of roots were found to have a strong hydroxyl scavenging activity with values of 14.00 ± 0.00 and 14.97 ± 0.00 µg/mL, respectively. In addition, the DecE of stems and both DecE and AqE of roots possessed a very high ion chelating activity with a strongest effect by AqE from roots ($IC_{50}$= 17.81 ± 0.00 µg/mL) in this case. Furthermore, AaE of stems had a strong reducing power scavenging effect and DecE of stems is highly reduced the oxidation of β-carotene (AA= 91.53 ± 0.98 %). This result suggests that stems and roots extracts of *P. scoparius* can attend as a free radical scavenger and it's obtained to inhibit the oxidation of β-carotene by compensating both the linoleate free and other liberal radicals generated in the reaction system.

The findings of the present study indicate that *Pituranthos scoparius* could be a new source of natural antioxidant drugs. The data highlights the good antioxidant proprieties of different extracts from stems and roots. This antioxidant potential is probably associated with the presence of various secondary metabolites which may have many benefits in treating oxidative stress-related diseases.

Regarding to the anti-inflammatory (*in vitro* and *in vivo*) activities of crude extracts of stems and roots, various tests models have been used to probe mediators of inflammation and screening of anti-inflammatory agents. The CrE of stems and roots present exhibited a significant inhibition of xylene and croton oil-induced ear edema in mice along to the anti-edematogenic effect in the carrageenan-induced rat paw edema.

In addition, findings from this investigation reveal that the isolated compounds and ButE exhibit significant anti-inflammatory effects in both *in vitro* and topical anti-inflammatory models, which justifies the use of this plant in the Algerian folk medicine.

Furthermore, results revealed that the crude extracts of stems and roots displays significantly analgesic effects against acute inflammation in mice.

Findings from the *in vivo* toxicological study revealed that CrE of stems and roots show no treatment-related sign of toxicity or mortality in rats. Thus, the extracts was not toxic even at 5 g/kg BW in rats. In addition, biochemical serum analysis and histopathology study did not show any marked effects.

Furthermore, the cytotoxicity of the four pure isolated compounds and the extract was examined using *in vitro* hemolytic activity against human red blood. Results revealed that the extract and isolated compounds were virtually non-toxicy as compared to the positive control.

Each step of this methodology, from the selection of the plant material to the extraction, separation and isolation of the active constituent, then, evaluated pharmacological properties is followed in pharmacology aspects.

In conclusion, this study allowed highlighting:

- An ethnomedicinal study confirmed the application of *P. scoparius* in traditional medicine by local Algerian for numerous pharmacological properties
- Qualitative and quantitative photochemical analysis revealed that the plant rich in bioactive compounds
- The isolated and structure elucidation of four compounds from stems of *P. scoparius* named: isorhamnetin-3-O-β-apiofuranosyl (1→2)-β-glucopyranoside, D-mannitol, isorhamnetin-3-O-β-glucoside and isorhamnetin-3-O-β-glucopyranosyl-(1→6)-β-glucopyranoside
- High antioxidant capacity of the stem and root extracts for various *in vitro* antioxidant models.
- Both *in vitro* and *in vivo* anti-inflammatory activities along to analgesic activity of CrE of stems and roots showed significantly effects, which may be related with different phytochemicals constituents
- Both *in vitro* and *in vivo* topical anti-inflammatory capacity of *n*-butanol stems extract and the isolated compounds present an excellent anti-inflammatory properties
- The *in vitro* cytotoxicity showed that the percentage lysis of the *n*-butanol stems extract and the isolated compounds was found to be nontoxic.
- The $LD_{50}$ of CrE from stems and roots of *P. scoparius* was higher than 5 g/kg.

The current study is the first report on the toxicological profile, antioxidants, anti-inflammatory and analgesic properties of stems and roots from *P. scoparius*. In addition, isolation of pure compound along the assessment of pharmacological activities was carried for the first time. However, more studies are required include:

- The future spread of the information about ethnopharmacological application of this plant helped people acceptance of phytoterapics

- ✓ The *in vitro* antioxidant results should be validated *in vivo* to develop a potent antioxidant agents from this plant in order to investigate whether its application in inflammation in traditional medicine is justified

- ✓ Other *in vivo* models of anti-inflammatory and analgesic assays are needed.

- ✓ More studies pertaining to the safety and efficacy of this plant are required.

- ✓ Present results lay the preparation for further studies on the molecular mechanisms underlying the pharmacological profile of these plant extracts, isolation and purification of more active principles in each extract as well as clarification of their mode of action

- ✓ These results provided an accurate data of the use of this plant in the folk medicine in Algeria in the quest of finding natural alternatives with increased bioactivity for therapeutic applications.

# REFERENCES

# PAPERS